\documentclass[prx,twocolumn,superscriptaddress]{revtex4-1}

\usepackage{graphicx}
\usepackage{color}
\usepackage[dvipsnames]{xcolor}
\usepackage{amsmath,amssymb,mathrsfs,ulem}
\usepackage{braket}
\usepackage{bbm}
\newcommand{\dunderline}[1]{\underline{\underline{#1}}}

\definecolor{mildblue}{rgb}{.2,0.7,.9}
\definecolor{warmred}{rgb}{.68,0.19,.14}

\definecolor{dante}{rgb}{.2,0.7,.9}

\definecolor{daniel}{rgb}{.0,1,0.5}

\date{\today}                  

\usepackage{array, multirow, bigdelim, makecell} 

\begin{document}



\title{Rational boundary charge in one-dimensional systems with interaction and disorder}

\author{Mikhail Pletyukhov}
\affiliation{Institut f\"ur Theorie der Statistischen Physik, RWTH Aachen, 
52056 Aachen, Germany and JARA - Fundamentals of Future Information Technology}
\author{Dante M. Kennes}
\affiliation{Institut f\"ur Theorie der Statistischen Physik, RWTH Aachen, 
52056 Aachen, Germany and JARA - Fundamentals of Future Information Technology}
\affiliation{Max Planck Institute for the Structure and Dynamics of Matter, Center for Free Electron Laser Science, 22761 Hamburg, Germany}
\author{Kiryl Piasotski}
\affiliation{Institut f\"ur Theorie der Statistischen Physik, RWTH Aachen, 
52056 Aachen, Germany and JARA - Fundamentals of Future Information Technology}
\author{Jelena Klinovaja}
\affiliation{Department of Physics, University of Basel, Klingelbergstrasse 82, 
CH-4056 Basel, Switzerland}
\author{Daniel Loss}
\affiliation{Department of Physics, University of Basel, Klingelbergstrasse 82, 
CH-4056 Basel, Switzerland}
\author{Herbert Schoeller}
\email[Email: ]{schoeller@physik.rwth-aachen.de}
\affiliation{Institut f\"ur Theorie der Statistischen Physik, RWTH Aachen, 
52056 Aachen, Germany and JARA - Fundamentals of Future Information Technology}

\begin{abstract}
We study the boundary charge $Q_B$ of generic semi-infinite one-dimensional insulators with translational invariance and show that non-local symmetries (i.e., including translations) lead to rational quantizations $p/q$ of $Q_B$. In particular, we find that (up to an unknown integer) the quantization of $Q_B$ is given in integer units of $\frac{1}{2}\bar{\rho}$ and $\frac{1}{2}(\bar{\rho}-1)$, where $\bar{\rho}$ is the average charge per site (which is a rational number for an insulator). This is a direct generalization of the known half-integer quantization of $Q_B$ for systems with local inversion or local chiral symmetries to any rational value. Quite remarkably, this rational quantization  remains valid even in the presence of short-ranged electron-electron interactions as well as static random disorder (breaking translational invariance). This striking stability can be traced back to the fact that local perturbations in insulators induce only local charge redistributions. We establish this result with complementary methods including  density matrix renormalization group calculations, bosonization methods, and exact solutions for particular lattice models. Furthermore, for the special case of half-filling $\bar{\rho}=\frac{1}{2}$, we present explicit results in single-channel and nearest-neighbor hopping models and identify Weyl semimetal physics at gap closing points. Our general framework also allows us to shed new light on the well-known rational quantization of soliton charges at domain walls.

\end{abstract}


\maketitle

\section{Introduction} 
\label{sec:introduction}
Charge fractionalization is a striking  phenomenon which emerges in a variety of condensed matter systems of high interest such as the fractional quantum Hall effect \cite{laughlin_prl_83,halperin_prl_84,arovas_schrieffer_wilczek_prl_84,saminadayar_etal_prl_97,picciotto_etal_nature_97}, Luttinger liquids \cite{pham_etal_prb_00,imara_etal_prb_02,steinberg_etal_natphys_08,kamata_etal_natnanotech_14,inoue_etal_prl_14}, and topological insulators \cite{bergholtz_karlhede_prl_05,seidel_etal_prl_05,seidel_lee_prl_06,bergholtz_etal_prb_06,lee_zhang_xiang_prl_07,liang_etal_nature_08,liang_etal_prb_08,rhy_etal_prb_09,goldman_etal_prl_10,klinovaja_loss_prb_15,przysiezna_etal_njp_15}. For its emergence a fundamental mechanism has been established via fractionally charged domain walls separating two systems with the same bulk spectrum but in different topological phases. This was analysed for a one-dimensional (1D) spinless Fermi gas coupled to a bosonic field with broken symmetry \cite{jackiw_rebbi_prd_76} and in polyacetylene chains due to electron-phonon coupling \cite{su_schrieffer_heeger,jackiw_schrieffer_nuclphys_81}. 
This mechanism was further analysed for more general setups \cite{rice_mele_prl_82,kivelson_prb_83} and a simple physical picture was proposed \cite{su_schrieffer_prl_81} to explain the fractional charge unit $\frac{1}{Z}$ via a $Z$-fold degenerate ground state generated by a charge-density wave (CDW) of wavelength $\lambda=Za$ ($a$ is the lattice constant) \cite{heeger_etal_review_RMP_88}. Within continuum field theories \cite{witten_physlett_79,takayama_etal_prb_80,goldstone_wilczek_prl_81,jackiw_semenoff_prl_83} the fractional part of the soliton (or interface) charge $Q_I$ was shown to be given by the Goldstone-Wilczek formula \cite{goldstone_wilczek_prl_81} $Q_I={\delta\alpha\over 2\pi}e\,\text{mod}(e)$, where $\delta \alpha$ is the phase difference between the two CDWs right and left to the interface. This interface charge is of purely topological nature, i.e., independent of the precise parameter values determining the domain wall. In addition, fluctuations of the soliton charges were analysed in continuum and lattice models showing that the fractional charge is a well-defined quantity \cite{kivelson_schrieffer_prb_82,bell_rajaraman_physlettB_82,kivelson_prb_82,bell_rajaraman_nucphysB_83,frishman_horovitz_prb_83,jackiw_etal_nucphysB_83,park_etal_prb_16}.

Besides quantized soliton charges, charge quantization has also been studied at the boundary of topological insulators. Previous works focused on the special case of local inversion or local chiral symmetry as well as on non-interacting and clean systems, where the boundary charge $Q_B$ is quantized in half-integer units. This was shown via the quantization of the Zak-Berry phase $\gamma$ in units of $\pi$ \cite{zak,schnyder_etal_njp_10}, which is related to the boundary charge by $Q_B=-e{\gamma\over2\pi}\,\text{mod}(e)$ \cite{kingsmith_vanderbilt_prb_93,vanderbilt_kingsmith_prb_93,resta,resta_revmodphys_94,marzari_etal_revmodphys_12,vanderbilt_book_2018,ortiz_martin_prb_94,rhim_etal_prb_17,miert_ortix_prb_17}. The quantization of the Zak-Berry phase in the presence of local inversion symmetry has led to the notion of topological crystalline insulators (TCI) \cite{hughes_etal_prb_83,chiu_etal_prb_88,shiozaki_sato_prb_90,alexandradinata_etal_prb_16,trifunovic_brouwer_prb_17,lau_etal_prb_16}, extending the standard classification schemes of topological insulators \cite{schnyder_etal_njp_10,schnyder_etal_prb_08,kitaev_advphys_09,slager_etal_natphys_12,jadaun_etal_prb_13,chiu_etal_prb_13,zhang_kane_mele_prl_13,benalcazar_etal_prb_14,morimoto_furusaki_prb_13,diez_etal_njp_15}, which are based on chiral, time reversal, and particle-hole symmetries only. In addition, combining local symmetries with translations (so-called non-local or non-symmorphic symmetries) new possibilities for TCIs have been predicted for 2D and 3D systems \cite{liu_etal_prb_14,young_kane_prl_15,wang_etal_nature_16,shiozaki_etal_prb_16,chang_etal_natphys_17,wieder_etal_science_18}. 

The central topic of the present work is the generalization of the half-integer quantization of the boundary charge $Q_B$ to any rational value ${p\over q}$ in generic 1D insulators. We will relate the occurrence of a rational boundary charge (RBC) to {\it non-local} symmetries, i.e. symmetries which can not be defined within the space of a single unit cell (see Appendix~\ref{app:local_nonlocal} for a summary of the precise definitions). Quite remarkably, in the presence of these symmetries, we will show that RBC can be easily understood in terms of the universal changes of $Q_B$ under translations and local inversion. Stability of RBC is demonstrated since these transformation laws are not violated in the presence of short-ranged electron-electron interaction or static random disorder. 

\begin{table}[t]
\centering
\begin{tabular}{l l l } 
 \toprule
 &&\\[-0.25cm]
  & $\overline{Q}_B\;\;  \text{mod}(e)$& Transformation  \\ [1ex] 
  \hline
   &&\\[-0.25cm]
$T_n$& $\phantom{-}{Q}_B +n\bar{\rho}$ &translation $\left|m\right\rangle\to\left|m+n\right\rangle$\\[1ex]  
$\Pi$& $-{Q}_B$ &unit-cell-local inversion\\[1ex]  
$U$& $\phantom{-}{Q}_B$ &site-local (anti-)unitary\\[1ex]  
${\color{mildblue}\Pi_n}=T_nU\Pi$\;\;\;\;\;\;\;& ${\color{mildblue}-{Q}_B+n\bar{\rho}}\;\;\;\;\;\;\;$ &unitary/time-reversal\\[1ex]  
${\color{warmred}S_n}=T_{-n}U$& $\phantom{-}{\color{warmred}{Q}_B-n\bar{\rho}}$ &chiral/particle-hole\\[1ex]  
 \hline
\end{tabular}
\caption{Transformations $Q_B\rightarrow\bar{Q}_B$ of the boundary charge under the elementary transformations $T_n$ (translation by $n$ lattice sites towards the boundary), $\Pi$ (local inversion within each unit cell, where the unit cell is defined as the one starting at the boundary of a semi-infinite system), $U$ (unitary or anti-unitary operations within the channel space of a single site), and combinations of these transformations defining the operations $\Pi_n=T_n U \Pi$ and $S_n=T_{-n}U$. Except $U$ all transformation rules are $\text{mod}(e)$ due to the possible occurrence of edge states. $\bar{\rho}=e\frac{\nu}{Z}$ is the average charge per site which is a rational number in the insulating regime. $\nu$ denotes the number of filled bands and $Z$ is the number of lattice sites of a unit cell (with $N_c$ channels per site). If $U$ is unitary (anti-unitary), $\Pi_n$ and $S_n$ are unitary (anti-unitary) operations. Highlighted in color are the transformation rules that need to be compared between Tables \ref{table:transformations} and \ref{table:symmetries} to obtain the rational boundary charge. } 
\label{table:transformations}
\end{table}

Our general and unified framework for the RBC is set up for generic 1D tight-binding models with any size $Za$ of the unit cell and any number $N_c$ of channels per site (like, e.g., described by spin, several orbitals, etc.). Importantly, our analytical study is not based on the representation of $Q_B$ in terms of the Zak-Berry phase. It relies exclusively on the fundamental property of insulators, namely that local perturbations by external fields lead only to local charge redistributions, i.e., the corrections beyond a typical length scale $\xi$ are exponentially small. This does not affect the fractional part of the boundary charge, since $Q_B$ is defined via a macroscopic average on scales much larger than $\xi$. In addition, bound states (localized at the boundary) crossing the chemical potential due to the local perturbation, can only lead to a change of $Q_B$ by an integer number. This local behavior, also known as the {\it nearsightedness principle} \cite{kohn_prl_76,prodan_kohn_pnas_05} (NSP), is responsible for the universal features of topological insulators and is also connected to the bulk-boundary correspondence \cite{fidkowski_etal_prl_11,mong_shivamoggi_prb_11,gurarie_prb_11,essin_gurarie_prb_11,fukui_etal_jphysjpn_12,yu_wu_xie_nucphysb_17,rim_bardarson_slager_prb_18,silveirinha_prx_19}. Furthermore, the same principle is responsible for charge pumping \cite{thouless_prb_83,niu_thouless_jphysA_84} and leads to exponential localization of the excess density at boundaries and interfaces \cite{kallin_halperin_prb_84}, such that $Q_B$ and $Q_I$ are well-defined quantities for insulators. Since the NSP is also valid for interacting and disordered systems, we can expect high
stability of our results against short-ranged electron-electron interactions and static random disorder, as long as the gap of the insulator in which the chemical potential lies is not closed. Besides the general expectation of stability we will also support the NSP by numerical calculations based on density matrix renormalization group (DMRG) methods in the presence of short-ranged electron-electron interactions, by exact diagonalizations for static random disorder, and by analytical results for the particular example of two coupled non-interacting single-channel and nearest-neighbor hopping models. In addition, we establish analytically the stability against short-ranged interactions in continuum models by using the bosonization method.

\begin{table}[t]
\centering
\begin{tabular}{l l l l } 
 \toprule
  &&\\[-0.25cm]
& Symmetry & $\overline{Q}_B$& Quantization  \\[1.ex] 
  \hline
   &&\\[-0.25cm]
${\color{mildblue}\Pi_n}$\;\;&$\Pi_nH\Pi_n^\dagger=H$\;\;\;\;\;\;\;& {\color{mildblue}${Q}_B $} &\hspace{-0.5em}\rdelim\}{3}{*}[${}\;$\fbox{${{Q}_B=\frac12 n\bar\rho\;\;  \text{mod}\left(\frac{e}{2}\right)}$ }]\\[1ex] 
${\color{warmred}S_n}$\;\;&$S_nHS_n^\dagger=-H$& \hspace{-2.1em}{\color{white}\rdelim\}{2}{*}[${}\;$ {\color{warmred}$-{Q}_B\; \text{mod}(e)$}]}  &\\
\phantom{$S_n$}&\;\;\;$\&\,\frac{1}{2}$-filling&\\[1ex] 
 \hline
\end{tabular}
\caption{Transformations $Q_B\rightarrow\bar{Q}_B$ of the boundary charge if the Hamiltonian $H$ fulfils a symmetry by either commuting with $\Pi_n$ or anti-commuting with $S_n$. For the symmetry $S_n$ one needs in addition half-filling $\bar{\rho}=e\frac{N_c}{2}$. If $U$ is a unitary (anti-unitary) operation, $\Pi_n$ is a unitary (time-reversal) symmetry and $S_n$ is a chiral (particle-hole) symmetry. For $n=0$ the operations $\Pi_0$ and $S_0$ are local symmetries acting within the space of a single unit-cell. For $n\ne 0$ they are non-local symmetries, see Appendix~\ref{app:local_nonlocal} for our conventions to distinguish between local and non-local symmetries. By identifying the values for $\bar{Q}_B$ from Tables~\ref{table:transformations} and \ref{table:symmetries} one obtains straightforwardly the rational quantization values $Q_B = \frac{1}{2}n\bar{\rho}\,\text{mod}(\frac{e}{2})$. Highlighted in color are the transformation rules that need to be compared between Tables \ref{table:transformations} and \ref{table:symmetries} to obtain the rational boundary charge} 
\label{table:symmetries}
\end{table}

To sketch our derivation of RBC we have summarized our main results in the two tables \ref{table:transformations} and \ref{table:symmetries}. Table \ref{table:transformations} lists the transformation of $Q_B$ under basic operations, in particular under translation $T_n$ by $n$ lattice sites towards the boundary and under local inversion $\Pi$ within each unit cell. Together with site-local transformations $U$ leaving the boundary charge invariant, we define the two central operations $\Pi_n=T_n U \Pi$ and $S_n=T_{-n}U$, which are non-local for $n\ne 0$ since they contain a translation. Table \ref{table:symmetries} states the change of $Q_B$ when the Hamiltonian has an explicit non-local symmetry by either commuting with $\Pi_n$ or anti-commuting with $S_n$ (for the symmetry $S_n$ one needs in addition half-filling). Comparing the transformations of $Q_B$ under $\Pi_n$ and $S_n$ stated in the two tables (marked in the same color) we arrive at the central result of RBC
\begin{align}
\label{eq:QB_rational}
Q_B = {1\over 2}n\bar{\rho}\quad\text{mod}\Big({e\over 2}\Big) \,.
\end{align} 
Here, $\bar{\rho}=e\frac{\nu}{Z}$ is the average charge per site which is a rational multiple of $e$ in the insulating regime, where $\nu$ is the number of filled bands and $Z$ is the number of sites of a unit cell. The trivial case of a local symmetry is $n=0$ leading to the well-known $\frac{1}{2}$-integer quantization of $Q_B$. Taking all integers $n\ne 0$ into account we find that $Q_B$ can take all rational quantization values. Due to the $\text{mod}(\frac{e}{2})$-part our quantization rule shows that $Q_B$ can always be written as a combination of multiples of two elementary quantization units: $\frac{1}{2}\bar{\rho}$ and $\frac{1}{2}(\bar{\rho}-1)$.

Besides the presence of a non-local symmetry of the Hamiltonian as stated in table~\ref{table:symmetries}, the central part of the proof of RBC is the transformation of $Q_B$ under the two elementary operations $T_n$ and $\Pi$ of translations and local inversion according to table~\ref{table:transformations}. They are the basic ingredients for the understanding of {\it all} universal properties of $Q_B$ and are given by  
\begin{align}
\label{eq:QB_trafo_translation}
Q_B &\xrightarrow{T_n} \bar{Q}_B= Q_B + n\bar{\rho} \quad \text{mod}(e) \,,\\
\label{eq:QB_trafo_inversion}
Q_B &\xrightarrow{\Pi} \bar{Q}_B= - Q_B \quad \text{mod}(e) \,.
\end{align} 
Both transformation laws will be shown in this work to be ultimately related to the NSP which demonstrates their stability under short-ranged electron-electron interaction and static random disorder. Eq.~(\ref{eq:QB_trafo_translation}) is a straightforward consequence of charge conservation since on average the charge $n\bar{\rho}$ is moved into the boundary when the translation is described via an adiabatic process (up to an integer charge arising from edge states crossing the chemical potential during the adiabatic process). It has been used in a variety of recent works on single-channel and nearest-neighbor tight-binding models to analyse the universal phase-dependence $Q_B(\varphi)$ as a function of a phase $\varphi$ describing a continuous shift of the lattice towards the boundary \cite{park_etal_prb_16,thakurathi_etal_prb_18,pletyukhov_etal_short,pletyukhov_etal_long,mara_citro_ortix_prb_15}. Eq.~(\ref{eq:QB_trafo_inversion}) is a fundamental transformation which is based on the simple observation that local inversion of a semi-infinite system with a left boundary turns it to the same semi-infinite system with a right boundary \cite{park_etal_prb_16}. Simple arguments based on the NSP will then show that the sum of these two boundary charges must be zero up to an integer charge. 

The fact that the two elementary transformations (\ref{eq:QB_trafo_translation}) and (\ref{eq:QB_trafo_inversion}) together with a  non-local symmetry property of the Hamiltonian under $\Pi_n$ or $S_n$ explain both the RBC and its stability under interactions and disorder in a straightforward way is the central result of this work. We note that the interaction and the disorder have to fulfil the non-local symmetry property as well for our proof to be valid. Whereas homogeneous density-density interaction terms are obviously invariant under translations $T_n$, local inversion $\Pi$, and site-local transformations $U$, it might not be the case for some fixed disorder configuration. However, for random disorder the symmetry will be fulfilled on average and our numerical results confirm that the RBC is stable in the presence of random disorder. In addition, when the density-density interaction is not homogeneous, it is expected that it follows precisely the symmetry constraints imposed by the modulation of the on-site potentials and hopping terms.

Interestingly, we will show that the two universal transformation laws (\ref{eq:QB_trafo_translation}) and (\ref{eq:QB_trafo_inversion}) shed also new light on the quantization of the interface charge. If the two lattices right and left to the interface have the same bulk spectrum and are only shifted relative to each other by $\delta n$ sites, they are connected by the transformation $T_{\delta n}\Pi$. Therefore, if the two lattices are not connected to each other, one finds from (\ref{eq:QB_trafo_translation}) and (\ref{eq:QB_trafo_inversion}) that the boundary charge $Q_B^L$ of the left lattice is related to the boundary charge $Q_B^R$ of the right lattice by $Q_B^L=-Q_B^R+\delta n\bar{\rho}\,\text{mod}(e)$. Using the NSP, turning on some local coupling between the two lattices does not change the fractional part of the interface charge such that $Q_I$ follows generically from
\begin{align}
\label{eq:QI_ICT}
Q_I &= Q_B^L + Q_B^R \quad \text{mod}(e) \\
\label{eq:QI_GW}
&=\delta n\bar{\rho}\quad\text{mod}(e)\,.
\end{align}
As a result, we have extended the Goldstone-Wilczek formula to a discrete lattice and, in addition, have shown that it is stable in the presence of short-ranged electron-electron interactions and static random disorder. 

We expect our results of RBC to be observable in experiments. Recent experiments in cold atom systems demonstrated that it is possible to get direct access to the boundary charge via the Zak-Berry phase \cite{atala_etal_nature_2013} and to measure soliton charges of the SSH model \cite{meier_etal_nature_comm_16}. In addition, concrete proposals for measuring topological solitons in solid state systems have been made such as carbon nanotubes \cite{efroni_etal_prl_17}, graphene nanoribbons \cite{yang_nanomat_19}, and Rashba nanowires \cite{klinovaja_stano_loss_prl_12,rainis_etal_prl_14}. Here, scanning single-electron transistor techniques allow for the direct measurement of local charges \cite{yoo_etal_science_97,tessmer_etal_nature_98,finkelstein_etal_science_00,ben-shach_etal_prb_15,xio_etal_natphys_15}. Moreover, the occurrence of interface states due to the quantization of the Zak phase has been measured in phononic crystals \cite{xio_etal_natphys_15}. Besides these materials promising candidates to measure the boundary charge are quantum dot arrays as proposed in Ref.~\cite{park_etal_prb_16}. Similiar to cold atom systems, quantum dot arrays have the particular advantage of control over all parameters to implement on demand the specific non-local symmetries needed for RBC.

As an interesting application of our general framework we will discuss the case of a single-channel (i.e., $N_c=1$) and nearest-neighbor hopping model. Of particular interest is the case of half-filling, $\bar{\rho}={e\over 2}$, where one obtains from (\ref{eq:QB_rational}) the two universal quantization classes $Q_B={e\over 2}\, \text{mod}({e\over 2})$ and $Q_B={e\over 4}\,\text{mod}({e\over 2})$. The first is the usual one present also for local inversion or local chiral symmetries. In contrast, the second was to the best of our knowledge not discussed before and is only possible for a {\it non-local} symmetry. We present  an explicit  realization of these classes in terms of a lattice model with equal hopping amplitudes and a harmonic modulation of the on-site potentials. Controlling the offset of the modulation by a phase-variable $\varphi$ this model is of relevance for the integer quantum Hall effect (IQHE) (where $\varphi$ corresponds to the transverse quasimomentum in a 2D quantum Hall setup) \cite{thakurathi_etal_prb_18}. At half-filling (where $Z$ must be even to open a gap), the model has the non-local chiral symmetry $S_{Z/2}=T_{-Z/2}U$ with $U|m\rangle=(-1)^m|m\rangle$ ($|m\rangle$ denotes the state at lattice site $m$). According to (\ref{eq:QB_rational}) this leads to the quantization values $Q_B={Z\over 8}e\,\text{mod}({e\over 2})$, i.e., the two quantization classes in terms of ${e\over 2}$ or ${e\over 4}$ are obtained for $Z=4,8,12,\dots$ and for $Z=2,6,10,\dots$, respectively. The model has the advantage that the chiral symmetry $S_{Z/2}$ holds for any phase $\varphi$ of the potential modulation. This pins $Q_B(\varphi)$ to quantized plateaus which change abruptly by $\pm{e\over 2}$ at gap closing points. This leads to Weyl semimetal physics since edge modes connecting the gap closing points play the role of Dirac arcs. Despite the fact that in this case the Chern number vanishes (leading to zero Hall current), we find a non-trivial quantization effect of the boundary charge $Q_B$. 

For single-channel and nearest-neighbor hopping models with very small gaps we will also set up a low-energy continuum theory via a Dirac model in $1+1$ dimensions with a complex gap parameter $\Delta=|\Delta|e^{i\alpha}$, in analogy to the study of interface charges via the Goldstone-Wilczek formula. For a semi-infinite system we obtain the following universal result for the boundary charge:
\begin{align}
\label{eq:QB_low_energy}
Q_B = {\alpha\over 2\pi}e + {e\over 4}\quad\text{mod}(e)\,.
\end{align}
Interestingly, the boundary charge is insensitive to the gap size and reveals a linear behavior as function of the phase of the gap parameter. If the original lattice model is at half-filling and pure potential modulation is realized (as discussed above for the Weyl case), we find the symmetry $S_{Z/2}$ for any phase $\varphi$ of the CDW \cite{com_CDW_phase}. In the corresponding continuum model we will show that the parameter $\alpha$ is obtained from the interference of two paths connecting right and left movers at the two Fermi points $\pm k_F$. For the two classes $Z=4,8,12,\dots$ and $Z=2,6,10,\dots$, we find a phase-locking effect pinning $\alpha$ to odd or even multiples of ${\pi\over 2}$, respectively. These two cases correspond to the two quantization classes of $Q_B$ in terms of ${e\over 2}$ or ${e\over 4}$, respectively, proving consistency of the continuum theory with our general framework. 

The paper is organized as follows. Section~\ref{sec:general} is devoted to the general framework to realize RBC in generic 1D insulators. We describe the model and the definition of boundary and interface charges in Section~\ref{sec:H_QB_QI}, and the RBC is analysed in Section~\ref{sec:frac_quant}. The basic transformation laws (\ref{eq:QB_trafo_translation}) and (\ref{eq:QB_trafo_inversion}) are derived in Sections~\ref{sec:QB_translation} and \ref{sec:QB_inversion}. Combining the two transformations we find the Goldstone-Wilczek formula (\ref{eq:QI_GW}) for the interface charge. In Section~\ref{sec:symmetries} we combine the transformation laws with non-local symmetries of the Hamiltonian and prove the central result (\ref{eq:QB_rational}). We proceed in Section~\ref{sec:application} with an application of our general framework to the case of single-channel and nearest-neighbor hopping models. In Section~\ref{sec:weyl} we describe Weyl semimetal physics at half-filling and discuss the connection to the IQHE by analysing the universal phase-dependence of the boundary charge, the Diophantine equation, and the Hall current in the presence of a gap closing. An effective low-energy description of boundary and interface charges in terms of a continuum Dirac model in $1+1$ dimensions is provided in Section~\ref{sec:low_energy}. The derivation of the model in the noninteracting and interacting case is given in Sections~\ref{sec:dirac_noninteracting} and \ref{sec:dirac_interacting}, respectively. The universal formula (\ref{eq:QB_low_energy}) for the boundary charge and the Goldstone-Wilczek formula for the interface charge are presented in Section~\ref{sec:dirac_QB_QI}. We close with a summary and outlook in Section~\ref{sec:summary}.

Throughout this work we use units such that $\hbar=e=a=1$.

\section{General framework} 
\label{sec:general}

In this section we describe the general framework to derive the central transformations (\ref{eq:QB_trafo_translation}) and (\ref{eq:QB_trafo_inversion}) of the boundary charge $Q_B$ under translations  and local inversion, respectively. We identify the non-local symmetries $\Pi_n$ and $S_n$ leading to the rational quantization values (\ref{eq:QB_rational}) of the boundary charge. In addition, we show that the Goldstone-Wilczek formula (\ref{eq:QI_GW}) for interface charges follows straightforwardly from the transformation laws.

\subsection{Hamiltonian, boundary and interface charges}
\label{sec:H_QB_QI}

\begin{figure}
\centering
\includegraphics[width= 0.75\columnwidth]{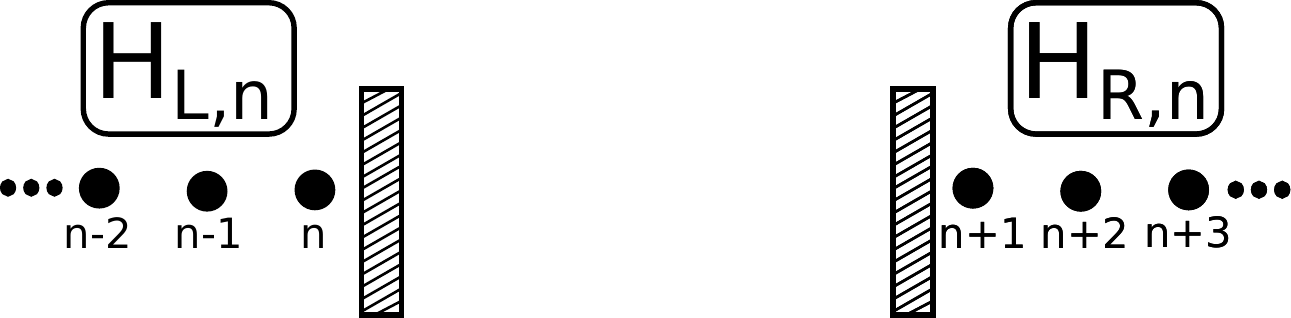} 
 \caption{ Illustration of the semi-infinite Hamiltonians $H_{R,n}$ (right figure) and $H_{L,n}$ (left figure). The Hamiltonian $H_{R,n}$ is obtained from the bulk Hamiltonian $H_{\text{bulk}}$ by starting it at site $m=n+1$, whereas $H_{L,n}$ is obtained by ending $H_{\rm bulk}$ at site $m=n$.} 
 \label{fig:hamiltonian}
\end{figure}

We consider a generic translationally invariant tight-binding model in 1D with arbitrary short-ranged hopping and $N_c$ channels per lattice site. For the infinite (bulk) case, the single-particle Hamiltonian reads
\begin{align}
\label{eq:H_bulk}
H_{\text{bulk}} = \sum_{m=-\infty}^\infty \sum_{\delta=-\delta_{\text{max}}}^{\delta_{\text{max}}}
\uline{c}^\dagger_{m+\delta}\,\dunderline{h}_m(\delta)\,\uline{c}_m \,.
\end{align}
Here, $m$ denotes the lattice site index and $\delta_{\text{max}}$ is the range of the hopping. The components $c_{m\sigma}$ of the $N_c$-dimensional vector $\uline{c}_m$ annihilate an electron on site $m$ in channel $\sigma=1,\dots,N_c$. $\dunderline{h}_m(\delta)$ is a generic $N_c\times N_c$-matrix describing the coupling between the channels of lattice site $m$ and $m+\delta$.  Translational invariance and hermiticity require the properties
\begin{align}
\label{eq:translational_invariance}
\dunderline{h}_m(\delta) &= \dunderline{h}_{m+Z}(\delta) \,,\\
\label{eq:hermiticity}
\dunderline{h}^\dagger_m(\delta) &= \dunderline{h}_{m+\delta}(-\delta)\,,
\end{align}
where $Z$ is the number of lattice sites of a unit cell. Semi-infinite systems extending to the right or left side are defined by the Hamiltonians $H_{R/L,n}$ by starting/ending the bulk Hamiltonian $H_{\text{bulk}}$ at site $m=n+1$ and $m=n$, respectively, see Fig.~\ref{fig:hamiltonian} for illustration. Since the numeration of the sites is arbitrary one can alternatively label the sites by $m=1,2,\dots$ for the right lattice and by $m=0,-1,-2,\dots$ for the left lattice, and formally define in compact notation
\begin{align}
\label{eq:H_R}
H_{R,n} &= \sum_{m=1}^\infty \sum_{\delta \atop m+\delta > 0 ,|\delta|\le \delta_{\text{max}}}
\uline{c}^\dagger_{m+\delta}\,\dunderline{h}_{m+n}(\delta)\,\uline{c}_m \,,\\
\label{eq:H_L}
H_{L,n} &= \sum_{m=-\infty}^0\sum_{\delta \atop m+\delta\le 0,|\delta|\le \delta_{\text{max}}}
\uline{c}^\dagger_{m+\delta}\,\dunderline{h}_{m+n}(\delta)\,\uline{c}_m \,.
\end{align}

If an interface is studied between the two semi-infinite systems $H_{R,n}$ on the right side and $H_{L,n'}$ on the left side, we take any short-ranged coupling $V_I$ defined within the interface region $M_L\le m \le M_R$ (with $|M_{L,R}|\sim O(Z)$) 
\begin{align}
\label{eq:V_I}
V_I = \sum_{M_L\le m,m' \le M_R} 
\uline{c}^\dagger_{m}\,\dunderline{v}_{mm'}\,\uline{c}_{m'} \,,
\end{align}
and define the total Hamiltonian by
\begin{align}
\label{eq:H_I}
H_{I,nn'} = H_{R,n} + H_{L,n'} + V_I \,.
\end{align}
For the study of the stability against short-ranged electron-electron interactions, we take a density-density interaction of the form 
\begin{align}
\label{eq:coulomb}
V_{ee} = {1\over 2} \sum_{m\ne m'} u(m-m') \hat{\rho}(m)\hat{\rho}(m')\,, 
\end{align}
where $u(m)=u(-m)$ is only nonzero for $|m|\le m_{\text{max}}\sim O(1)$ and $\hat{\rho}(m)=\uline{c}^\dagger_m \uline{c}_m$. If a semi-infinite lattice is studied, the sum over the sites is restricted to the corresponding regions. In the presence of an interface the short-ranged electron-electron interaction between the sites of the left and right lattices is also a coupling term which can be included alternatively in $V_I$ by adding these many-particle terms to the interface coupling. 

We will also test stability with respect to static random disorder. In this case we include a quenched onsite disorder 
\begin{align}
\label{eq:disorder}
H_{\rm dis} =  \sum_{m} d(m) \hat{\rho}(m)\,, 
\end{align}
where $d(m)$ is drawn from a uniform distribution $d(m)\in[-d/2,d/2)$. Of course, more complicated (channel-dependent or hopping) forms of disorder can be considered, which, however, will not change the main thrust of the arguments presented here.

\begin{figure}
\centering
\includegraphics[width= \columnwidth]{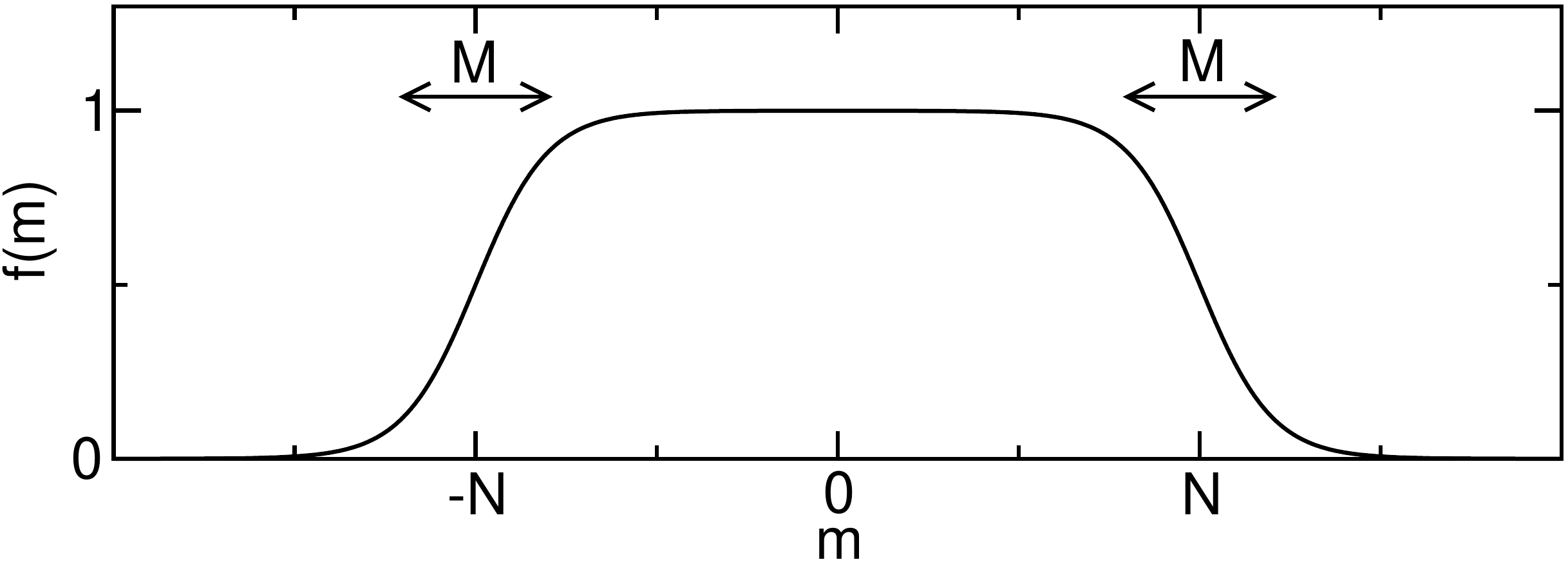} 
 \caption{Sketch of the envelope function $f(m)$ with $N\gg M \gg Z,\xi$.}
\label{fig:envelope}
\end{figure}

In the insulating regime, where the excess density falls off exponentially on scale $\xi$ into the bulk \cite{kallin_halperin_prb_84,pletyukhov_etal_long} starting from a boundary or interface, the observables of interest are the boundary charges $Q_{B,n}^R$ and $Q_{B,n'}^L$ of the semi-infinite systems described by $H_{R,n}$ and $H_{L,n'}$, respectively, and the interface charge $Q_{I,nn'}$ of the Hamiltonian $H_{I,nn'}$, defined by a macroscopic average on scales much larger than $Z$ and $\xi$ 
\begin{align}
\label{eq:QB_R} 
Q_{B,n}^R &= \sum_{m=1}^\infty (\rho(m)-\bar{\rho}) f(m) \,,\\
\label{eq:QB_L} 
Q_{B,n'}^L &= \sum_{m=-\infty}^0 (\rho(m)-\bar{\rho}) f(m) \,,\\
\label{eq:QB_I} 
Q_{I,nn'} &= \sum_{m=-\infty}^\infty (\rho(m)-\bar{\rho}) f(m) \,.
\end{align}
Here, $\rho(m)=\langle\hat{\rho}(m)\rangle$ is the total charge at site $m$ in a grand canonical ensemble with respect to the corresponding Hamiltonians $H_{R,n}$, $H_{L,n'}$, and $H_{I,nn'}$, respectively. We assume zero temperature and the chemical potential $\mu$ to lie in some given band gap of the insulator. Further, $\bar{\rho}$ is the average charge per site for the translationally invariant bulk Hamiltonian (\ref{eq:H_bulk}) defined by
\begin{align}
\label{eq:rho_bar}
\bar{\rho} = {1\over Z}\sum_{j=1}^Z \rho_{\text{bulk}}(j) = {\nu\over Z}\,,
\end{align}
where $\nu$ is the filling factor defined as the number of occupied bands. The function $f(m)$ is the envelope function of a charge measurement probe falling off smoothly from unity to zero, see Fig.~\ref{fig:envelope}. 

To simplify notations, if no index $n$ is displayed, we assume implicitly $n=0$, i.e.,
\begin{align}
\label{eq:n_zero_H}
& H_{R/L}\equiv H_{R/L,0} \,\,,\,\, H_I\equiv H_{I,00}\,,\\
\label{eq:n_zero_Q}
& Q_B^{R/L}\equiv Q_{B,0}^{R/L} \,\,,\,\, Q_I\equiv Q_{I,00} \,.
\end{align}
We note that we have not used this convention in the introductory part (to avoid too many sub-indices at the beginning) where we denoted by $Q_B^{R/L}$ and $Q_I$ the boundary and interface charges for the systems $H_{R/L,n}$ or $H_{I,nn'}$ under consideration. Furthermore, we used implicitly the convention $Q_B\equiv Q_B^R$ in the introductory part.  

Due to the NSP, it is expected that $Q_{I,nn'}$ is independent of the coupling $V_I$ (up to an integer). For particular examples we demonstrate this in Appendix~\ref{app:NSP_DMRG} via exact diagonalization and DMRG calculations in the presence of static random disorder or short-ranged electron-electron interaction. Moreover, we show in Appendix~\ref{app:NSP_1_channel} analytically that $Q_{I,nn'}$ is independent of the size of a single link between two noninteracting and 1-channel nearest-neighbor hopping models. As a result we find [see Eq.~(\ref{eq:QI_ICT})] that the interface charge is the independent sum of the boundary charges of the left and right lattice
\begin{align}
\label{eq:QI_ICT_nn'}
Q_{I,nn'} = Q_{B,n}^R + Q_{B,n'}^L \quad \text{mod}(1)\,.
\end{align}
In addition, we conclude that also the boundary charge does not change, when a local perturbation is added close to the boundary. In Section~\ref{sec:frac_quant} we will furthermore show that the interface charge depends only on the relative difference $\delta n=n'-n$, i.e.,
\begin{align}
\label{eq:QI_relative}
Q_{I,nn'} = Q_{I,\delta n} \quad,\quad \delta n=n-n'\,,
\end{align}
see the Goldstone-Wilczek formula Eq.~(\ref{eq:QI_GW}) or Eq.~(\ref{eq:goldstone_wilczek}) below.

\subsection{Rational quantization of boundary charge}
\label{sec:frac_quant}

In this section we show how the boundary charge transforms under translations and local inversion according to Eqs.~(\ref{eq:QB_trafo_translation}) and (\ref{eq:QB_trafo_inversion}). Together with certain symmetry requirements of the Hamiltonian we will determine the rational quantization values for the boundary charge given by Eq.~(\ref{eq:QB_rational}). In addition, we show that the Goldstone-Wilczek formula Eq.~(\ref{eq:QI_GW}) for the interface charge is based only on the fundamental transformation laws of the boundary charge. For the special case of noninteracting and clean systems, we show in the Supplemental Material (SM) \cite{SM} that the same results can also be obtained from the transformation of the Zak-Berry phase under translations and local inversion.

\subsubsection{Translations}
\label{sec:QB_translation}

\begin{figure}
\centering
\includegraphics[width= \columnwidth]{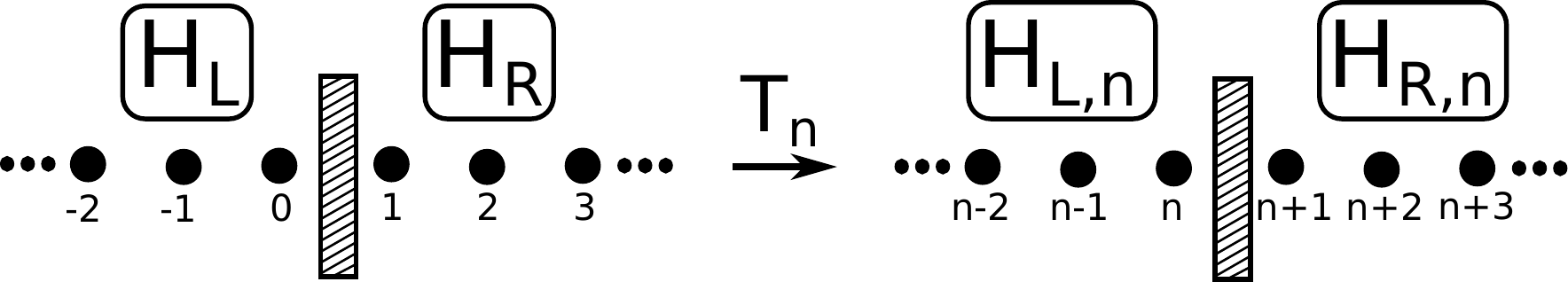} 
 \caption{Illustration of the translation by $n$ lattice sites for the left and right semi-infinite lattice. In both cases the lattice is moved to the left. This means that for $H_R\rightarrow H_{R,n}$ the lattice is moved by $n$ sites {\it towards} the boundary whereas, for $H_L\rightarrow H_{L,n}$, the lattice is moved by $n$ sites {\it away} from the boundary. As a consequence, the boundary charge increases or decreases by $n\bar{\rho}$, respectively.}
 \label{fig:translation}
\end{figure}

A translation $T_n=T_{-n}^\dagger$ of the lattice by $n$ sites to the left is defined by the operation 
\begin{align}
\label{eq:translation}
\dunderline{h}_m(\delta)\xrightarrow{T_n} \dunderline{h}_{m+n}(\delta)\,.
\end{align}
Applying this to Eqs.~(\ref{eq:H_R}) and (\ref{eq:H_L}), we evidently get 
\begin{align}
\label{eq:H_RL_translation}
H_R \xrightarrow{T_n} H_{R,n}\quad,\quad
H_L \xrightarrow{T_n} H_{L,n}\,,
\end{align}
as illustrated in Fig.~\ref{fig:translation}. Performing the transformation (\ref{eq:translation}) via an adiabatic process, the lattice is shifted as a whole by $n$ sites to the left, i.e., due to charge conservation, the average charge $n\bar{\rho}$ will move into the left boundary of the right lattice and will move out of the right boundary of the left lattice. Since the boundary charges $Q_{B,n}^{R/L}$ are defined as a macroscopic average via Eqs.~(\ref{eq:QB_R}) and (\ref{eq:QB_L}), we get 
\begin{align}
\label{eq:QB_RL_translation}
Q_{B,n}^{R/L} = Q_B^{R/L} \pm n\bar{\rho} \quad\text{mod}(1)\,.
\end{align}
Together with Eq.~(\ref{eq:H_RL_translation}) this provides the following transformation of the boundary charges under translation:
\begin{align}
\label{eq:QB_RL_translation_trafo}
Q_B^{R/L} \xrightarrow{T_n} \bar{Q}_B^{R/L} = Q_B^{R/L} \pm n\bar{\rho} \quad\text{mod}(1)\,,
\end{align}
which proves Eq.~(\ref{eq:QB_trafo_translation}). These relations are exact and do not depend on the presence or absence of short-ranged electron-electron interaction or random disorder, see Appendix~\ref{app:NSP_DMRG} and bosonization studies in Section~\ref{sec:dirac_QB_QI}. They are based on the same arguments as charge pumping \cite{thouless_prb_83,niu_thouless_jphysA_84} and have been extensively used recently for noncyclic adiabatic processes to analyse the universal average slope of the phase-dependence of the boundary charge \cite{park_etal_prb_16,thakurathi_etal_prb_18,pletyukhov_etal_short,pletyukhov_etal_long}. The unknown integer arises since bound states (at the boundaries) can cross the chemical potential during the adiabatic process leading to discrete integer jumps of the boundary charge. 

Alternatively, Eq.~(\ref{eq:QB_RL_translation}) can also be derived directly from the NSP. Since local perturbations at the boundary do not change $Q_B^R$ (up to an integer) we can add to $H_R$ an infinitely high potential on the first $n$ sites such that $\rho(m)=0$ for $m=1,2,\dots,n$. This leaves for the boundary charge from Eq.~(\ref{eq:QB_R}) only the contribution $-n\bar{\rho}$ for the first $n$ sites and $Q_{B,n}^R$ for the rest. Using the invariance due to the NSP this gives $Q_B^R=Q_{B,n}^R-n\bar{\rho}$ leading to Eq.~(\ref{eq:QB_RL_translation}). In an analogous way one can prove Eq.~(\ref{eq:QB_RL_translation}) for $Q_B^L$ by starting from $H_{B,n}^L$ and putting an infinite potential on the last $n$ sites. 

\subsubsection{Local inversion}
\label{sec:QB_inversion}

\begin{figure}
\centering
\includegraphics[width= \columnwidth]{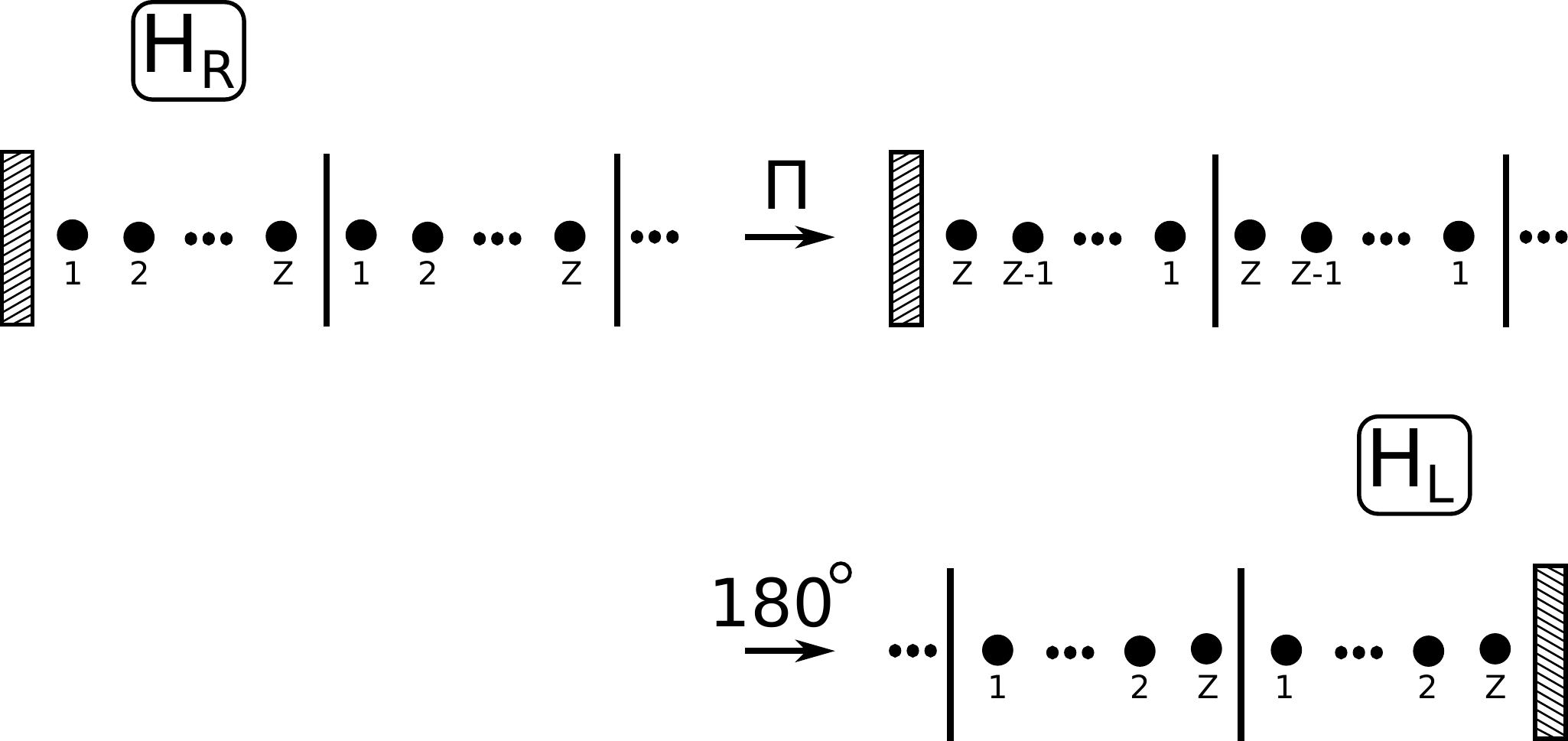} 
 \caption{Illustration of local inversion by indicating the lattice sites of the unit cells. Inverting the lattice sites within a unit cell via the interchange $m\leftrightarrow Z-m+1$ and turning the system by $180^\circ$ one finds that $H_R$ is transformed to $H_L$.} 
 \label{fig:inversion}
\end{figure}

Local inversion $\Pi$ is defined as a symmetry operation performing an inversion in each unit cell separately, where the unit cell is defined such that it starts at the boundary. Since the unit cell defined in this way depends on the index $n$ of the Hamiltonians $H_{R/L,n}$ we discuss the case $n=0$ in the following (see the discussion at the end of Section~\ref{sec:symmetries} for transformations defined with respect to other choices of the unit cell). In this case the unit cell starting at the boundary consists of the sites $m=1,\dots,Z$ and is identical to the one for the bulk Hamiltonian $H_{\text{bulk}}$. In this subspace the transformation $\Pi$ is formally defined by
\begin{align}
\label{eq:inversion}
\dunderline{h}_m(\delta)\xrightarrow{\Pi} \dunderline{h}_{Z-m+1}(-\delta)
= [\dunderline{h}_{Z-m+1-\delta}(\delta)]^\dagger\,,
\end{align}
where we have used the hermiticity condition (\ref{eq:hermiticity}) in the last equality. Using the periodicity condition (\ref{eq:translational_invariance}), this defines $\Pi$ for all $m$. Applying an inversion turns the semi-infinite system $H_R$ with a left boundary obviously to the semi-infinite system $H_L$ with a right boundary
\begin{align}
\label{eq:H_RL_inversion}
H_R \xrightarrow{\Pi} H_L\,,
\end{align}
as illustrated in Fig.~\ref{fig:inversion}. Taking the Hamiltonian (\ref{eq:H_I}) without any coupling $V_I=0$ between the left and right lattice, the interface charge is obviously given by $Q_I=Q_B^L+Q_B^R$. On the other hand, taking for $V_I$ exactly the coupling corresponding to the bulk Hamiltonian (\ref{eq:H_bulk}) we get a translational invariant lattice everywhere with a periodic bulk density $\rho_{\text{bulk}}(m)=\rho_{\text{bulk}}(m+Z)$. The corresponding interface charge $Q_{I,\text{bulk}}$ vanishes since
\begin{align}
\nonumber
Q_{I,\text{bulk}}&=\sum_m(\rho_{\text{bulk}}(m)-\bar{\rho})f(m)\\
\nonumber
&\hspace{-1cm}
=\sum_n\sum_{j=1}^Z(\rho_{\text{bulk}}(j)-\bar{\rho})f(Zn+j)\\
\nonumber
&\hspace{-1cm}
=\sum_nf(Zn)\sum_{j=1}^Z(\rho_{\text{bulk}}(j)-\bar{\rho}) \\
\label{eq:QI_bulk}
&\hspace{-0.5cm}
+ \sum_nf'(Zn)\sum_{j=1}^Z(\rho_{\text{bulk}}(j)-\bar{\rho})j = 0\,, 
\end{align}
where we have used $f(Zn+j)\approx f(Zn)+f'(Zn)j$ in the second step and $\sum_nf'(Zn)\approx \int dx f'(Zx)=0$ together with the definition (\ref{eq:rho_bar})
of $\bar{\rho}$ in the third step. The approximations become exact in the limit of infinite parameters $N$ and $M$ defining the smoothness of the envelope function via Fig.~\ref{fig:envelope}. Finally, due to the NSP, the interface charge can only change $\text{mod}(1)$ when switching on $V_I$, leading to 
\begin{align}
\label{eq:QB_RL_relation}
Q_B^L + Q_B^R = 0 \quad \text{mod}(1)\,.
\end{align}
Together with (\ref{eq:H_RL_inversion}) this provides the following transformation of the boundary charges under inversion 
\begin{align}
\label{eq:QB_RL_inversion_trafo}
Q_B^{R/L} \xrightarrow{\Pi} \bar{Q}_B^{R/L} = - Q_B^{R/L} \quad\text{mod}(1)\,,
\end{align}
which proves Eq.~(\ref{eq:QB_trafo_inversion}). The universal relation (\ref{eq:QB_RL_relation}) has also been found recently in Ref.~\cite{pletyukhov_etal_long} for the special case of a noninteracting single-channel and nearest-neighbor hopping model, where the unknown integer has been specified for a single band. Again, we emphasize that it is also valid in the presence of short-ranged electron-electron interaction and random disorder since we only used the NSP to derive it, see also Appendix~\ref{app:NSP_DMRG} and bosonization studies in Section~\ref{sec:dirac_QB_QI}.

We note that, by inserting the relations (\ref{eq:QB_RL_translation}) and  (\ref{eq:QB_RL_relation}) into the formula (\ref{eq:QI_ICT_nn'}) for the interface charge, we obtain straightforwardly the Goldstone-Wilczek formula (\ref{eq:QI_GW}) for a discrete lattice
\begin{align}
\label{eq:goldstone_wilczek}
Q_{I,nn'} = (n-n') \,\bar{\rho} \quad \text{mod}(1)\,,
\end{align}
with $\delta n\equiv n-n'$. We conclude that the charge quantization at interfaces separating regions in different topological phases with the same bulk spectrum is fundamentally related to the NSP and the transformation laws of the boundary charge under translation and local inversion. This provides a very elegant proof of the Goldstone-Wilczek formula and shows at the same time that it is stable under short-ranged electron-electron interaction and static random disorder.

\subsubsection{Non-local symmetries}
\label{sec:symmetries}

If the transformations (\ref{eq:QB_RL_translation_trafo}) and (\ref{eq:QB_RL_inversion_trafo}) of the boundary charges $Q_B^{R/L}$ under translations and local inversion are combined with explicit symmetries of the Hamiltonian, one can straightforwardly prove the rational quantization of the boundary charge. We note that symmetries are always defined with respect to the bulk Hamiltonian $H_{\text{bulk}}$ and we discuss the consequences for the boundary charges $Q_B^{R/L}$, where the unit cell starting/ending at the boundary is the same as the one chosen for $H_{\text{bulk}}$. We define {\it local} symmetries by transformations acting only in the space of a single unit cell as they are used in the usual $10$-fold classification schemes of TIs \cite{schnyder_etal_njp_10,schnyder_etal_prb_08,kitaev_advphys_09,slager_etal_natphys_12,jadaun_etal_prb_13,chiu_etal_prb_13,zhang_kane_mele_prl_13,benalcazar_etal_prb_14,morimoto_furusaki_prb_13,diez_etal_njp_15} in terms of local chiral, time-reversal, and particle-hole symmetries. The case when the local symmetry is defined with respect to another choice of the unit cell is always discussed separately if relevant, see also the detailed discussion at the end of this section. For a summary of the precise definitions and our conventions to distinguish between local and non-local symmetries we refer to Appendix~\ref{app:local_nonlocal}.

First, we note that all (anti-)unitary transformations $U_m=U_{m+Z}$ acting only in the channel space of lattice site $m$ commute with the operator $\uline{c}^\dagger_m\uline{c}_m$ and, therefore, leave the boundary charge invariant. This are transformations defined by 
\begin{align}
\label{eq:U_trafo}
\dunderline{h}_m(\delta) \quad &\xrightarrow{U} \quad U_{m+\delta}^\dagger\,\dunderline{h}_m(\delta)\,U_m\,.
\end{align}

Secondly, we define two fundamental operations $\Pi_n$ and $S_n$ by combining $U$ with local inversion and translations
\begin{align}
\label{eq:Pi_S_n}
\Pi_n = T_n U \Pi \quad,\quad S_n = T_{-n} U \,.
\end{align}
$\Pi_n$ and $S_n$ are defined in such a way that the boundary charge transforms according to 
\begin{align}
\label{eq:QB_trafo_Pi_n}
Q_B^{R/L} &\xrightarrow{\Pi_n} \bar{Q}_B^{R/L} = 
- Q_B^{R/L} \pm n\bar{\rho} \quad \text{mod}(1) \,,\\
\label{eq:QB_trafo_S_n}
Q_B^{R/L} &\xrightarrow{S_n} \bar{Q}_B^{R/L} =  
Q_B^{R/L} \mp n\bar{\rho} \quad \text{mod}(1) \,,
\end{align} 
where we have used (\ref{eq:QB_RL_translation_trafo}) and (\ref{eq:QB_RL_inversion_trafo}).

\begin{figure}
\centering
\includegraphics[width= \columnwidth]{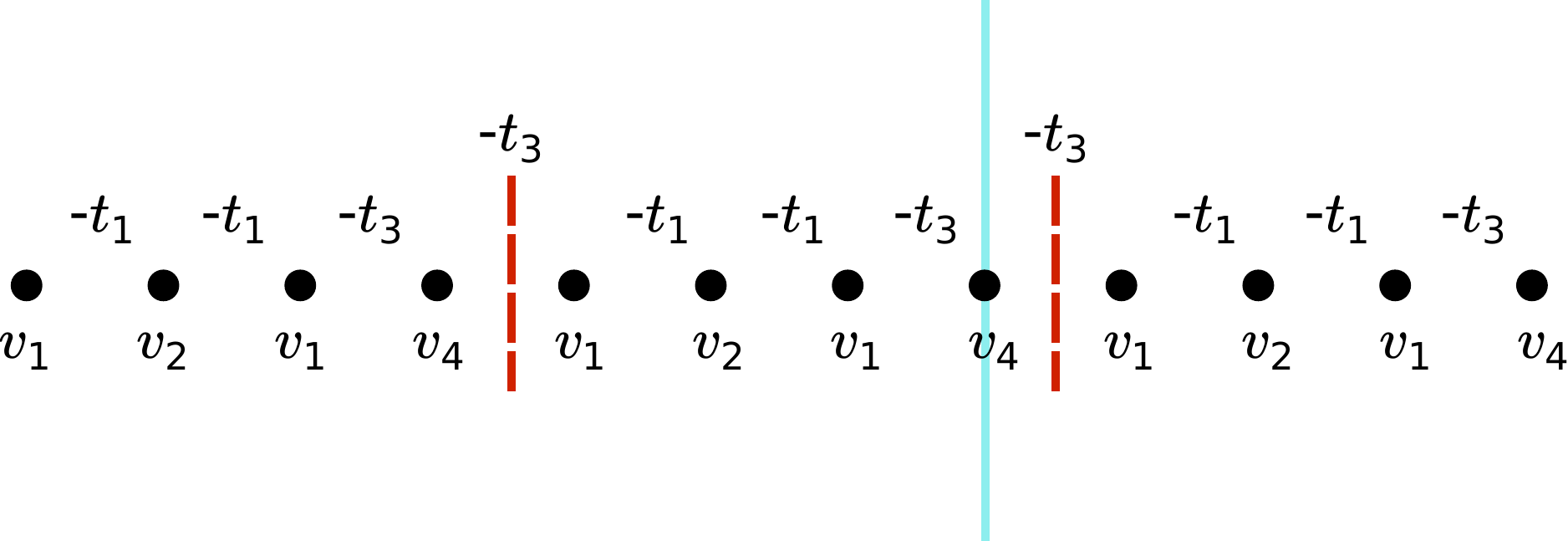} 
\includegraphics[width= \columnwidth]{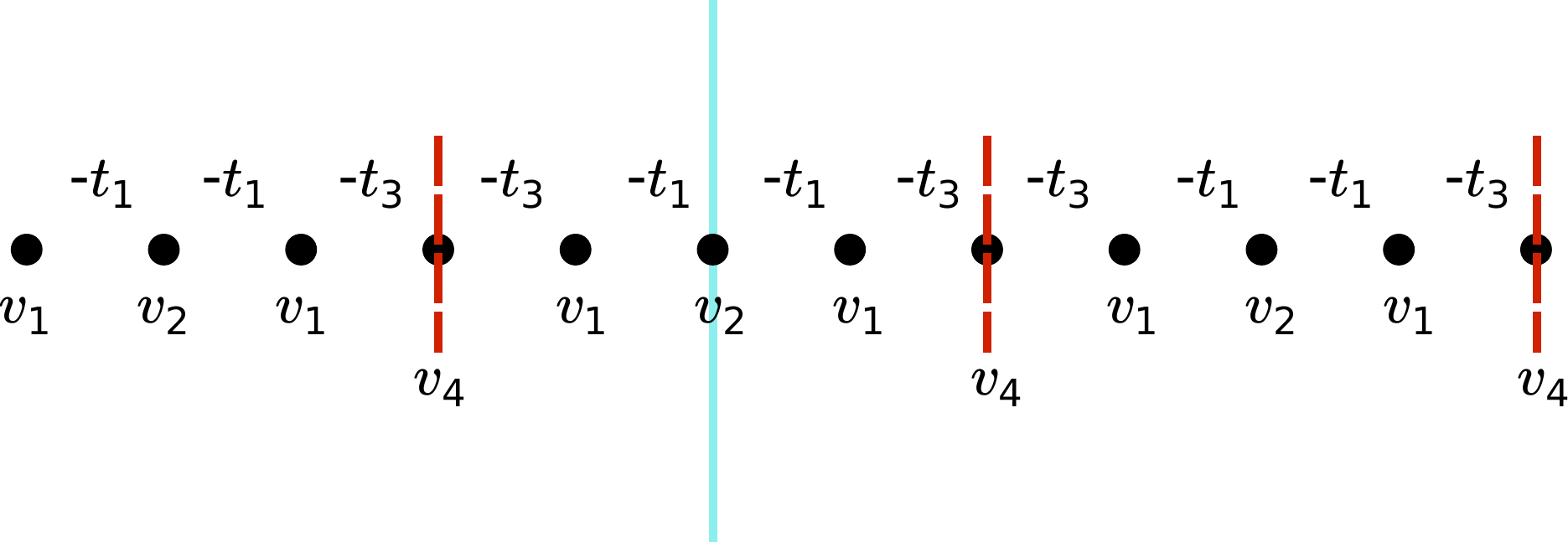} 
 \caption{Visualization of the nonlocal inversion symmetry $\Pi_n$ for $n=1$ and $Z=4$ for a nearest-neighbor hopping model with one channel per site. $v_j$ denote the on-site potentials and $-t_j$ are the hoppings. The red dashed vertical lines indicate the boundaries between the unit cells and the blue solid vertical line is the symmetry axis of inversion symmetry. In the lower figure the unit cell is redefined such that the symmetry becomes a local one. However, this is not possible for a tight-binding model since the unit cell contains only one half of a site at the boundaries.}
\label{fig:nonlocal}
\end{figure}

Finally, we assume that the Hamiltonian either commutes with $\Pi_n$ or anticommutes with $S_n$
\begin{align}
\label{eq:symmetry}
\dunderline{h}_m(\delta) \xrightarrow{\Pi_n} \dunderline{h}_m(\delta)\quad\text{or}\quad
\dunderline{h}_m(\delta) \xrightarrow{S_n} - \dunderline{h}_m(\delta)\,.
\end{align}
Using Eqs.~(\ref{eq:translation}), (\ref{eq:inversion}), and (\ref{eq:U_trafo}) this requires one of the following symmetry conditions
\begin{align}
\nonumber
\Pi_n:\quad\dunderline{h}_m(\delta) &=U_{Z-m-n+1-\delta}^\dagger\cdot\\
\label{eq:Pi_n_condition}
&\cdot [\dunderline{h}_{Z-m-n+1-\delta}(\delta)]^\dagger\,U_{Z-m-n+1}\,,\\
\label{eq:S_n_condition}
S_n:\quad\dunderline{h}_m(\delta) &=
-U_{m-n+\delta}^\dagger\,\dunderline{h}_{m-n}(\delta)\,U_{m-n}\,.
\end{align}
In contrast to all previous relations this defines a certain {\it non-local} symmetry which the Hamiltonian has to fulfil. As we will show below the symmetry implies rational quantization values of the boundary charge. When $U$ is a unitary transformation, (\ref{eq:Pi_n_condition}) denotes a unitary symmetry $\Pi_n$ and (\ref{eq:S_n_condition}) a chiral symmetry $S_n$. Similarly, when $U$ is an anti-unitary transformation, (\ref{eq:Pi_n_condition}) denotes a time-reversal symmetry $\Pi_n$ and (\ref{eq:S_n_condition}) a particle-hole symmetry $S_n$. Both symmetries are {\it non-local} for $n\ne 0$ since they involve translations. A special case is $\Pi_n$ which, except for $Z$ even and $n$ odd, can be turned into a local symmetry but with respect to another choice of the unit cell. This follows since $\Pi_n$ is an inversion symmetry around the axis $m={1\over 2}(Z-n+1)$ (for $n$ even) or $m=Z-{1\over 2}(n-1)$ (for $n$ odd).  This leads to a local site-inversion symmetry for $Z$ odd, and to a local bond-inversion symmetry for both $Z$ and $n$ even. However, for $Z$ even and $n$ odd, we obtain a site-inversion symmetry but the corresponding unit cell contains only half of the sites at the boundaries which is not possible for tight-binding models, see Fig.~\ref{fig:nonlocal} for illustration. Using the terminology of symmorphic and non-symmorphic symmetries one should call $\Pi_n$ a symmorphic symmetry depending on the quasimomentum $k$ (for $Z$ even and $n$ odd, sometimes also called unconventional non-symmorphic symmetry) since it returns to the same lattice site when applied twice \cite{yang_etal_prb_17,brzezicki_cuoco_prb_17,zhang_etal_prm_18,malard_etal_prb_18}, whereas $S_n$ is a non-symmorphic symmetry \cite{zhao_schnyder_prb_16}. 

For the (anti-)unitary symmetry $\Pi_n$, the boundary charge can obviously not change since the Hamiltonian is invariant. For the particle-hole or chiral symmetry $S_n$ each eigenstate $|\psi\rangle$ of the Hamiltonian with energy $\epsilon$ has a corresponding eigenstate $S_n|\psi\rangle$ with negative energy $-\epsilon$. Therefore, the operation $S_n$ transforms the boundary charge $Q_B^-$ (with $Q_B\equiv Q_B^{R/L}$) from all states with negative energy to the boundary charge $Q_B^+$ from all states with positive energy. When all states are filled we get $\rho(m)=\bar{\rho}=N_c$ giving zero boundary charge, i.e., $Q_B^+ + Q_B^- = 0$.  At half-filling $\bar{\rho}={N_c\over 2}$, the chemical potential is located somewhere in the gap near zero energy.  In this case the boundary charge is given by $Q_B=Q_B^-\,\text{mod}(1)$, where the integer arises from edge states contributing an integer number to the boundary charge. As a consequence, up to an integer, we find that the boundary charge changes sign under $S_n$ at half-filling. Therefore, we get the following relations
\begin{align}
\label{eq:QB_trafo_Pi_n_symmetry}
Q_B^{R/L} &\xrightarrow{\Pi_n} \bar{Q}_B^{R/L} = Q_B^{R/L} \,,\\
\label{eq:QB_trafo_S_n_symmetry}
Q_B^{R/L} &\xrightarrow{S_n} \bar{Q}_B^{R/L} =  - Q_B^{R/L} \, \text{mod}(1),\,\, \text{for}\quad\bar{\rho}={N_c\over 2}\,.
\end{align} 
Taking these equations together with (\ref{eq:QB_trafo_Pi_n}) and (\ref{eq:QB_trafo_S_n}), we arrive for both symmetries at the following rational quantization values of the boundary charge
\begin{align}
\label{eq:QB_RL_rational}
Q_B^{R/L} = \pm {1\over 2}n\bar{\rho}\quad\text{mod}\Big{(}{1\over 2}\Big{)} \,,
\end{align}
which proves our central result (\ref{eq:QB_rational}). 

Eq.~(\ref{eq:QB_RL_rational}) shows that the quantization of $Q_B^{R/L}$ can always be written as
\begin{align}
\label{eq:QB_RL_rational_units}
Q_B^{R/L} = n_1{1\over 2}\bar{\rho} + n_2{1\over 2}(\bar{\rho}-1) \,,
\end{align}
with some integers $n_1$ and $n_2$. As a consequence, the quantization units of the boundary charge are given by ${1\over 2}\bar{\rho}$ and ${1\over 2}(\bar{\rho}-1)$ in contrast to the quantization unit $\bar{\rho}$ of interface charges, see Eq.~(\ref{eq:goldstone_wilczek}). Furthermore, the quantization of $Q_B^{R/L}$ requires certain symmetries of the bulk Hamiltonian, whereas the quantization of $Q_I$ is only related to a symmetry relation between the lattices left and right to the interface, which are connected by $T_{\delta n}\Pi$. For the special case $n=0$, one recovers from (\ref{eq:QB_RL_rational}) the known quantization of the boundary charge in half-integer units \cite{zak,schnyder_etal_njp_10} in the presence of local symmetries.

One can ask the delicate question what happens if a local symmetry is not defined with respect to the unit cell starting at the boundary but with respect to any choice of the unit cell. This is equivalent to the question how the boundary charge changes when the semi-infinite system is cut off at  a different site such that it starts with site $n'+1$ (for $H_R$) or ends with site $n'$ (for $H_L$). According to the transformation (\ref{eq:QB_RL_translation_trafo}) of $Q_B^{R/L}$ under translations this leads to a shift of $Q_B^{R/L}$ by $\pm n'\bar{\rho}\,\text{mod}(1)$. Therefore, for systems with local symmetries with respect to any definition of the unit cell, one can realize all quantization values 
\begin{align}
\label{eq:QB_RL_rational_local}
Q_B^{R/L} = \pm n'\bar{\rho}\quad\text{mod}\Big{(}{1\over 2}\Big{)} \,,
\end{align}
where the $\text{mod}({1\over 2})$ contribution stems from local symmetries defined with respect to the unit cell starting at the boundary. This provides the quantization units $\bar{\rho}$ and ${1\over 2}$. In contrast, the presence of symmetries which are {\it non-local} with respect to any choice of the unit cell  provides the interesting new possibility of a realization of the quantization unit ${1\over 2}\bar{\rho}$. 

For given $Z$ but arbitrary filling factor $\nu=1,\dots,N_c Z-1$, one can also analyse the conditions if $n{1\over 2}\bar{\rho}\,\text{mod}({1\over 2})$ can lead to new rational quantization values compared to $n'\bar{\rho}\,\text{mod}({1\over 2})$. This is the case if the equation 
\begin{align}
\label{eq:nonlocal_vs_local}
{1\over 2}n \bar{\rho} = n'\bar{\rho} - m'{1\over 2}
\end{align}
can not be solved for any integers $n'$ and $m'$. To analyse this we insert $\bar{\rho}={\nu\over Z}$ and find that (\ref{eq:nonlocal_vs_local}) is equivalent to 
\begin{align}
\label{eq:nonlocal_vs_local_1}
n \nu = 2n'\nu - m'Z
\end{align}
For $n$ even, this equation is solved by $n'={n\over 2}$ and $m'=0$. For $n$ odd and $Z$ odd, it is solved by $n'={1\over 2}(Z+n)$ and $m'=\nu$. However, for $n$ odd and $Z$ even, (\ref{eq:nonlocal_vs_local_1}) can not be solved for $\nu$ odd. Since the equation does not change when $Z$ and $\nu$ are multiplied with the same integer $l$, we conclude that new quantization classes occur for non-local symmetries if 
\begin{align}
\label{eq:new_classes}
Z = 2q l\quad,\quad \nu = (2p-1)l \,,
\end{align}
with two integers $q$ and $p$. Since ${1\over 2}\bar{\rho}$ changes by ${1\over 2}$ if $\nu$ changes by $Z$, we conclude that the new quantization values of the boundary charge due to non-local symmetries are given by 
\begin{align}
\label{eq:new_units}
Q_B\rightarrow n{2p-1\over 4q}\,,
\end{align}
with $n$ odd, $q=1,2,\dots$, and $p=1,2,\dots,q$.

\section{Application to single-channel and nearest-neighbor hopping models}
\label{sec:application}

In this section we discuss a concrete realization of the rational quantization values of the boundary charge in the special case of single-channel and nearest-neighbor hopping models. We present the discussion of discrete lattice models in Section~\ref{sec:weyl}, where we will concentrate on Weyl semimetal physics occurring at half-filling and discuss the relevance for the IQHE. In Section~\ref{sec:low_energy} we will develop a low-energy theory in terms of a Dirac model in $1+1$ dimensions and discuss the universal properties of the boundary and interface charge exactly for the noninteracting case and via the bosonization method in the presence of short-ranged electron-electron interactions. Throughout this section we concentrate on the boundary charge $Q_B^R$ and denote it by $Q_B$.

\subsection{Weyl semimetal physics at half-filling}
\label{sec:weyl}

In Refs.~\cite{pletyukhov_etal_short,pletyukhov_etal_long} the case of single-channel and nearest-neighbor tight-binding models has been studied analytically for any value of $Z$ and for generic modulations of the on-site potentials 
$v_m=h_m(0)$ and the hoppings $t_m=-h_m(1)$ (which can all be chosen real and positive $t_m>0$, see Appendix A of Ref.~\cite{pletyukhov_etal_long}). In these references the universal phase-dependence of the boundary charge 
$Q_B(\varphi)$ has been studied for arbitrary $2\pi$-periodic functions $v_m(\varphi)$ and $t_m(\varphi)$. The special case
\begin{align}
\label{eq:potential_modulation}
v_m(\varphi) = -2V \cos\left({2\pi\over Z}m + \varphi\right) \quad,\quad t_m(\varphi) = t 
\end{align}
has been considered in Refs.~\cite{park_etal_prb_16,thakurathi_etal_prb_18}, in particular due to the relevance for the IQHE, see the detailed discussion in Ref.~\cite{thakurathi_etal_prb_18}. In this case, $\varphi$ can be interpreted as the transverse quasimomentum in a 2D quantum Hall setup \cite{thakurathi_etal_prb_18} and $Z$ corresponds to the magnetic length. Whereas Ref.~\cite{thakurathi_etal_prb_18} has discussed small filling factors $\nu$ with finite Chern number, the particular interest in this work is the case of half-filling, $\nu={Z\over 2}$ and $\bar{\rho}={1\over 2}$, where $Z$ is even to open a gap. Due to (\ref{eq:QB_RL_rational}), the following two universal quantization classes are possible for $Q_B\equiv Q_B^R$ at half-filling in case certain symmetries are fulfilled 
\begin{align}
\label{eq:QB_quantization_half_filling}
 Q_B= {1\over2}\,\text{mod}\Big{(}{1\over 2}\Big{)}\quad\text{or}\quad Q_B={1\over4}\,\text{mod}\Big{(}{1\over 2}\Big{)}\,.
\end{align}
The first ${1\over 2}$-class is the usual known one which occurs also in the presence of local inversion or local chiral symmetries. The second ${1\over 4}$-class is a novel one which requires essentially non-local symmetries. For half-filling, we will explain in the following that it is possible that the quantization of $Q_B(\varphi)$ persists for all phases $\varphi$ and Weyl semimetal physics occurs with $Q_B$ jumping by $\pm {1\over 2}$ at gap closing points. This is of relevance for the IQHE. Whereas the Chern number and, therefore, the Hall current vanishes, the boundary charge shows an interesting quantization feature. 

In Appendix~\ref{app:single_channel_nearest_neighbor} the symmetry conditions (\ref{eq:Pi_n_condition}) and (\ref{eq:S_n_condition}) are explicitly evaluated for the special case of single-channel and nearest-neighbor hopping models. For the symmetry $\Pi_n=T_nU\Pi$ one obtains $U_m=1$ and the condition
\begin{align}
\label{eq:Pi_condition_special}
v_m = v_{Z-m-n+1}\quad,\quad t_m = t_{Z-m-n} \,,
\end{align}
whereas for $S_n=T_{-n}U$ one finds $U_m=(-1)^m$ together with
\begin{align}
\label{eq:S_condition_special}
v_m = - v_{m-n}\quad,\quad t_m = t_{m-n} \,.
\end{align}
The non-local chiral symmetry $S_n$ has the advantage that it can be fulfilled for all phases $\varphi$, whereas $\Pi_n$ leads to a rational quantization of the boundary charge only at certain values of $\varphi$. Therefore, we concentrate in the following on $S_n$, where an interesting application relevant for the IQHE at half-filling can be formulated. 

Applying the symmetry condition (\ref{eq:S_condition_special}) twice one finds $v_m=v_{m-2n}$ implying that $Z=2n$ defines the wavelength of the modulation which is identical to the number of sites of the unit cell (the hopping has the wavelength $Z/2$). Therefore, the translation $T_{-n}=T_{Z/2}$ shifts the lattice by half of the unit cell length, typical for non-symmorphic symmetries.
For $n={Z\over 2}$ and $\bar{\rho}={1\over 2}$, we get from (\ref{eq:QB_RL_rational}) the quantization values
\begin{align}
\label{eq:QB_quantization_Zhalf}
Q_B = {Z\over 8}\quad\text{mod}\Big{(}{1\over 2}\Big{)}\,,
\end{align}
leading to the ${1\over 2}$-class for $Z=4,8,12,\dots$ and to the novel ${1\over 4}$-class for $Z=2,6,10,\dots$. 

\begin{figure}
\centering
 \includegraphics[width=\columnwidth]{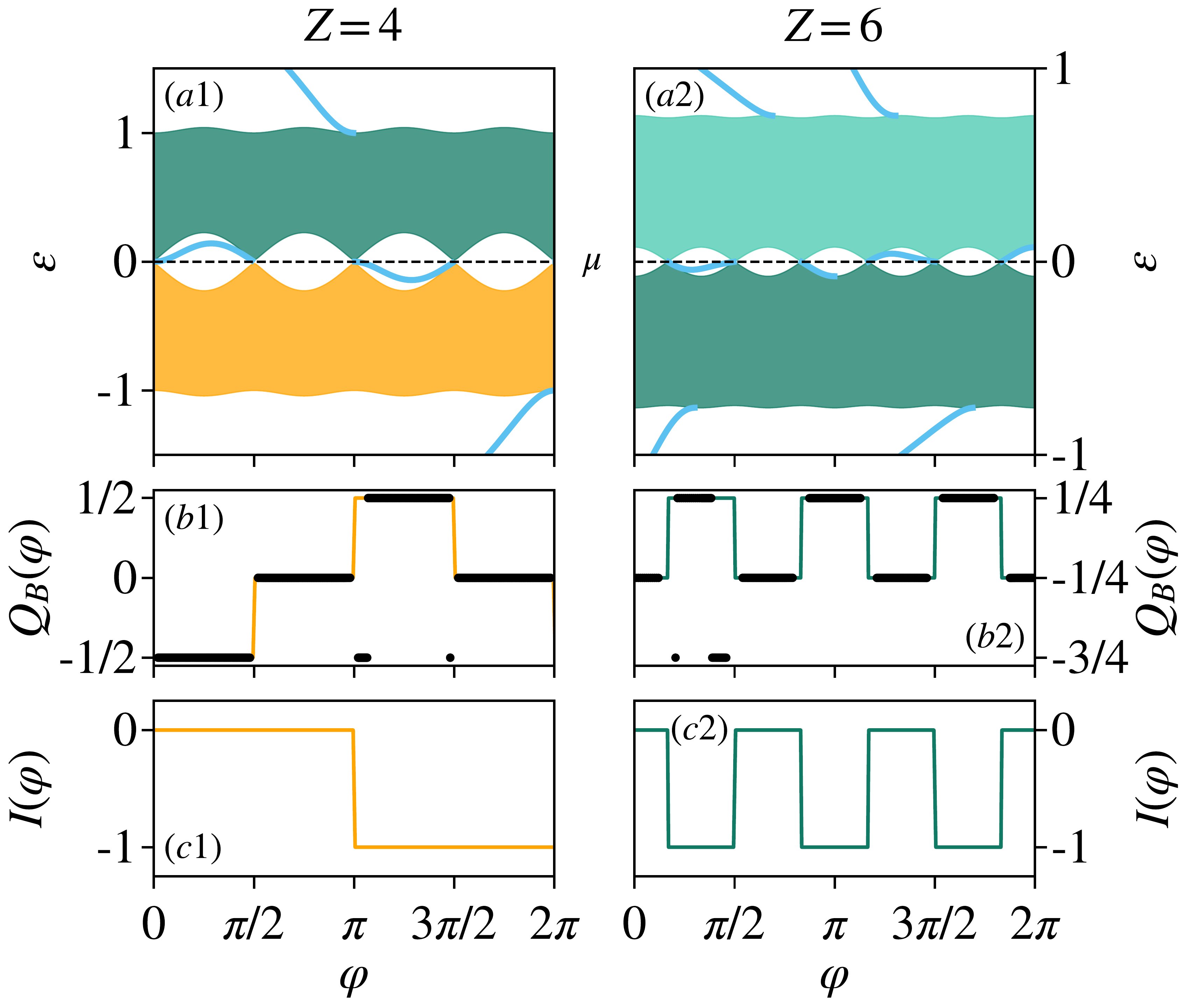}
 \caption{
   Phase-dependence of (a) band structure, (b) boundary charge $Q_B$, and (c) invariant $I(\varphi)=Q_B(\varphi+{2\pi\over Z})-Q_B(\varphi)-\bar{\rho}$ for the model (\ref{eq:potential_modulation}) with $V=0.5$ and $t=1$ at half-filling $\bar{\rho}={1\over 2}$, and $Z=4,6$ (left/right panel), calculated for a semi-infinite system. In (a) the bands $\alpha={Z\over 2},{Z\over 2}+1$ are shown together with edge modes connecting the gap closing points, where $Q_B$ jumps by $\pm {1\over 2}$. The invariant is quantized to $I\in\{-1,0\}$. Due to the nonlocal chiral symmetry $v_m(\varphi)=-v_{m\pm {Z\over 2}}(\varphi)$, quantization of $Q_B$ occurs in units of ${1\over 2}$ for $Z=4$, whereas for $Z=6$ we get $Q_B={1\over 4} \text{mod}(1/2)$. The black symbols (mainly overlaying the lines) in (b) show the case with random disorder for a finite system for additional staggered onsite-disorder drawn from a uniform distribution $(-0.025,0.025]$ for a finite system of $Z\cdot 10^5$ lattice sites.}
\label{fig:weyl}
\end{figure}

For $n={Z\over 2}$, a concrete realization of (\ref{eq:S_condition_special}) for all phases is given by the pure potential modulation model (\ref{eq:potential_modulation}) with constant hopping. Other more complicated realizations are also possible but do not lead to qualitative differences. This model has the advantage that a phase shift of $\varphi$ by ${2\pi\over Z}$ shifts the lattice by one site towards the boundary, i.e., $Q_B$ changes by $\bar{\rho}\,\text{mod}(1)$ according to (\ref{eq:QB_RL_translation}). This must be a half-integer for $\bar{\rho}={1\over 2}$. Furthermore, since $Q_B(\varphi)$ is quantized for all $\varphi$ and since edge states crossing the chemical potential during the shift can change $Q_B$ only by an integer value, we conclude that there must be necessarily a gap closing point in any phase interval of size ${2\pi\over Z}$. Between the gap closing points $Q_B$ is quantized due to the symmetry and edge modes connecting the gap closing points play the role of Dirac arcs, see Figs.~\ref{fig:weyl}(a1,a2). Therefore, we call this the {\it Weyl case}. At a gap closing point $Q_B$ jumps by $\pm{1\over 2}$ such that (\ref{eq:QB_RL_translation}) is fulfilled, see Figs.~\ref{fig:weyl}(b1,b2). This is also demonstrated in Figs.~\ref{fig:weyl}(c1,c2), where we show the integer invariant $I(\varphi)$, defined by
\begin{align}
\label{eq:invariant}
I(\varphi) &= \Delta Q_B(\varphi)- {\nu\over Z} \in\{-1,0\}\,,\\
\label{eq:phase_change}
\Delta Q_B(\varphi) &= Q_B\left(\varphi+{2\pi\over Z}\right) - Q_B(\varphi)\,.
\end{align}
According to Refs.~\cite{pletyukhov_etal_short,pletyukhov_etal_long} this invariant fulfils the topological constraint $I\in\{-1,0\}$ due to charge conservation of particles and holes. We note that this property is not changed when a gap closing point appears during the shift of the lattice by one site. The phase-dependence of the model parameters can always be chosen such that no gap closing appears during the shift without changing the parameters before and after the shift, see Ref.~\cite{pletyukhov_etal_long}.

One can also generalize the universal form of $Q_B(\varphi)$, derived in Refs.~\cite{park_etal_prb_16,thakurathi_etal_prb_18,pletyukhov_etal_short,pletyukhov_etal_long}, to the case of gap closings. When the gap is non-zero for all phases, the form is given by 
\begin{align}
\label{eq:QB_form}
Q_B(\varphi) = f(\varphi)+{M_--M_+\over 2\pi}\varphi + F(\varphi)\,,
\end{align}
where $f(\varphi)$ is a nonuniversal smooth ${2\pi\over Z}$-periodic function, and 
\begin{align}
\label{eq:F}
F(\varphi)=\sum_{\sigma=\pm}\sum_{i=1}^{M_\sigma}\sigma\Theta(\varphi-\varphi_i^{\sigma})
\end{align} 
describes the discrete jumps of $Q_B$ by $\pm 1$ when edge states cross the chemical potential at $\varphi_i^{\pm}$ from above or below, respectively. $M_\pm$ denotes the total number of edge states moving below/above the chemical potential when the phase changes by $2\pi$. The average slope from the linear term determines the Chern number $C_\nu=M_--M_+$ 
\cite{thouless_etal_prl_82,dana_jpc_85,kohmoto_prb_89_jpsj_92,hatsugai_prb_93,pletyukhov_etal_long}. Moreover, Eqs.~(\ref{eq:invariant}) and (\ref{eq:QB_form}) imply the relation 
\begin{align}
\label{eq:diophantine}
C_\nu = \nu - s_\nu Z\,,
\end{align}
where $s_\nu=\Delta F(\varphi) - I(\varphi)$ is a phase-independent integer characteristic for each gap. This relation is equivalent to the Diophantine equation \cite{dana_jpc_85,kohmoto_prb_89_jpsj_92,hatsugai_prb_93}, a central relation for the bulk-boundary correspondence of the IQHE.  

The universal form (\ref{eq:QB_form}) remains valid in the presence of gap closing points for the Weyl case, with the only difference that the jumps of $Q_B$ have size $\pm {1\over 2}$ at a gap closing since the charge of the edge state is distributed symmetrically among the two bands \cite{pletyukhov_etal_long}. Thus, Eq.~(\ref{eq:QB_form}) remains the same, we only have to add a factor ${1\over 2}$ in Eq.~(\ref{eq:F}) for the terms in the sum corresponding to the jumps at gap closings and, correspondingly, 
count only the contribution $\pm {1\over 2}$ to $M_\pm$. As a consequence, the Diophantine equation (\ref{eq:diophantine}) remains also valid in the Weyl case, but $s_\nu$ becomes half-integer. E.g., for Fig.~\ref{fig:weyl}, we have $s_\nu={1\over 2}$ 
which gives with $\nu={Z\over 2}$ a vanishing Chern number $C_\nu=0$. 
\begin{figure*}
\centering
  \includegraphics[width= \textwidth]{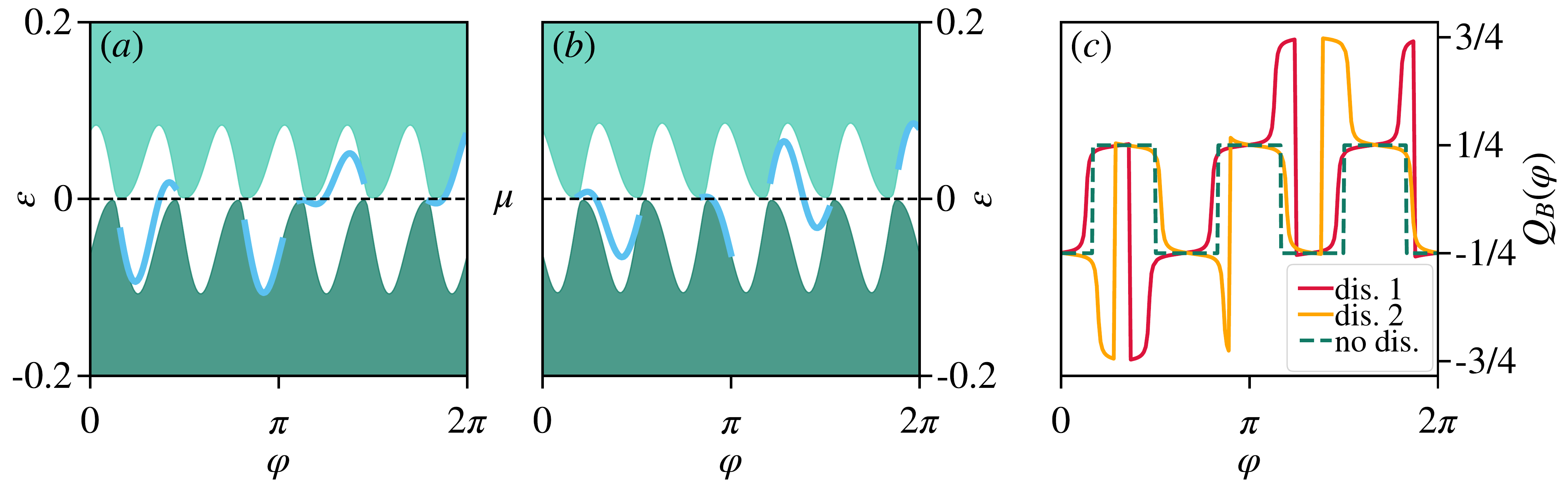} 
 \caption{(a) and (b) show the phase-dependence of the band structure and the edge states for $Z=6$ and the same parameters as in Fig.~\ref{fig:weyl} but with additional periodic disorder for the on-site potentials on the scale $0.2$ (which is chosen rather strong to visualize the gap openings), calculated for a semi-infinite system. As a result, $3$ edge states run either from the valence to the conduction band or vice versa. In (c) the phase-dependence of the boundary charge $Q_B(\varphi)$ is shown for the two disordered configurations and compared to the clean case, demonstrating stability of the quantization of $Q_B$ between the jumps.
 } 
\label{fig:spectrum}
\end{figure*}

As shown in Ref.~\cite{thakurathi_etal_prb_18} the Hall current $I_x$ for a $2D$ quantum Hall system (with periodic boundary conditions in $y$-direction, described by the azimuthal direction of a cylinder topology) along the direction $x$ of the effective 1D system in response to a perpendicularly applied voltage bias $V_y$ is related to $Q_B(\varphi)$ by
\begin{align}
\label{eq:Hall_current}
I_x = -{d\over dt} Q_{\text{edge}}^{(2D)}(t) = 
-{d\over dt}\sum_{n=1}^M Q_B\Bigg{(}{2\pi\over M}\Bigg{(}n + {\Phi(t)\over\Phi_0}\Bigg{)}\Bigg{)}\,,
\end{align}
where $Q_{\text{edge}}^{(2D)}$ is the charge along the edge of the physical $2D$ system, $\Phi_0=h/e$ denotes the flux quantum, and $V_y=-{d\over dt}\Phi(t)$ is generated by a time-dependent magnetic flux $\Phi(t)$ applied through the cylinder. The discrete values $k_y={2\pi\over M}n$ describe the perpendicular quasimomentum along the azimuthal  direction of the cylinder (with $M$ lattice sites around the cylinder). Inserting (\ref{eq:QB_form}) one finds that the ${2\pi\over Z}$-periodic and smooth part $f(\varphi)$ and the piecewise constant function $F(\varphi)$ provide a $\Phi$-independent contribution to the sum (up to discrete jumps at particular values of $\Phi$ from $F(\varphi)$). Therefore, the Hall current probes only the linear term of
(\ref{eq:QB_form}), leading to a quantized Hall conductance $\sigma_{xy}$ in terms of the Chern number 
\begin{align}
\label{eq:Hall_conductance}
I_x = \sigma_{xy} V_y \quad,\quad \sigma_{xy}={e\over h}C_\nu\,.
\end{align}
Therefore, for the Weyl case discussed above, the Hall conductance vanishes. 

A delicate question concerns the unknown function $f(\varphi)$ in (\ref{eq:QB_form}). When $Q_B(\varphi)$ is quantized for all $\varphi$ we get $M_+=M_-$ and $f(\varphi)$ is a constant determining the quantization value. However, when a small symmetry-breaking term in the form of periodic disorder is added to the on-site potentials, the gap will open at the Weyl points, and one obtains a discontinuous jump to a finite Chern number $C_\nu=M_--M_+$, see Figs.~\ref{fig:spectrum}(a,b) for $Z=6$. The gap opens slightly and ${Z\over 2}$ edge states move either from the valence to the conduction band or vice versa, giving rise to two different Chern numbers $\pm {Z\over 2}$, with a corresponding jump of the Hall current. In addition to the linear term ${C_\nu\over 2\pi}\varphi$, also the functions $f(\varphi)$ and $F(\varphi)$ jump discontinuously such that all three terms on the right hand side of Eq.~(\ref{eq:QB_form}) are unstable against small symmetry-breaking terms for {\it all} phases. However, the boundary charge determined by the sum of all three terms remains a stable quantity between the jumps as shown in Fig.~\ref{fig:spectrum}(c). This shows that the quantization values of $Q_B(\varphi)$ at fixed $\varphi$ between the gap closing points are well-defined and stable quantities accessible by experiments for 1D systems.

\begin{figure}
\centering
 \includegraphics[width=\columnwidth]{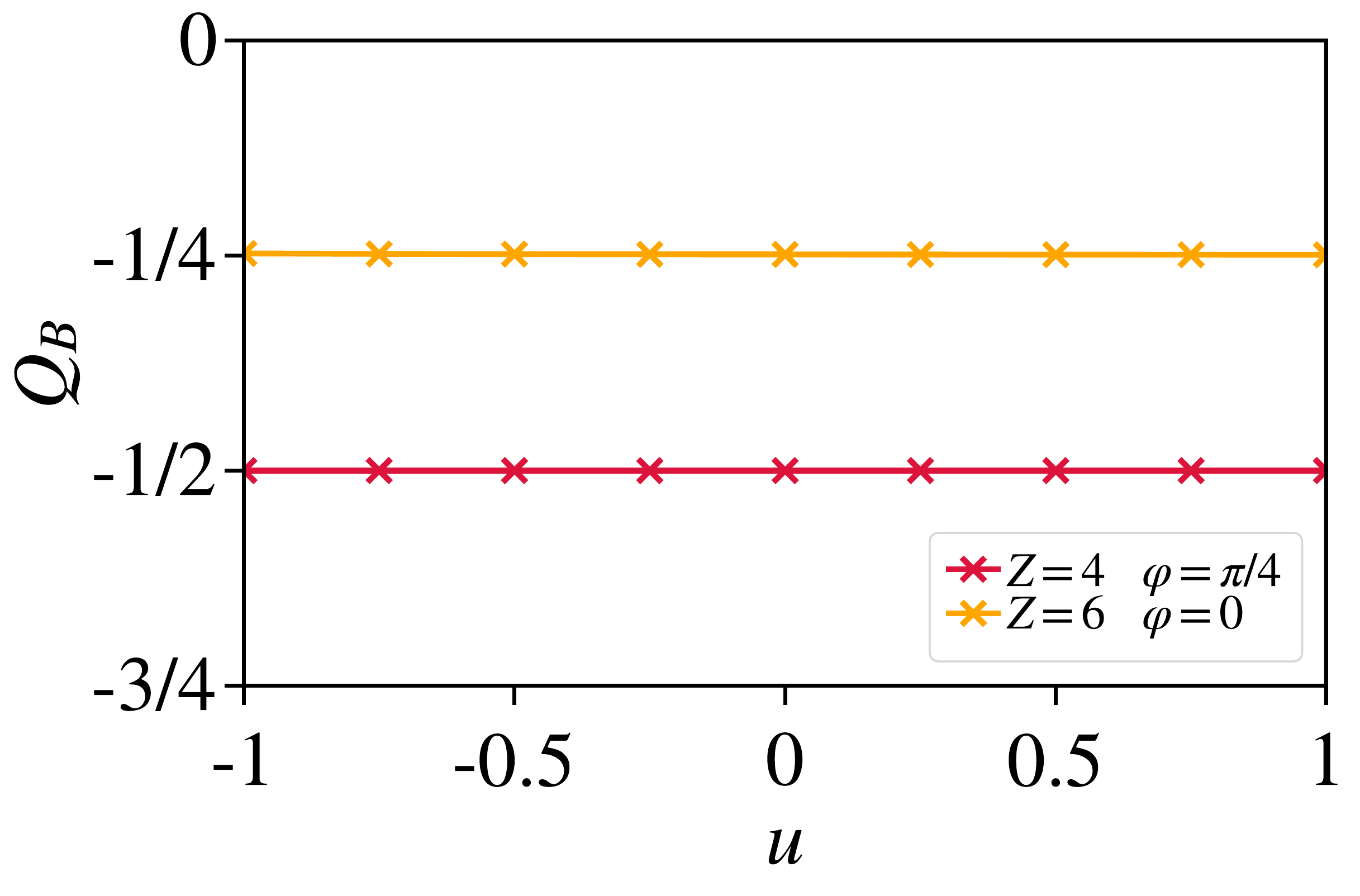}
 \caption{Stability of the rational quantization of the boundary charge upon inclusion of nearest neighbor interaction. The figure shows the boundary charge $Q_B$ for the same parameters (but finite size $N=1000$) as used in Fig.~\ref{fig:weyl} at two values of $\varphi=\pi/4$ and $\varphi=0$, where the gap is maximal, for $Z=4$ and $Z=6$, respectively. }
\label{fig:weyl_U}
\end{figure}

In Fig.~\ref{fig:weyl} we add an analysis of the effects of non-periodic disorder in the panels showing $Q_{ B}(\varphi)$. We overlay the results obtained without disorder (lines) with those calculated in the presence of a quenched onsite disorder (symbols) of quite moderate strength $d/t=0.05$. The two results mainly overlap, besides of at values of $\varphi$ close to gap closings  where small energy shifts and gap openings induced by the disorder might change the value of quantization $\text{mod}(1)$, similar to the effects of weak periodic disorder as shown in Fig.~\ref{fig:spectrum}(c) (where the disorder strength is chosen much stronger to visualize the effects on the spectrum in Figs.~\ref{fig:spectrum}(a,b)). Therefore, the overall quantization of $1/2$ and $1/4$ for $Z=4$ and $Z=6$, respectively, remains perfectly intact $\text{mod}(1)$. 

Finally, we study the robustness of the quantization with respect to adding short-ranged electron-electron interaction at half-filling ($\mu=0$ in the non-interacting case above). We add the particle-hole symmetric version of Eq.~\eqref{eq:coulomb} as the interaction
\begin{equation}
\label{eq:coulomb2}
    V_{ee}=u\sum_{m}\left(\hat\rho_m-\frac{1}{2}\right)\left(\hat\rho_{m+1}-\frac{1}{2}\right)
\end{equation}
to the Hamiltonian and study the resulting interacting quantum many-body system using DMRG. This particle-hole symmetric formulation is chosen for convenience of implementation only. The results for $\varphi$ being at the maximal single particle gap are summarized in Fig.~\ref{fig:weyl_U} for $Z=4$ and $Z=6$ and the other parameters as in  Fig.~\ref{fig:weyl}. As shown the quantization of $Q_{B}$ remains perfectly intact upon including this local interaction up to rather large values of the interaction strength $u$.

\subsection{Low-energy theory}
\label{sec:low_energy}

In this section we discuss the low-energy theory in terms of a Dirac model in $1+1$ dimensions, following closely the treatment in Refs.~\cite{gangadharaiah_etal_prl_12,park_etal_prb_16,thakurathi_etal_prb_18}, where low-energy models have been derived from lattice models for the special case of potential modulation with constant hopping. Here, we discuss the general derivation of Dirac models and present an exact formula for the complex gap parameter entering the low-energy model. We discuss in detail the restrictions for the phase of the gap parameter in the presence of non-local symmetries or for the special case of half-filling. We present exact formulas for the boundary and interface charge for the noninteracting Dirac model and prove the consistency with our general framework. Furthermore, we discuss the stability under short-ranged electron-electron interaction by using the bosonization method. Whereas the gap is significantly renormalized by interactions \cite{kivelson_thacker_wu_prb_85,horovitz_solyom_prb_85,wu_kivelson_prb_86,gangadharaiah_etal_prl_12}, it turns out that the boundary charge is insensitive to the gap size but depends only the phase of the gap parameter in a linear fashion which is not influenced by interactions. Since the low-energy model is most conveniently written in continuum space, we explicitly write the lattice spacing $a$ at the appropriate places and do not set it to one in this section.

\subsubsection{Noninteracting Dirac model}
\label{sec:dirac_noninteracting}

We first split the noninteracting single-channel nearest-neighbor hopping model with on-site potentials $v_m$ and hoppings $-t_m$ into two parts by writing $t_m = t + \delta t_m$
\begin{align}
\label{eq:H_bulk_splitting}
H_{\text{bulk}} &= H_0 + H'\,,\\
\label{eq:H_0}
H_0 &= - t \sum_m c_{m+1}^\dagger c_m + h.c. \,,\\
\label{eq:H'}
H' &= \sum_m v_m c_m^\dagger c_m - \sum_m \delta t_m (c_{m+1}^\dagger c_m + h.c.) \,.
\end{align}
$H'$ describes the external modulation which, just for illustration (other cases can be treated analogously), we take to be harmonic with wavelength $\lambda_{\text{ex}} = Za$ 
\begin{align}
\label{eq:v_modulation}
v_m &= - 2\delta v \cos(k_{\text{ex}} ma + \varphi_v) \,,\\
\delta t_m &= - \delta t \cos(k_{\text{ex}} ma + \varphi_t)\,,
\end{align}
where $k_{\text{ex}} = {2\pi\over \lambda_{\text{ex}}}$ is the wavevector of the external modulation. $H_0$ can be easily diagonalized in terms of plane waves $|k\rangle = \sqrt{a\over 2\pi}\sum_m e^{ikma}|m\rangle$, leading to 
\begin{align}
\label{eq:H_0_diagonal}
H_0 = \int_{-\pi/a}^{\pi/a} dk \,\epsilon_k^{(0)} \tilde{c}_k^\dagger \tilde{c}_k \,,
\end{align}
with 
\begin{align}
\label{eq:cos_dispersion}
\epsilon_k^{(0)} &= - 2t \cos(ka) \quad,\quad -{\pi\over a} < k < {\pi\over a} \,,\\
\label{eq:c_k}
\tilde{c}_k^\dagger &= \sqrt{a\over 2\pi} \sum_m e^{ikma} c_m^\dagger \,.
\end{align}
In the representation $|k\rangle$ of the exact single-particle eigenstates of $H_0$, the matrix elements of $H'$ can be straightforwardly calculated and one obtains
\begin{align}
\nonumber 
\langle k|H'|k'\rangle &=\\
\nonumber
&\hspace{-1cm}
= \delta(k-k'-k_{\text{ex}}) \left\{-\delta v e^{i\varphi_v} + 
{\delta t\over 2} (e^{-ika} + e^{ik'a}) e^{i\varphi_t}\right\} \\
\label{eq:H'_k_representation}
&\hspace{-1cm}
= \delta(k-k'+k_{\text{ex}}) \left\{-\delta v e^{-i\varphi_v} + 
{\delta t\over 2} (e^{-ika} + e^{ik'a}) e^{-i\varphi_t}\right\} \,.
\end{align}
These matrix elements lead to $Z-1$ gap openings labeled by $\nu=1,2,\dots,Z-1$ at wavevectors $\pm k_F^{(\nu)}$ with $2k_F^{(\nu)}=\nu k_{\text{ex}}$. The gaps are generated in $\nu'$-th order perturbation theory in $H'$ with $\nu'=\text{min}\{\nu,Z-\nu\}$ \cite{klinovaja_loss_epb_14} and are of the order 
\begin{align}
\label{eq:gap_order}
|\Delta_\nu| \sim t \left({\text{max}\{|\delta v|,|\delta t|\}\over t}\right)^{\nu'} \,.
\end{align} 

In the following we will concentrate on a certain gap with index $\nu$ and write $k_F\equiv k_F^{(\nu)}$ and $\Delta\equiv\Delta_\nu$ for brevity. Our aim is to develop an effective low-energy model for energies close to the Fermi energy $\epsilon_F=\epsilon_{k_F}^{(0)}=-2t\cos(k_Fa)$. Using Brillouin-Wigner perturbation theory the coupling between the states close to the two Fermi points $\pm k_F$ can be described by an effective Hamiltonian 
\begin{align}
\label{eq:H'_eff}
H'_{\text{eff}} = P\Bigg{(}H'+H'Q{1\over \epsilon_F-QH_{\text{bulk}}Q}QH'\Bigg{)}P\,,
\end{align}
where $P$ projects on the low-energy sector and $Q=1-P$. It is then straightforward to see that for $|k|,|k'|\ll k_F$ one obtains a coupling between the two Fermi points via $\nu'-1$ virtual intermediate states described by the matrix element
\begin{align}
\label{eq:H'_matrix_element}
\langle k_F + k| H'_{\text{eff}} |-k_F + k'\rangle = \Delta \delta(k-k')\,,
\end{align}
where
\begin{align}
\label{eq:delta}
\Delta = |\Delta| e^{i\alpha}
\end{align}
is a complex gap parameter with negligible $k$-dependence. Using Eq.~(\ref{eq:H'_k_representation}) one finds after a straightforward calculation 
\begin{align}
\label{eq:delta_value}
\Delta\equiv\Delta_\nu = 
\begin{cases} \Delta_\nu^- \quad \text{for} \,\,\nu<{Z\over 2} \\
\Delta_\nu^+ \quad \text{for} \,\,\nu>{Z\over 2} \\
\Delta_\nu^- + \Delta_\nu^+ \quad \text{for} \,\,\nu={Z\over 2} \end{cases}\,,
\end{align}
with
\begin{align}
\label{eq:delta_relation}
\Delta_{\nu}^{+} = - (-1)^{Z-\nu}
\Delta_{Z-\nu}^{-}\Big{|}_{\varphi_{v}\rightarrow -\varphi_{v}\atop \varphi_{t}\rightarrow -\varphi_{t} + k_{\text{ex}}+\pi}\,,
\end{align}
and 
\begin{align}
\nonumber
\Delta_\nu^- &= \prod_{s=1}^{\nu-1}
{1\over \Big{|}\epsilon_F-\epsilon_{-k_F+sk_{\text{ex}}}^{(0)}\Big{|}}\prod_{l=1}^\nu\Big\{-\delta v e^{i\varphi_v}+ \\
\label{eq:delta_minus}
& +{\delta t\over 2}\left[e^{i(k_F-lk_{\text{ex}})a} + e^{-i(k_F-lk_{\text{ex}})a}e^{-ik_{\text{ex}}a}\right]e^{i\varphi_t}\Big\}\,.
\end{align}  
Most importantly, the gap parameter $\Delta_{Z/2}$ at half-filling (only possible for $Z$ even) is determined by an interference of two processes. This will become important for the quantization values of the boundary charge (see below). 

Splitting the field operator $\psi(ma)\equiv {1\over \sqrt{a}} c_m$ in slowly varying right and left moving fields $R(x)$ and $L(x)$ via
\begin{align}
\label{eq:RL_splitting}
\psi(x) = R(x) e^{ik_Fx} + L(x) e^{-ik_Fx}\,,
\end{align}
and inserting this decomposition in the effective Hamiltonian 
$H_{\text{eff}}=H_0 + H'_{\text{eff}}$, one finds after neglecting strongly oscillating terms the final result for the low-energy Hamiltonian in the form of a Dirac Hamiltonian in $1+1$ dimensions
\begin{align}
\nonumber
H_{\text{eff}} &= \int dk \\
\label{eq:H_eff}
&\hspace{0cm}
\uline{\psi}_k^\dagger \left\{v_F k \sigma_z + |\Delta| \cos{\alpha} \sigma_x
- |\Delta| \sin{\alpha} \sigma_y\right\} \uline{\psi}_k\,,
\end{align}
where $v_F=2ta\sin(k_Fa)$ and 
\begin{align}
\label{eq:psi_k_operator}
\uline{\psi}_k={1\over\sqrt{2\pi}}\int dx \,e^{-ikx}\uline{\psi}(x) \,\,,\,\,
\uline{\psi}(x)= \left(\begin{array}{c} R(x) \\ L(x) \end{array}\right)\,.
\end{align}

For equal phases $\varphi_v=\varphi_t=\varphi$ we can directly see from 
(\ref{eq:delta_relation}) and (\ref{eq:delta_minus}), that the phase $\alpha$ of the
gap parameter is a linear function of $\varphi$ for $\nu\ne{Z\over 2}$ 
\begin{align}
\label{eq:alpha_varphi}
\alpha = \pm \nu \varphi + \text{const} \quad\text{for}\quad \nu\lessgtr{Z\over 2}\,,
\end{align}
where the constant term is non-universal but independent of $\varphi$.
As we will see in Section~\ref{sec:dirac_QB_QI} this leads to a universal linear behaviour of the boundary charge as function of $\varphi$. 

For the case of half-filling $\nu={Z\over 2}$ (only possible for $Z$ even) it is more complicated due to the interference effect from two paths. If we take potential modulation only, i.e., $\delta t=0$, as considered in Section~\ref{sec:weyl}, we get from (\ref{eq:delta_relation}) and
(\ref{eq:delta_minus}) 
\begin{align}
\label{eq:delta_minus_half_filling}
\Delta_{Z/2}^- &= |\Delta_{Z/2}^-| (-1)^{Z/2} e^{i{Z\over 2}\varphi}\,,\\
\label{eq:delta_plus_half_filling}
\Delta_{Z/2}^+ &= -|\Delta_{Z/2}^-| e^{-i{Z\over 2}\varphi}\,,
\end{align}
which gives for the sum
\begin{align}
\label{eq:delta_half_filling}
\Delta_{Z/2} = |\Delta_{Z/2}^-|
\begin{cases} 2i\sin({Z\over 2}\varphi) \,\,\,\quad \text{for} \,\,{Z\over 2}\,\text{even} \\
-2\cos({Z\over 2}\varphi) \quad \text{for} \,\,{Z\over 2}\,\text{odd} 
\end{cases}\,.
\end{align}
Therefore, due to the interference of the two paths, one obtains for the phase $\alpha$ at half-filling an interesting phase-locking effect 
\begin{align}
\label{eq:alpha_half_filling}
\alpha = 
\begin{cases} {\pi\over 2} + \pi \Theta[-\sin({Z\over 2}\varphi)] \,\quad \text{for} \,\,{Z\over 2}\,\text{even} \\
\pi\Theta[\cos({Z\over 2}\varphi)] \quad\quad\quad\quad \text{for} \,\,{Z\over 2}\,\text{odd} 
\end{cases}\,,
\end{align}
which, as we will see in Section~\ref{sec:dirac_QB_QI}, explains the pinning of the boundary charge to certain quantization values. 

Finally, we analyse the restrictions for the values of $\alpha$ in the presence of the non-local symmetries $\Pi_n$ or $S_n$ discussed in Section~\ref{sec:symmetries}. This can be done directly by using the general definition of $\Delta$ via (\ref{eq:H'_matrix_element}) without using some special form for $v_m$ and $t_m$. As shown in Appendix~\ref{app:single_channel_nearest_neighbor} the two symmetries act on the lattice sites as
\begin{align}
\label{eq:Pi_lattice}
\Pi_n|m\rangle &= |Z-m-n+1\rangle\,,\\
\label{eq:S_lattice}
S_n|m\rangle &= (-1)^m|m-n\rangle\,,
\end{align}
which implies the following transformation in quasimomentum space
\begin{align}
\label{eq:Pi_momentum}
\Pi_n|k\rangle &= e^{ik(Z-n+1)a}|-k\rangle\,,\\
\label{eq:S_momentum}
S_n|k\rangle &= e^{ik(n+\pi)a}|k+\pi\rangle\,.
\end{align}
Together with (\ref{eq:H'_matrix_element}) this implies for the unitary symmetry $\Pi_n$ and the chiral symmetry $S_n$ the following condition for the gap parameter
\begin{align}
\label{eq:Pi_delta_condition}
\Pi_n: \Delta &= e^{-2ik_F(Z-n+1)a}\Delta^* \,,\\
\label{eq:S_delta_condition}
S_n: \Delta &= -e^{-2ik_Fna}\Delta^* \,.
\end{align}
Using $2k_Fa=\nu k_{\text{ex}}a=2\pi\bar{\rho}$ and, for the symmetry $S_n$ (which requires half-filling), $k_Fa={\pi\over 2}$, we get for both symmetries the following pinning of $\alpha$
\begin{align}
\label{eq:alpha_pinning_symmetries}
\alpha = (n-1)\pi\bar{\rho} \quad \text{mod}(\pi)\,.
\end{align}
As shown in Section~\ref{sec:dirac_QB_QI}, this will prove consistency of the low-energy approach with the quantization of the boundary charge according to our general framework.

\subsubsection{Interacting Dirac model}
\label{sec:dirac_interacting}

Let us now proceed to include electron-electron interaction effects within the short-ranged density-density interaction form (\ref{eq:coulomb}). Using standard bosonization techniques \cite{gogolin_etal_98,delft_schoeller_98,giamarchi_10} we find within the low-energy model
\begin{align}
\nonumber
H&=H_{eff}+V_{ee} =\frac{v}{2}\int{dx}\Big\{ K [\hat{\Pi}(x)]^{2}+\frac{1}{K}[\partial_{x}\hat{\varphi}(x)]^{2}\Big\}\\
\label{eq:H_bosonized}
&\hspace{0.5cm}
+(-1)^p \frac{|\Delta|}{2\pi a}\int{dx}\cos(\sqrt{4\pi}\hat{\varphi}(x)-\alpha-\pi/2),
\end{align}
where the Luttinger liquid parameter $K$ and the renormalized Fermi velocity $v$ are defined by
\begin{align}
\label{eq:K}
K = \Bigg{(}1+{2u_1-u_2\over \pi v_F}\Bigg{)}^{-1/2}\quad,\quad v = v_F / K \,,
\end{align}
with $u_1=a\sum_m u(m)$ and $u_2=a\sum_m u(m) e^{2ik_F ma}$ corresponding to forward and backward scattering processes, respectively. Here, $u(m-m')$ describes the short-ranged Coulomb interaction between the densities at site $m$ and $m'$, see Eq.~(\ref{eq:coulomb}).
$\hat{\Pi}(x)=\partial_x[\hat{\varphi}_{+}(x)-\hat{\varphi}_{-}(x)]$ and $\hat{\varphi}(x)=\hat{\varphi}_{+}(x)+\hat{\varphi}_{-}(x)$ are canonically conjugate momentum and field variables. The chiral boson fields $\hat{\varphi}_{\pm}(x)$ are related to the right and left movers $\hat{\psi}_+(x)\equiv R(x)$ and $\hat{\psi}_-(x)\equiv L(x)$ via
\begin{align}
\label{eq:psi_bosonization}
\hat{\psi}_{p}(x)&=\frac{1}{\sqrt{4\pi a}}\,e^{ip\sqrt{4\pi}\hat{\varphi}_{p}(x)}\,,
\end{align}
where ${1\over a}$ is used for the momentum cutoff (or $v/a$ for the high-energy cutoff). 
A subtlety is the phase shift by ${\pi\over 2}$ in the cosine term of (\ref{eq:H_bosonized}) and the undetermined prefactor $(-1)^p$. This is related to the commutator $[\hat{\varphi}_+(x),\hat{\varphi}_-(x')]={i\over 4}(2p+1)$ which arises from the zero mode phases to ensure the anticommutation relation of left and right movers. Here, the value $p$ is an arbitrary integer which will be finally determined by comparing the boundary charge with the exact solution for the noninteracting Dirac model. 

In order to get an insight into the physics of the interacting system, it is instructive to perform a perturbative renormalization group analysis \cite{gangadharaiah_etal_prl_12} using standard operator product expansion techniques \cite{delft_schoeller_98}. Reducing the high-energy cutoff $\Lambda$ (with intial value $\Lambda_0=v/a$) we obtain the following flow equations for the gap $|\Delta|$ and the Luttinger parameter $K$:
\begin{align}
\label{eq:RG_gap}
\frac{d|\Delta|}{d\ell}&=(1-K)|\Delta|, \quad d\ell=-\frac{d\Lambda}{\Lambda},\\
\label{eq:RG_K}
\frac{dK}{d\ell}&=-c\frac{a_\Lambda^{2}K^{2}|\Delta|^{2}}{2\pi v^{2}}\,,
\end{align}
with $a_\Lambda=v/\Lambda$ and some unimportant constant $c\sim O(1)$ which depends on the RG procedure. As one can see, for repulsive interactions ($K<1$) the cosine term in (\ref{eq:H_bosonized}) is a relevant perturbation and the system flows into the gapped phase while the interaction grows under the RG. This fact allows one to conclude that the fluctuations of the cosine term are getting effectively frozen in the renormalized theory and is of crucial importance in the determination of the boundary charge in the interacting theory in the following. Although it seems that the gap grows to infinity under the RG flow (with $K$ shrinking to zero), we note that the flow equations can only be trusted until the cutoff $\Lambda$ reaches a critical scale  $\Lambda_c\sim |\Delta|_c$ (or $a_c\sim v/|\Delta|_c$), with $|\Delta|_c=|\Delta|_{\Lambda=\Lambda_c}$, at which the flow has to be truncated.

At half-filling $k_{F}a=\frac{\pi}{2}$, in principle, one also has to consider the Umklapp scattering process in the bosonized Hamiltonian. However, the Umklapp term is RG relevant only for strong two-particle interactions with $K<1/2$ \cite{giamarchi_10}. Bearing this in mind, we are not going to focus on the Umklapp process in the following by confining ourselves to moderate electron-electron interactions.

\subsubsection{Boundary and interface charge}
\label{sec:dirac_QB_QI}

Let us turn to the discussion on the boundary and interface charge quantization in the low-energy description. We start with the boundary charge $Q_B$ of the noninteracting Dirac model on the half-line $x>0$ with vanishing boundary condition $\psi(0)=R(0)+L(0)=0$, given by
\begin{align}
\nonumber
H_{\text{eff}} &= \int_0^\infty dx\, \uline{\psi}^\dagger(x) \left\{v_F (-i\partial_x) \sigma_z \right.\\
\label{eq:H_eff_semi_infinite}
&\hspace{0cm}
\left. + |\Delta| \cos{\alpha} \sigma_x
- |\Delta| \sin{\alpha} \sigma_y\right\} \uline{\psi}(x)\,.
\end{align}
In order to study the quantization of $Q_B$, one has to construct the single-particle eigenstates of the Hamiltonian.  As shown in Appendix~\ref{app:QB_dirac} there are two distinct types of admissible eigenstates. First of all, there are two scattering states (labeled by the nonnegative momentum $k$) living at energies $\epsilon_{k,\pm}=\pm\sqrt {v_ {F}^{2}k^{2}+|\Delta|^{2}} $, corresponding to conduction and valence bands, respectively. In addition, under the condition $\sin\alpha>0$, inside the gap, there exists a single edge state sitting at an energy $-|\Delta|\cos\alpha$ \cite{gangadharaiah_etal_prl_12}. Assuming that the chemical potential of the system lies at the bottom of the conduction band, we show in Appendix~\ref{app:QB_dirac} that the respective contributions of the edge and scattering states are
\begin{align}
\label{eq:QB_edge}
Q_{B}^{\text{edge}}(\alpha)&=\Theta_{0<\alpha<\pi} \equiv\begin{cases} 1 \quad \text{for} \ 0<\alpha<\pi \\  
0 \quad\text{otherwise} \end{cases}\,,\\
\label{eq:QB_scattering}
Q_{B}^{\text{scatt}}(\alpha)&=\frac{\ln(-e^{i\alpha})}{2\pi i}-\frac{1}{4}\,.
\end{align}
Combining both contributions, one immediately arrives at the following universal result 
\begin{align}
\label{eq:QB_universal}
Q_{B}(\alpha)=\frac{\alpha}{2\pi}+\frac{1}{4}\,
\end{align}
for $-\pi < \alpha < \pi$, and periodic continuation to the other intervals. For arbitrary position of the chemical potential in the gap one has to use the result $Q_{B}^{\text{edge}}(\alpha)=\theta(\mu+|\Delta|\cos\alpha)\Theta_{0<\alpha<\pi}$ for the edge state charge, i.e., one has to add the integer term $-\theta(-\mu-|\Delta|\cos\alpha)\Theta_{0<\alpha<\pi}$ to (\ref{eq:QB_universal}). This proves Eq.~(\ref{eq:QB_low_energy}) stated in the introduction.

Let us emphasize that the low energy result accurately reproduces the conclusions based on the microscopic theory. For instance, consider the case $\nu\gtrless\frac{Z}{2}, \quad \varphi_{v}=\varphi_{t}=\varphi$, so that according to (\ref{eq:alpha_varphi}), $\alpha=\pm\nu \varphi + \text{const}$, leading to the universal linear behaviour of $Q_{B}$ as a function of $\varphi$ \cite{park_etal_prb_16,thakurathi_etal_prb_18,pletyukhov_etal_short,pletyukhov_etal_long}. At half-filling $\nu={Z\over 2}$, where $\alpha$ is pinned to the values given by (\ref{eq:alpha_half_filling}), one arrives at the following quantization of $Q_B$
\begin{align}
\label{eq:DQ4_II}
Q_{B}(\varphi)=\begin{cases}\frac{1}{2}+\frac{1}{2}\Theta(-\sin({Z\over 2}\varphi)), \quad \text{for} \ \frac{Z}{2} \ \text{even} \\  
\frac{1}{4}+\frac{1}{2}\Theta(\cos({Z\over 2}\varphi)), \quad\quad \text{for} \ \frac{Z}{2} \ \text{odd}
 \end{cases}\,,
\end{align}
again showing complete agreement with the microscopic prediction (\ref{eq:QB_quantization_Zhalf}). Finally, in the presence of the symmetries $\Pi_n$ or $S_n$, where $\alpha$ is pinned to $\alpha=(n-1)\pi\bar{\rho}\,\text{mod}(\pi)$ according to (\ref{eq:alpha_pinning_symmetries}), one finds
\begin{align}
\label{eq:QB_exact} 
Q_B = n{1\over 2}\bar{\rho} -{1\over 2}\bar{\rho} + {1\over 4} \quad\text{mod}\Big({1\over 2}\Big)\,.
\end{align}
Comparing with the exact solution (\ref{eq:QB_RL_rational}) there is a difference given by the constant $-{1\over 2}\bar{\rho} + {1\over 4}$ which vanishes only at half-filling. This difference can be traced back to the fact that the Dirac model contains an infinite set of high-energy states which are unphysical. Interestingly, this constant can be shown to be given by the negative boundary charge of the original lattice model $H_0$ on a half line at zero gap, see Eq.~(\ref{eq:QB_zero_gap}) in Appendix~\ref{app:zero_gap}. Thus, we get the following relation between the boundary charge of the Dirac model and the exact one
\begin{align}
\label{eq:QB_dirac_exact}
Q_B^{\text{Dirac}} = Q_B^{\text{exact}} - Q_B^{\text{exact}}|_{\Delta=0}\,.
\end{align}

Next we study the interface charge quantization for the noninteracting Dirac model. Now the diagonalization problem is formulated on the entire real line, however, $\alpha$ is now allowed to be a function of position $\alpha=\alpha(x)$ with Hamiltonian 
\begin{align}
\nonumber
H_{\text{eff}} &= \int dx\, \uline{\psi}^\dagger(x) \left\{v_F (-i\partial_x) \sigma_z \right.\\
\label{eq:H_eff_interface}
&\hspace{0cm}
\left. + |\Delta| \cos[\alpha(x)] \sigma_x
- |\Delta| \sin[\alpha(x)] \sigma_y\right\} \uline{\psi}(x)\,.
\end{align}
In particular, we make the following choice $\alpha(x)=\Theta(x)\alpha_{R}+\Theta(-x)\alpha_{L}$ and define $\delta\alpha=\alpha_R-\alpha_L$. As shown in Appendix~\ref{app:QI_dirac}, one concludes that there are two different types of states to consider, the scattering states, as well as the in-gap states localized at the interface. The bound state is present for $\sin(\delta\alpha/2)>0$ with energy $-|\Delta|\cos(\delta\alpha/2)$ and contributes unity to the total interface charge (if occupied). As opposed to the semi-infinite problem, the valence and conduction ($\epsilon_{k,\pm}=\pm\sqrt{v_{F}^{2}k^{2}+|\Delta|^{2}}$) band states are now two-fold degenerate. Indeed, for a given energy, one always has two distinct scattering channels, the one where particles scatter from left to right, and the opposite one, where particles scatter from right to left, and hence the degeneracy. Assuming that the chemical potential is located at the bottom of the conduction band, we show in Appendix~\ref{app:QI_dirac} that the interface charge follows the Goldstone-Wilczek formula
\begin{align}
\label{eq:DQ5_II}
&Q_{I}=\frac{\delta\alpha}{2\pi} 
\end{align}
for $\delta\alpha \in (0,2 \pi)$. Values of $Q_I$ on other intervals are to be found from its periodic dependence on $\delta\alpha$. Similiar to the boundary charge one has to add the integer term 
$-\theta(-\mu-|\Delta|\cos(\delta\alpha/2))\Theta_{0<\delta\alpha/2<\pi}$ 
for a chemical potential with arbitrary position in the gap. 

Let us now proceed by studying the effects of the electron-electron interaction on the quantization of the boundary charge. We take the bosonized Hamiltonian (\ref{eq:H_bosonized}) on the semi-infinite part $x>0$, together with the boundary condition
\begin{align}
\label{eq:boundary_condition}
0 = \psi(0) = {1\over \sqrt{4\pi a}}\Big(e^{i\sqrt{4\pi}\hat{\varphi}_+(0)} + e^{-i\sqrt{4\pi}\hat{\varphi}_-(0)}\Big)\,. 
\end{align}
This requires $i\sqrt{4\pi}\hat{\varphi}_-(0)=i\pi(2q-1)-i\sqrt{4\pi}\hat{\varphi}_+(0)$ with some integer $q$, leading to the following boundary condition for $\hat{\varphi}(0)=\hat{\varphi}_+(0)+\hat{\varphi}_-(0)$ (see also Refs.~\cite{mattsson_etal_prb_97, gangadharaiah_etal_prl_12})
\begin{align}
\label{eq:varphi_zero}
\hat{\varphi}(0) = {1\over 2\sqrt{\pi}}(2q-1)\pi\,.
\end{align}
As we have seen in Section~\ref{sec:dirac_interacting}, the gap $|\Delta|$ increases under the RG flow, effectively freezing the quantum fluctuations of $\hat{\varphi}$ such that the cosine term in (\ref{eq:H_bosonized}) is minimized in the bulk. This leads to the following asymptotic value  
\begin{align}
\label{eq:varphi_asymptotic}
\hat{\varphi}(\infty) = {1\over 2 \sqrt{\pi}}\Bigg(\alpha + {\pi\over 2} - p\pi + (2s-1)\pi\Bigg)\,,
\end{align}
with another integer $s$. With the help of bosonization identities, we deduce that the boundary charge may be related to the difference of the  values of $\hat{\varphi}(x)$ at $x=\infty$ and $x=0$:
\begin{align}
\nonumber
Q_{B}&=\frac{1}{\sqrt{\pi}}\int_{0}^{\infty}dx\Braket{\partial_{x}\hat{\varphi}(x)}\\
\label{eq:QB_varphi}
&={1\over \sqrt{\pi}} [\hat{\varphi}(\infty)-\hat{\varphi}(0)].
\end{align}
Inserting (\ref{eq:varphi_zero}) and (\ref{eq:varphi_asymptotic}) we thus conclude 
\begin{align}
\label{eq:DQ13_II}
&Q_{B}=\frac{\alpha}{2\pi}+\frac{1}{4}-{1\over 2}p \quad\text{mod}(1).
\end{align}
Comparing this result with the exact solution (\ref{eq:QB_universal}) without interaction we find that we have to choose $p=0$. This proves the stability of the boundary charge under short-ranged electron-electron interaction within the low-energy model. 

A similar calculation may be done in the case of an interface charge quantization by using an arbitrary function $\alpha(x)$ describing the interface. In this case we get
\begin{align}
\nonumber
Q_{I}&=\frac{1}{\sqrt{\pi}}\int{dx}\Braket{\partial_{x}\hat{\varphi}}\\
\label{eq:QI_1}
&={1\over \sqrt{\pi}}[\hat{\varphi}(\infty)-\hat{\varphi}(-\infty)]\,.
\end{align}
By using the same procedure as above, we arrive at the following result
\begin{align}
\label{eq:QI_interacting}
Q_I=\frac{\alpha(\infty)-\alpha(-\infty)}{2\pi}\quad\text{mod}(1)\,.
\end{align}
which in the case of $\alpha(x)=\alpha_{R}\Theta(x)+\alpha_{L}\Theta(-x)$, reduces to the non-interacting result, and thus again shows the robustness of the interface charge quantization. A similiar expression has been found in Ref.~\cite{goldstone_wilczek_prl_81}.

\noindent

\section{Summary and outlook}
\label{sec:summary}

For generic 1D insulators we have provided in this work a complete analysis of symmetry conditions to realize rational quantizations of the boundary charge. We obtained two interesting results: (a) {\it Any} rational quantization ${p\over q}$ can be realized if non-local symmetries involving translations are taken into account. (b) Besides the quantization unit ${1\over 2}$ known from local symmetries we identified a new quantization unit ${1\over 2}\bar{\rho}$, where $\bar{\rho}$ is the average charge per site. This has to be contrasted to the known quantization unit $\bar{\rho}$ for interface charges. 

Both the quantization of the boundary and the interface charge were shown to follow straightforwardly from the transformation laws of the boundary charge under translations and local inversion. These fundamental principles are physically very intuitive and were rigorously related to the intriguing property of insulators that local perturbations lead only to local charge redistributions. Therefore, all our results were proven to be stable against static random disorder and short-ranged electron-electron interaction. We demonstrated this explicitly by using exact diagonalization, DMRG methods, and bosonization calculations. In addition, the stability of the quantization of the boundary charge was recently analysed via functional renormalization group (fRG) studies for the interacting Rice-Mele model and the same conclusions were found \cite{lin_etal_prb_20}. Besides the boundary charge also other quantities were studied with fRG for this model like the full density profile and the precise form of edge states, where interaction effects have a more subtle effect. In the future it will be of interest to study also other quantities like density-density correlation functions in the presence of a boundary. In addition, fluctuations of the boundary charge are of relevance. While the overall size of fluctuations is expected to be small \cite{kivelson_schrieffer_prb_82,bell_rajaraman_physlettB_82,kivelson_prb_82,bell_rajaraman_nucphysB_83,frishman_horovitz_prb_83,jackiw_etal_nucphysB_83,park_etal_prb_16} when the gap is finite, it will be of interest to reveal universal properties of the fluctuations and to study their topological nature \cite{piasotski_etal_preprint_20}. 

In addition to the general framework we have provided in this work an interesting application to identify a novel quantization class ${e\over 4}$ in the special case of single-channel and nearest-neighbor hopping models at half-filling. As function of the phase variable controlling the offset of the potential modulation we found Weyl physics close to gap closing points and demonstrated the stability of the quantization of the boundary charge in contrast to the Hall current. We suggest such systems to be realizable in cold atom systems \cite{atala_etal_nature_2013,meier_etal_nature_comm_16}, in carbon based materials \cite{efroni_etal_prl_17,yang_nanomat_19} or phononic crystals \cite{xio_etal_natphys_15}. Other promising candidates could be quantum dot arrays as outlined in Ref.~\cite{park_etal_prb_16}, where control over all model parameters is possible. As shown in Appendix~\ref{app:finite_dot_array}, the quantization of the boundary charge is already visible for an array size of $\sim 20-30$ dots, which is within experimental reach. 

As shown in Refs.~\cite{park_etal_prb_16,thakurathi_etal_prb_18,pletyukhov_etal_short,pletyukhov_etal_long} the transformation law of the boundary charge under translations is also responsible for the quantization of the average linear slope of the boundary charge which is of fundamental importance for the understanding of the integer quantum Hall effect \cite{thakurathi_etal_prb_18}. For a finite system of size commensurable with the unit cell size, it was found in Ref.~\cite{park_etal_prb_16} that the sum of the boundary charges at the left and right end of the system is zero (up to an integer). This is equivalent to the result proven rigorously in this work that the boundary charge changes sign under local inversion. The fact that the transformation laws are also responsible for rational quantization values of the boundary charge demonstrates the topological nature of the boundary charge and its usefulness for the characterization of topological insulators. This is of particular advantage compared to other topological indices, since the transformation laws are perfectly valid in the presence of disorder and interactions, as demonstrated in the present work. 

Of further interest is the specification of the unknown integers in the transformation laws. They are related in a subtle way to bound states occurring at boundaries and interfaces. Therefore, their knowledge is of importance to establish a link between the boundary charge and the appearance of
bound states. This question has been analysed recently in Refs.~\cite{pletyukhov_etal_short,pletyukhov_etal_long} for the special case of noninteracting single-channel and nearest-neighbor hopping models. If only one band is occupied (i.e., $\bar{\rho}={1\over Z}$) it was shown that the difference $(n-n'){1\over Z}-(Q_{B,n}^R-Q_{B,n'}^R)$ is a quantized topological index related to the winding number of the gauge-invariant phase difference of the Bloch wave function between site $m=n$ and $m=n'$. The same index describes the quantity $(n-n'){1\over Z} + (Q_{B,n-1}^L-Q_{B,n'-1}^L)$. In addition, it was found that the sum of the boundary charges left and right to a common boundary is given by the winding number of the phase-difference of the Bloch wave function between the first and last site of the unit cell starting at the boundary \cite{com_LR}. As a result, the topological index defined via the winding number of the phase difference of the Bloch wave function between different sites has a direct physical meaning and controls the transformation laws of the boundary charge in a unique way. Therefore, it will be of high interest in the future to find analogous rules for multi-channel systems via non-abelian versions of these winding numbers \cite{mueller_etal_preprint}. 

The framework developed
in the present work can be straightforwardly generalized to other systems with a conserved quantity like, e.g., the boundary spin occurring in superconducting systems \cite{serina_etal_prb_18} or spin systems \cite{estarellas_etal_scirep_17}. The underlying foundation 
for the transformation laws of the boundary charge is charge conservation and the presence of a gap. Therefore, if the spin in a certain direction is a conserved quantity analogous quantizations of the boundary spin are expected for insulating materials in the presence of symmetries. The same applies for the quantization at interfaces. Moreover, via dimensional reduction, we expect our results to be also of relevance for higher-dimensional systems. 

\section*{Acknowledgments}
We thank A. Altland, V. Meden, and A. Palyi for fruitful discussions. This work was supported by the Deutsche Forschungsgemeinschaft via RTG 1995, the Swiss National Science Foundation (SNSF) and NCCR QSIT and by the Deutsche Forschungsgemeinschaft (DFG, German Research Foundation) under Germany's Excellence Strategy - Cluster of Excellence Matter and Light for Quantum Computing (ML4Q) EXC 2004/1 - 390534769. We acknowledge support from the Max Planck-New York City Center for Non-Equilibrium Quantum Phenomena. Simulations were performed with computing resources granted by RWTH Aachen University. Funding was received from the European Union's Horizon 2020 research and innovation program (ERC Starting Grant, grant agreement No 757725).

M.P. and D.M.K. contributed equally to this work.

\begin{appendix}

\section{Local vs. non-local symmetries}
\label{app:local_nonlocal}

In this Appendix we provide a summary of our conventions to distinguish between local and non-local symmetries. Although this being standard (see, e.g., Ref.~\cite{schnyder_etal_prb_08}), conventions sometimes differ in the literature and the material might be helpful for readers not so familiar with the precise definitions of the various symmetries. 

For a given Hamiltonian $H$, there are four kinds of symmetries, depending on whether the symmetry operation commutes/anticommutes with $H$ and whether it is unitary or anti-unitary
\begin{align}
\label{eq:unitary}
U H U^\dagger = H \quad &,\quad S H S^\dagger = -H\,,\\
\label{eq:anti_unitary}
T H T^\dagger = H \quad &,\quad C H C^\dagger = -H\,.
\end{align}
Here, $U$ and $S$ are unitary operators, whereas $T$ and $C$ are anti-unitary operators. $S$ is called a chiral symmetry, $T$ a time-reversal symmetry, and $C$ a particle-hole (or charge conjugation) symmetry. The anti-unitary symmetries $T$ and $C$ consist of a combination of unitary operations $U_T$ and $U_C$ with complex conjugation $K$: $T=U_T K$ and $C=U_C K$. The operation $K$ of complex conjugation requires a basis in which it is defined. Here, we take always the real-space representation in terms of $|m\sigma\rangle$, where $m$ is the lattice site index and $\sigma$ the channel index. 

To distinguish local from non-local symmetries one needs to specify the unit cell and write the total Hilbert space as a direct product of the space within the unit cell (labeled by the site index $j=1,\dots,Z$ and the channel index $\sigma=1,\dots,N_c$ for each site) and the space of all unit cells labeled by the integer $n$. In the $1$-particle subspace, the tight-binding model (\ref{eq:H_bulk}) can then be alternatively written as
\begin{align}
\label{eq:tight_binding_nj}
H =\sum_{n,\tau}\uuline{h}(\tau) \otimes |n+\tau\rangle\langle n|\,, 
\end{align}
where $\uuline{h}(\tau)$ are $ZN_c\times ZN_c$-matrices describing the coupling of unit cell $n$ with unit cell $n+\tau$ (the lattice site index $m$ used in (\ref{eq:H_bulk}) is related to $n$ and $j$ by $m=Z(n-1)+j$; note that $\tau$ has a different meaning compared to $\delta$ used in (\ref{eq:H_bulk}), the same applies for the symbol $\uuline{h}$). A local symmetry is then defined by a symmetry with respect to the Hamiltonian $\uuline{h}(\tau)$ (i.e., it acts only within the space of a single unit cell) and, in addition, is independent of $\tau$
\begin{align}
\label{eq:unitary_local}
U \uuline{h}(\tau) U^\dagger = \uuline{h}(\tau) \quad &,\quad S \uuline{h}(\tau) S^\dagger = -\uuline{h}(\tau)\,,\\
\label{eq:anti_unitary_local}
T \uuline{h}(\tau) T^\dagger = \uuline{h}(\tau) \quad &,\quad C \uuline{h}(\tau) C^\dagger = -\uuline{h}(\tau)\,.
\end{align}
Using the Fourier transform $\uuline{\tilde{h}}(k)=\sum_\tau \uuline{h}(\tau)e^{-ik\tau}$, with real quasimomentum $-\pi < k < \pi$, this can also be written as
\begin{align}
\label{eq:unitary_local_k}
U \uuline{\tilde{h}}(k) U^\dagger = \uuline{\tilde{h}}(k) \quad &,\quad 
S \uuline{\tilde{h}}(k) S^\dagger = -\uuline{\tilde{h}}(k)\,,\\
\label{eq:anti_unitary_local_k}
T \uuline{\tilde{h}}(-k) T^\dagger = \uuline{\tilde{h}}(k) \quad &,\quad 
C \uuline{\tilde{h}}(-k) C^\dagger = -\uuline{\tilde{h}}(k)\,.
\end{align}
Within our convention, a non-local symmetry can not be written in this form. There are three possibilities: (1) The non-local symmetry can be written as a local one by taking another choice of the unit cell. (2) The non-local symmetry acts within the space of a single unit cell but depends on $\tau$ (or, equivalently, on the quasimomentum $k$). (3) The non-local symmetry does not act within the space of a single unit cell whatever choice one takes for the unit cell, i.e., can only be written with respect to the total Hamiltonian $H$. For $n\ne 0$, the symmetries $\Pi_n$ and $S_n$ defined in Eq.~(\ref{eq:Pi_S_n}) are non-local symmetries within our definition. Examples for cases (1) and (2) are discussed in the paragraph following Eq.~(\ref{eq:S_n_condition}) via special cases for the symmetry $\Pi_n$. The case (1) is discussed extensively at the end of Section~\ref{sec:symmetries} when the local symmetries $\Pi_0$ or $S_0$ are present but not with respect to the unit cell starting at the boundary of the semi-infinite system. For $n\ne 0$, the symmetry $S_n$ is an example for case (3).

\section{Stability of NSP: DMRG analysis}
\label{app:NSP_DMRG}

In this Appendix we analyse the influence of static random disorder and short-ranged electron-electron interaction on the boundary and interface charge by using exact diagonalization and DMRG. For particular examples we demonstrate that the interface charge (\ref{eq:QI_ICT_nn'}) is independent of the interface coupling $V_I$ (up to an integer), and we show that Eqs.~(\ref{eq:QB_RL_translation}) and (\ref{eq:QB_RL_relation}) for the boundary charge are generically valid. 

\begin{figure}
\centering
 \includegraphics[width=\columnwidth]{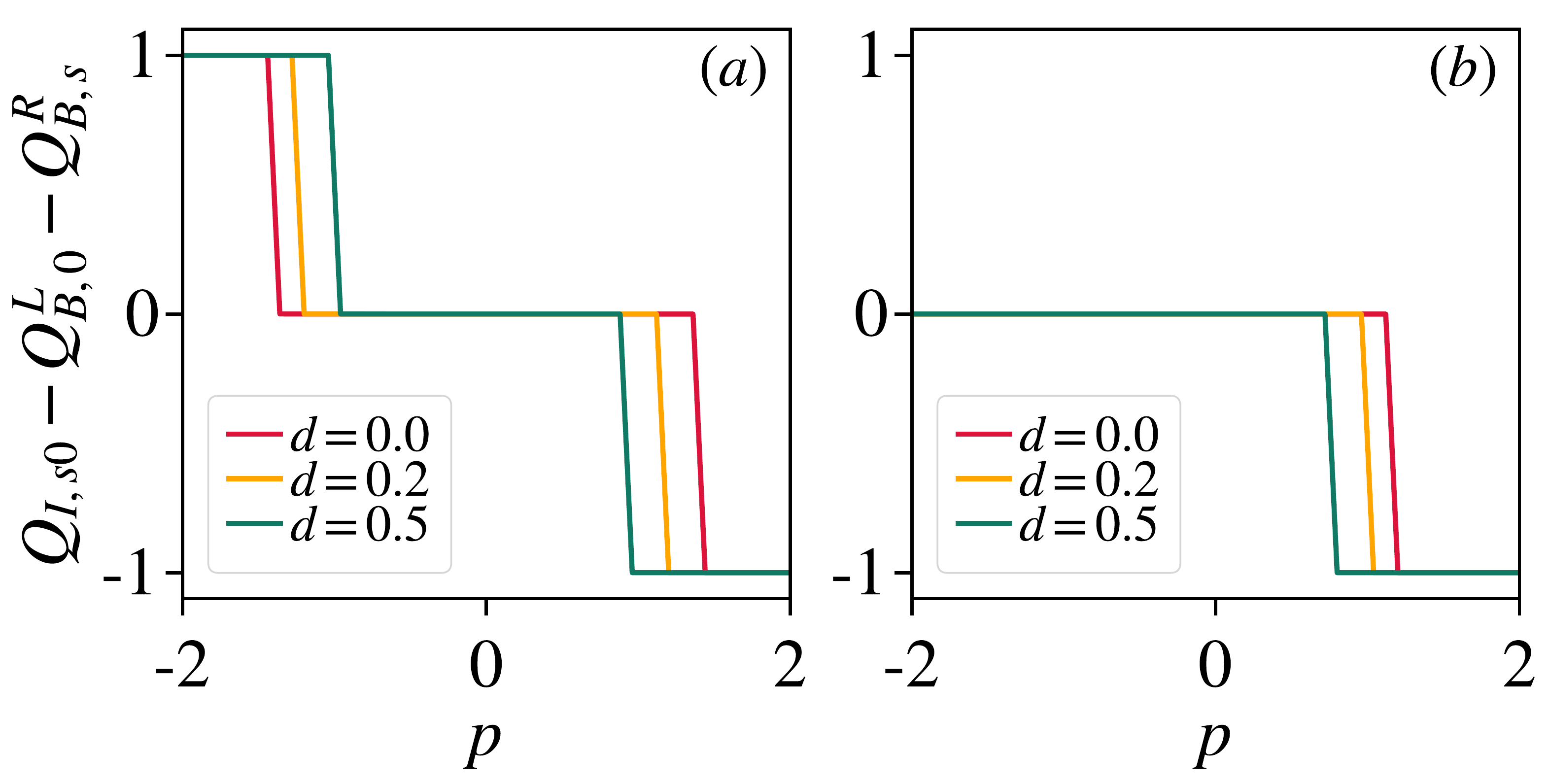}
 \caption{Stability of Eq.~(\ref{eq:QI_ICT_nn'}) with respect to random disorder. The figure shows $Q_{I,s0}-Q_{B,0}^L-Q_{B,s}^R$ for the same parameters as used in the left column ($Z=4$) of  Fig.~\ref{fig:weyl} at  $\varphi=\pi/4$, where the gap is maximal (but at finite size $N=1000$). The relative shift of the chains left and right to the interface are (a) $s=0$, and (b) $s=1$. We take half-filling instead of $\mu=0$ here. Random disorder drawn from a uniform distribution $[-d/2,d/2)$ is added to the onsite potentials and $p$ describes changes to the interface properties (see main text for details). As the properties of the interface are swept through $Q_{I,s0}-Q_{B,0}^L-Q_{B,s}^R$ only changes $\text{mod}(1)$.  }
\label{fig:Inter_dis}
\end{figure}

\begin{figure}
\centering
 \includegraphics[width=\columnwidth]{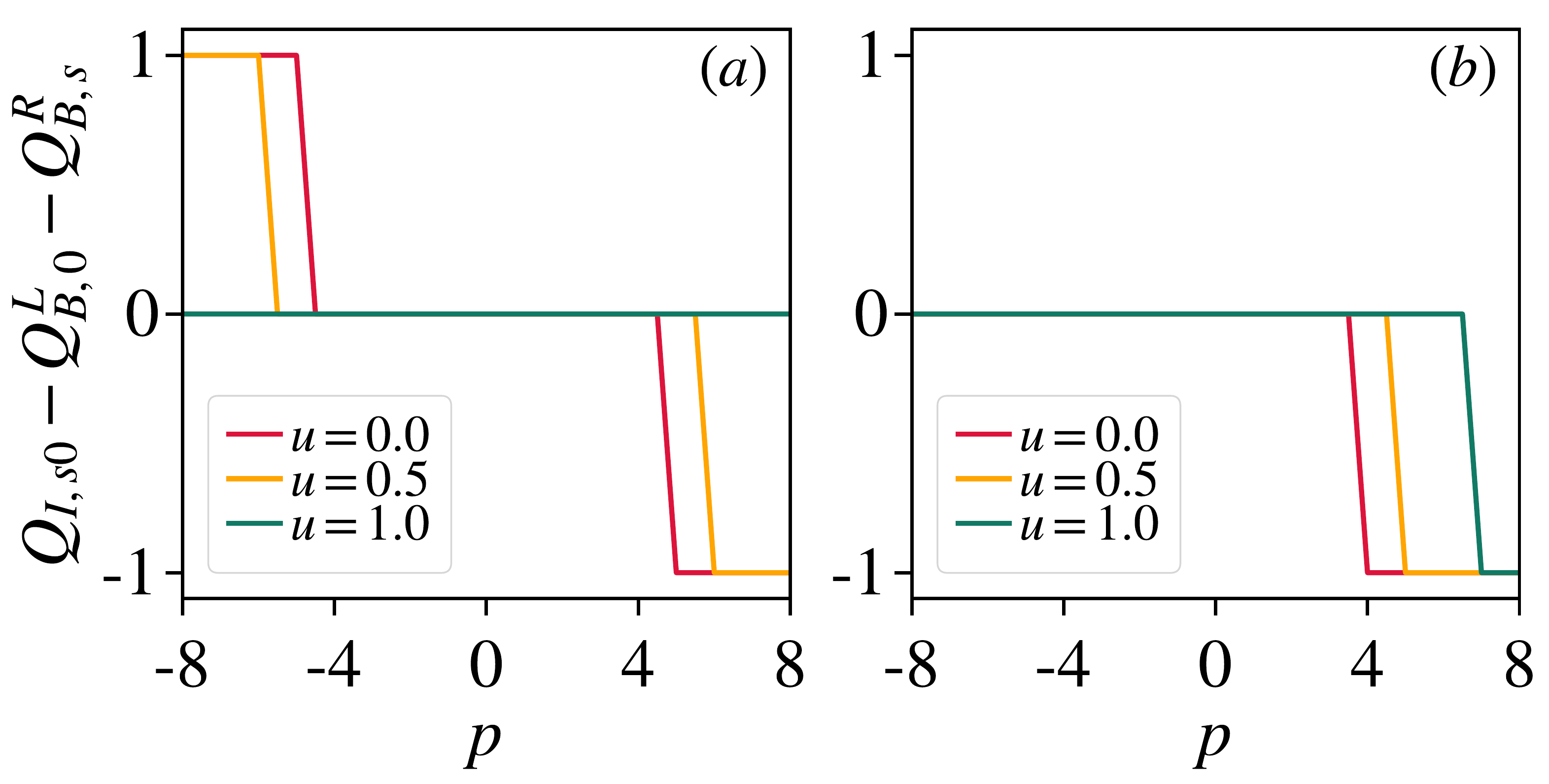}
 \caption{Stability of Eq.~(\ref{eq:QI_ICT_nn'}) with respect to interactions. The figure shows $Q_{I,s0}-Q_{B,0}^L-Q_{B,s}^R$ for the same parameters as used in the left column ($Z=4$) of  Fig.~\ref{fig:weyl} at  $\varphi=\pi/4$, where the gap is maximal (but at finite size $N=200$ and with larger $V/t=1.2$). The relative shift of the chains left and right to the interface are (a) $s=0$, and (b) $s=1$.  We take half-filling instead of $\mu=0$ here.  $p$ describes changes to the interface properties (see main text for details). As the properties of the interface are swept through $Q_{I,s0}-Q_{B,0}^L-Q_{B,s}^R$ only changes $\text{mod}(1)$. }
\label{fig:Inter_U}
\end{figure}

We start with the interface charge and demonstrate in Figs.~\ref{fig:Inter_dis} and Figs.~\ref{fig:Inter_U} that Eq.~(\ref{eq:QI_ICT_nn'}) holds even in the presence of random disorder as well as short-ranged electron-electron interaction, respectively. We consider an interface of the following form: Take initially two decoupled chains of the form as defined by Eq.~\eqref{eq:potential_modulation}. We want to include changes where the potential form of the right chain is shifted in the variable $\varphi$ with respect to the left one by an integer multiple of $\frac{2\pi}{Z}$. This means $\varphi\to \varphi+ s\frac{2\pi}{Z}$, which effectively shifts the  right lattice by $s$ sites compared to the left one. In Eq.~(\ref{eq:QI_ICT_nn'}) this means that $n=s$ and $n'=0$. 

To define a single parameter $p$ which changes the interface's properties continuously we consider the link between the rightmost site of the left lattice to the leftmost site of the right lattice to be $t_{\rm link}=|p|/2$ and add an onsite potential of size $p$ to both of these sites. Therefore, $p=0$ is the decoupled case of two chains without an additional onsite potential at the edge and for negative $p$ charges tend to get trapped at the interface, while for positive $p$ they are pushed out. We add a quenched disorder following Eq.~\eqref{eq:disorder} for the results in Fig.~\ref{fig:Inter_dis} and a electron-electron interaction following Eq.~\eqref{eq:coulomb2} for the results in Fig.~\ref{fig:Inter_U}. Since we concentrate on nearest-neighbor interaction we additionally scale the interaction over the interface bond by $p$, such that $p=0$ is the limit of two decoupled chains. Clearly Eq.~(\ref{eq:QI_ICT_nn'}) remains valid in both cases. 

\begin{figure}
\centering
 \includegraphics[width=0.49\columnwidth]{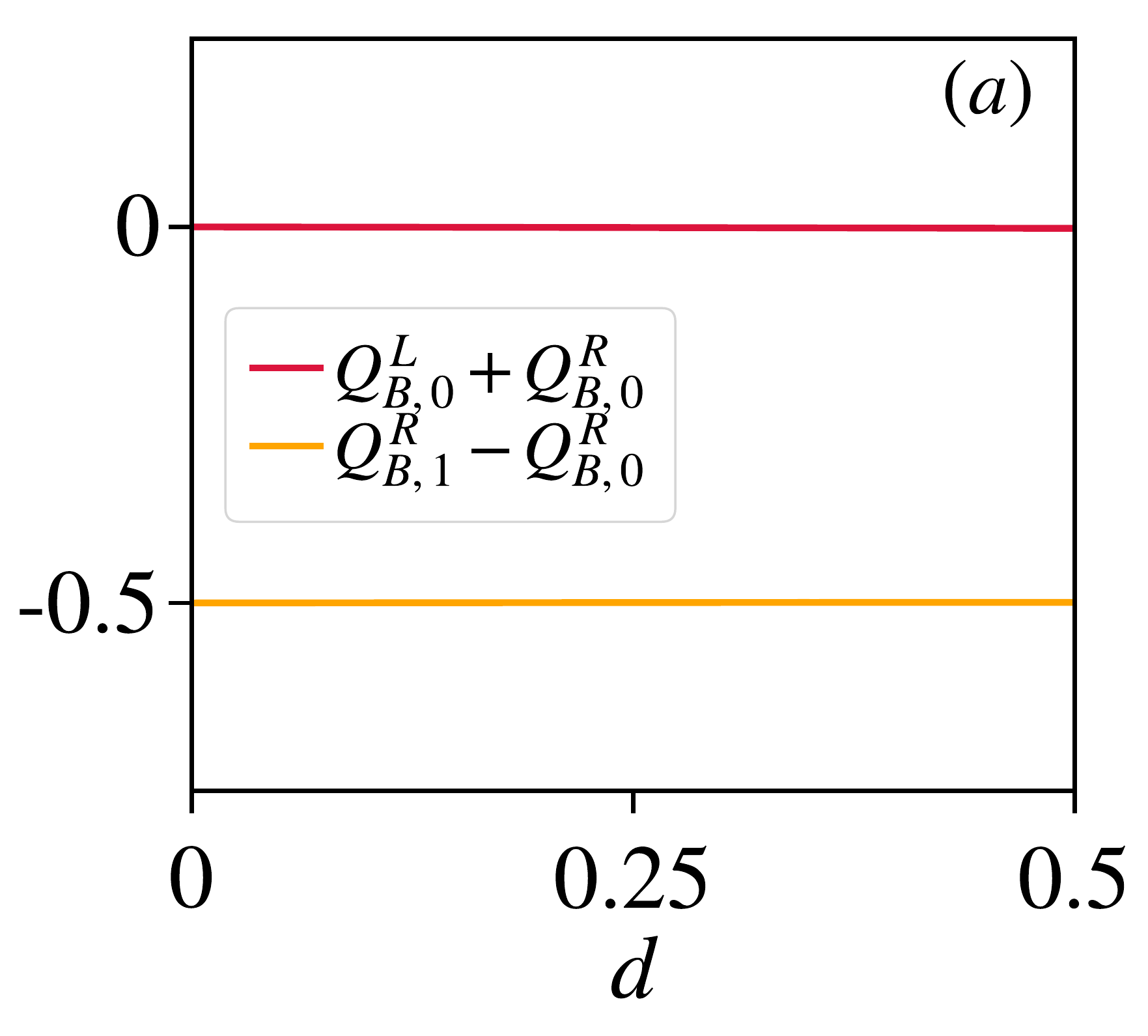}
 \includegraphics[width=0.48\columnwidth]{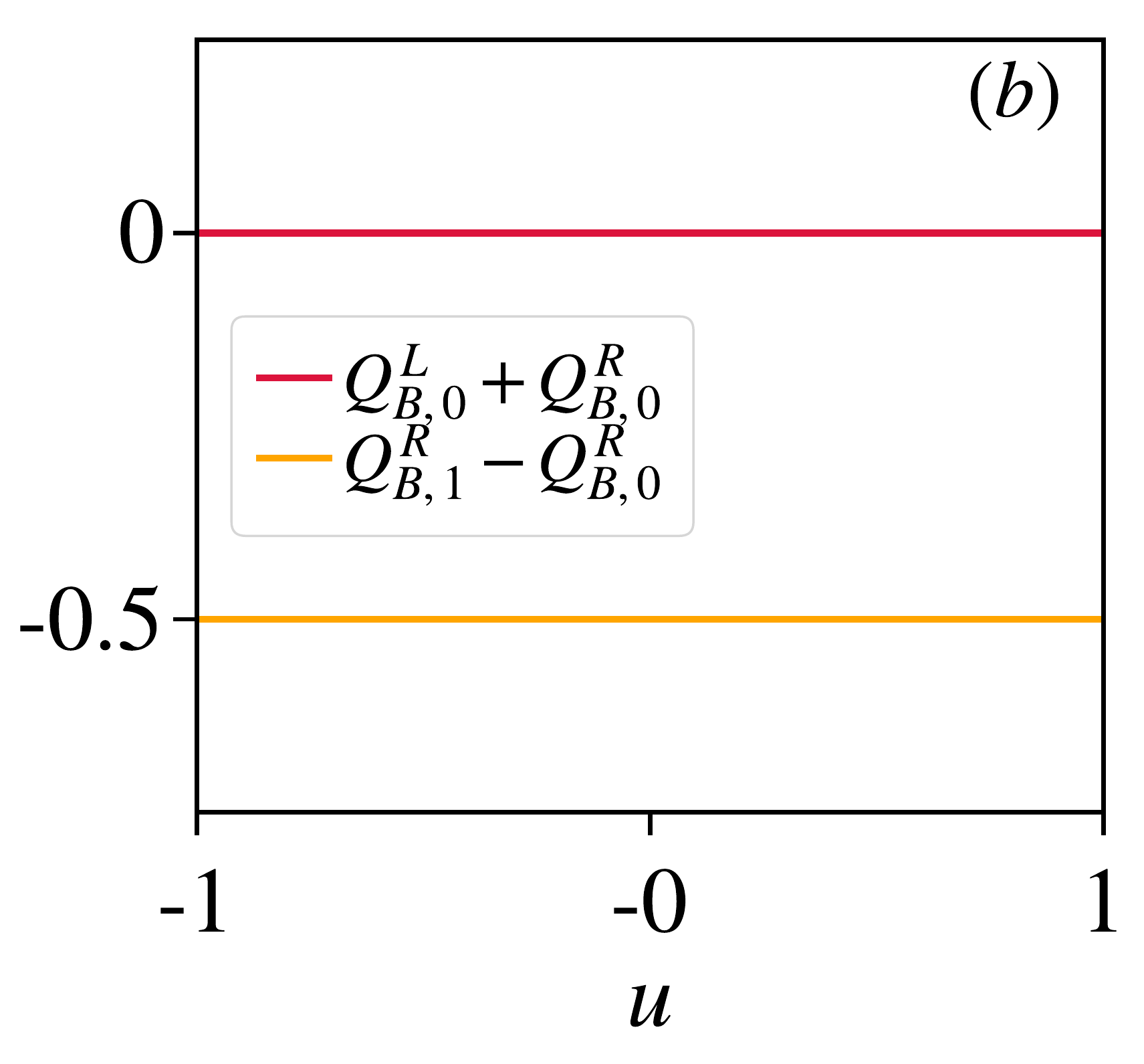}
 \caption{Stability of Eqs.~(\ref{eq:QB_RL_translation}) and (\ref{eq:QB_RL_relation})  with respect to (a) random disorder and (b) interaction.  We work at half filling such that $\bar \rho=1/2$. $Q_{B,1}^R-Q_{B,0}^R=\bar \rho$ $\text{mod}(1)$ is shown to demonstrate Eq.~(\ref{eq:QB_RL_translation}), while  $Q_{B,0}^L+Q_{B,0}^R=0$ $\text{mod}(1)$ illustrates Eq.~(\ref{eq:QB_RL_relation}). The parameters are the same as in Fig.~\ref{fig:Inter_dis}. }
\label{fig:Inter_dis2}
\end{figure}

Next we study the influence of static random disorder and short-ranged electron-electron interaction on the transformation laws (\ref{eq:QB_RL_translation}) and (\ref{eq:QB_RL_relation}) of the boundary charge under translations and local inversion, see Figs.~\ref{fig:Inter_dis2}(a,b). Up to rather large values of the disorder and the electron-electron interaction both transformations laws remain perfectly valid, as expected from the NSP.

\section{Stability of NSP: 1-channel systems}
\label{app:NSP_1_channel}

In this Appendix we demonstrate the validity of Eq.~(\ref{eq:QI_ICT_nn'}) (with $n=n'=0$) for a model of two non-interacting single-channel nearest-neighbor chains coupled with each other via a tunable hopping amplitude. It is explicitly shown that \eqref{eq:QI_ICT_nn'} holds for any strength of the link.

Let us consider the Hamiltonian of the infinite chain $H=H_R + H_L + V_I$ consisting of the three parts
\begin{align}
H_R &= \sum_{n=1}^{\infty} \left\{ |n \rangle \langle n | \otimes h (0)
\right. \nonumber \\
&\quad \left. + |n+1 \rangle \langle n| \otimes h (1) + |n \rangle \langle n+1| \otimes h (-1)\right\}, \\
H_L &= \sum_{n=-\infty}^{0} \left\{ |n \rangle \langle n | \otimes h (0)
\right. \nonumber \\
& \quad \left. + |n \rangle \langle n-1| \otimes h (1) + |n-1 \rangle \langle n| \otimes h (-1) \right\}, \\
V_I &= \lambda [ |n=1\rangle \langle n=0 | \otimes h (1) \nonumber \\
&\quad + |n=0\rangle \langle n=1 | \otimes h (-1)],
\end{align}
which describe the right semi-infinite chain, the left semi-infinite chain, and the tunneling between them, respectively. Here, in contrast to the lattice site index $m$, the index $n$ enumerates unit cells. Both $H_R$ and $H_L$ describe the lattices with the same structure of a unit cell, which is encoded in $h(0)= \sum_{j=1}^Z v_j |j\rangle \langle j| - \sum_{j=1}^{Z-1} t_j (| j \rangle \langle j+1|+ |j+1 \rangle \langle j|)$, $h(1)= - t_Z |j=1\rangle \langle j=Z|$, and $h(-1)= h^{\dagger} (1)$, i.e. characterized by $Z$ sites $j=1,\dots,Z$ per unit cell, a single orbital (channel) per site, and by the same values for hoppings $t_j$ and onsite potentials $v_j$. The tunneling amplitude $\lambda t_Z$ between the two subsystems is quantified by the real-valued parameter $\lambda \geq 0$. Its special values $\lambda=0$ and $\lambda=1$ correspond to the cases of the two decoupled semi-infinite chains and the translationally invariant infinite chain, respectively. A restoration of the translational symmetry in the latter case is guaranteed by the perfect matching of the unit cells touching each other at the interface.

Due to the same structure, the Hamiltonians $H_R$ and $H_L$ appear to be isospectral, and their extended eigenstates can be therefore labeled by the same band index $\alpha$ and quasimomentum $k$ on the both sides from the interface. Moreover, these quantum numbers can be also used for a construction of scattering eigenstates of the coupled system, since eigenenergies $\epsilon_k^{(\alpha)}$ of the extended states remain independent of $\lambda$, and they can be ultimately evaluated from the bulk Hamiltonian, i.e. at $\lambda=1$. 
On the basis of this observation, we make the following ansatz for the two distinct scattering eigenstates additionally labeled by either $r$ or $l$:
\begin{align}
 \psi_k^{(\alpha,r)} (n,j) &= \frac{\Theta_{n \leq 0}}{\sqrt{2 \pi}} [\chi_k^{(\alpha)} (j) e^{i k n} + r_k^{(\alpha)} \chi_{-k}^{(\alpha)} (j) e^{-i k n} ]  \nonumber \\
&+  \frac{\Theta_{n \geq 1}}{\sqrt{2 \pi}} t_k^{(\alpha)} \chi_k^{(\alpha)} (j) e^{i k n} , \label{rss} \\
 \psi_k^{(\alpha,l)} (n,j) &=    \frac{\Theta_{n \geq 1}}{\sqrt{2 \pi}} [\chi_{-k}^{(\alpha)} (j) e^{-i k n} +   r'_k \phantom{}^{(\alpha)} \chi_{k}^{(\alpha)} (j) e^{i k n} ] \nonumber \\
&+ \frac{\Theta_{n \leq 0}}{\sqrt{2 \pi}} t'_k \phantom{}^{(\alpha)} \chi_{-k}^{(\alpha)} (j) e^{-i k n},
\label{lss}
\end{align}
with $k \in [0, \pi]$. Both $\psi_k^{(\alpha,r)}$ and $\psi_k^{(\alpha,l)}$ as well as the bulk Bloch states $\chi_{\pm k}^{(\alpha)}$ correspond to the eigenenergy $\epsilon_k^{(\alpha)}$.
In the following, we focus on the band $\alpha$ and omit the band index for brevity.

Inserting the ansatz \eqref{rss} and \eqref{lss} into the eigenvalue problem, we establish the scattering matrix 
\begin{align}
S_k = \left( \begin{array}{cc} t_k & r'_k \\ r_k & t'_k \end{array}\right).
\end{align}
Its components read
\begin{align}
t_k &= t'_k = \lambda \frac{e^{2 i \phi_k}-1}{e^{2 i \phi_k}-\lambda^2}, \\
r_k &= \frac{\lambda^2-1}{e^{2 i \phi_k}-\lambda^2} e^{2 i \varphi_k (Z)}, \\
r'_k &= \frac{\lambda^2  - 1}{e^{2 i \phi_k} - \lambda^2}  e^{2 i \phi_k} e^{-2 i \varphi_k (Z)},
\end{align}
where $\phi_k = \varphi_k (Z) - \varphi_k (1) -k$ is a gauge-invariant phase difference expressed in terms of the gauge-dependent phases $\varphi_k (j)$ of the complex-valued components $\chi_k (j)$. By a direct calculation one can confirm the unitarity property $S_k^{\dagger} S_k =1$, which implies both the orthogonality of $\psi_k^{(\alpha,r)}$ and $\psi_k^{(\alpha,l)}$ as well as their proper normalization.

The interface charge $Q_I^{(\alpha)}$ associated with the band $\alpha$ consists of the Friedel part $Q_F^{(\alpha)}$ and the polarization part $Q_P^{(\alpha)}$ by using the following splitting based on Eq.~(\ref{eq:QB_I})
\begin{align}
\label{eq:QI_splitting}
Q_I^{(\alpha)} &= Q_F^{(\alpha)} + Q_P^{(\alpha)}\,,\\
\label{eq:QF}
Q_F^{(\alpha)} &= \sum_{m=-\infty}^\infty \Big[\rho^{(\alpha)}(m)-\rho^{(\alpha)}_{\text{bulk}}(m)\Big] f(m) \,,\\
\label{eq:QP}
Q_P^{(\alpha)} &= \sum_{m=-\infty}^\infty \Big[\rho^{(\alpha)}_{\text{bulk}}(m)-{1\over Z}\Big] f(m) \,,
\end{align}
with $m=Z(n-1)+j$. Here, $\rho^{(\alpha)}_{\text{bulk}}(m)={1\over 2\pi}\int_{-\pi}^\pi dk |\chi_k^{(\alpha)}(j)|^2$ is the contribution from band $\alpha$ to the charge at site $m$ from the bulk Hamiltonian. As shown in Eq.~(\ref{eq:QI_bulk}), the polarization part to the interface charge vanishes $Q_P^{(\alpha)}=0$. In turn, the Friedel part $Q_F^{(\alpha)} \equiv Q_F$ amounts to 
\begin{align}
Q_F   &=  \sum_{n=-\infty}^0 \int_{-\pi}^{\pi} \frac{d k}{2 \pi} r_k^* \sum_{j=1}^Z \chi_k^2 (j) e^{2 i (k-i \eta) n} \\
&+  \sum_{n=1}^{\infty} \int_{-\pi}^{\pi} \frac{d k}{2 \pi} r_k' \sum_{j=1}^Z \chi_k^2 (j) e^{2 i (k+ i \eta) n} \\
&= -1+ \int_{-\pi}^{\pi} \frac{d k}{2 \pi} (r_k' - r_k^*) \sum_{j=1}^Z \chi_k^2 (j) \frac{i e^{ik}}{2 \sin k},
\end{align}
where $\eta \to 0^+$ is a convergence factor. The last equality is only valid for $\lambda^2 \neq 1$, since the limits $\lambda \to 1$ and $\eta \to 0^+$ do not commute. In the translationally invariant case $\lambda =1$ there is no reflection at the interface, and one simply gets $Q_F =0$.

In the following we prove that in general $Q_F$ takes only integer values for arbitrary $\lambda$.

Let us introduce the two gauges: I) $\chi_k^{\textup{I}}$ with $e^{2 i \varphi_k (Z)}=1$, i.e. the last component is real; II) $\chi_k^{\textup{II}}$ with $e^{2 i \varphi_k (1)}=1$, i.e. the first component is real. Apparently, $\chi_k^{\textup{II}} = e^{i \phi_k + i k} \chi_k^{\textup{I}}$. 

Next, we express the quantity $Q_F  +1$ in the mixed form
\begin{align}
  & \int_{-\pi}^{\pi} \frac{d k}{2 \pi}  \left[ -\frac{\lambda^2  - 1 }{ \lambda^2 - e^{2 i \phi_k}   }  e^{2 i \phi_k } \sum_{j=1}^Z [\chi_k^{\textup{I}} (j)]^2  \frac{i e^{ik}}{2 \sin k}  \right. \nonumber \\
  & \left. \qquad +  \frac{\lambda^2  - 1 }{ \lambda^2 - e^{-2 i \phi_k} }  e^{-2 i \phi_k}  \sum_{j=1}^Z [\chi_k^{\textup{II}} (j)]^2  \frac{i e^{-ik}}{2 \sin k} \right].
  \label{qf_init}
\end{align}

In Ref.~\cite{pletyukhov_etal_long} we established that the components of the Bloch state in the gauge I have the form
\begin{align}
\chi_k^{\textup{I}} (j) &= \frac{f_j^{\textup{I}} e^{-ik} + g_j^{\textup{I}}}{\sqrt{N_k^{\textup{I}}}} , \quad 1 \leq j \leq Z-1, 
\label{chi_all} \\ 
\chi_k^{\textup{I}} (Z) &=\frac{s}{\sqrt{N_k^{\textup{I}}}}, 
\label{chiZ}
\end{align}
where $f_j^{\textup{I}}$, $g_j^{\textup{I}}$, and $s$ are real-valued polynomial functions of $\epsilon_k$, and $N_k^{\textup{I}} = s^2 + \sum_{j=1}^{Z-1} |f_j^{\textup{I}} e^{-ik} + g_j^{\textup{I}}|^2$. In that paper we also noted the following relations
\begin{align}
 & -  \textup{Im} [\chi_k^{\textup{I} \, \dagger} \frac{d}{d k} \chi_k^{\textup{I}}] =  \sum_{j=1}^{Z-1} \frac{(f^{\textup{I}}_j)^2 + f^{\textup{I}}_j g^{\textup{I}}_j \cos k}{N_k^{\textup{I}}}, \label{bcI} \\
& \sum_{j=1}^Z [\chi_k^{\textup{I}}  (j)]^2 \frac{i e^{i k}}{2 \sin k}  = \frac{i e^{ik}}{2 \sin k} + \sum_{j=1}^{Z-1} \frac{(f_j^{\textup{I}})^2 + f_j^{\textup{I}} g_j^{\textup{I}} e^{i k}}{N_k^{\textup{I}}} \nonumber \\
&= -  \textup{Im} [\chi_k^{\textup{I} \, \dagger} \frac{d}{d k} \chi_k^{\textup{I}}]  +  \frac{i e^{ik}}{2 \sin k}  +   \frac{i \sin k}{N_k^{\textup{I}}} \sum_{j=1}^{Z-1}  f_j^{\textup{I}} g_j^{\textup{I}} .
\label{qbI}
\end{align}
In addition, it is also possible to derive the relation
\begin{align}
\frac{ \sin k}{N_k^{\textup{I}}}\sum_{j=1}^{Z-1} f_j^{\textup{I}} g_j^{\textup{I}} = -  \frac{1}{2 s} \frac{d s}{d k} .
\end{align}
Hence,
\begin{align}
     \sum_{j=1}^Z [\chi_k^{\textup{I}}  (j)]^2 \frac{i e^{i k}}{2 \sin k} &= -  \textup{Im} [\chi_k^{\textup{I} \, \dagger} \frac{d}{d k} \chi_k^{\textup{I}}]  - \frac12 \nonumber \\
     &+  \frac{i \cos k}{2 \sin k}  -   \frac{i}{2 s} \frac{d s}{d k} .
\label{qbIa}
\end{align}

Let us now establish similar relations for $\chi_k^{\textup{II}}$. We note that the Bloch state
\begin{align}
    \bar{\chi}_k^{\textup{II}} = \left( \begin{array}{cc} 0 & e^{-i k}  1_{(Z-1)\times (Z-1)} \\ 1 & 0 \end{array} \right) \chi_k^{\textup{II}}
    \label{bar_chi2}
\end{align}
is the eigenstate corresponding to the re-defined unit cell, which begins with the site 2, has the pre-last site $Z$, and ends with the site 1. Moreover, the component $\bar{\chi}_k^{\textup{II}} (Z)$ is real, and then by analogy with \eqref{qbIa} it holds
\begin{align}
 \sum_{j=1}^Z [\bar{\chi}_k^{\textup{II}}  (j)]^2 \frac{i e^{i k}}{2 \sin k}  &= -  \textup{Im} [\bar{\chi}_k^{\textup{II} \, \dagger} \frac{d}{d k} \bar{\chi}_k^{\textup{II}}]  - \frac12 \nonumber \\
&+  \frac{i \cos k}{2 \sin k} - \frac{i}{2 \bar{s}} \frac{d \bar{s}}{d k} ,
\label{qbIIa}
\end{align}
where $\bar{s}$ is a part of the representation for $\bar{\chi}_k^{\textup{II}}$, which is analogous to \eqref{chi_all}, \eqref{chiZ}.

From \eqref{bar_chi2} it follows that
\begin{align}
     \textup{Im} [\bar{\chi}_k^{\textup{II} \, \dagger} \frac{d}{d k} \bar{\chi}_k^{\textup{II}}]= \textup{Im} [\chi_k^{\textup{II} \, \dagger}  \frac{d}{d k}  \chi_k^{\textup{II}} ] + |\bar{\chi}_k^{\textup{II}} (Z)|^2 -1,
\end{align}
and thus we find that
\begin{align}
    & \sum_{j=1}^Z [\chi_k^{\textup{II}} (j)]^2  \frac{i e^{-ik}}{2 \sin k} = \sum_{j=1}^Z [\bar{\chi}_k^{\textup{II}} (j)]^2  \frac{i e^{ik}}{2 \sin k} + |\bar{\chi}_k^{\textup{II}} (Z)|^2 \nonumber \\
    &  = -\textup{Im} [\chi_k^{\textup{II} \, \dagger}  \frac{d}{d k}  \chi_k^{\textup{II}} ]  +\frac12+  \frac{i \cos k}{2 \sin k}  -   \frac{i}{2 \bar{s}} \frac{d \bar{s}}{d k}.
    \label{qbIIb}
\end{align}

Inserting \eqref{qbIa} and \eqref{qbIIb} into \eqref{qf_init} and accounting the symmetry properties of integrands under the reflection $k \to -k$, we obtain
\begin{align}
& Q_F  +1 =
   \int_{-\pi}^{\pi} \frac{d k}{2 \pi}  \left\{ \textup{Re} \left[  \frac{\lambda^2  - 1 }{ \lambda^2 - e^{2 i \phi_k}   }  e^{2 i \phi_k } \right] \right. \nonumber \\
   & \times \left( \textup{Im} [\chi_k^{\textup{I} \, \dagger} \frac{d}{d k} \chi_k^{\textup{I}}]  -\textup{Im} [\chi_k^{\textup{II} \, \dagger}  \frac{d}{d k}  \chi_k^{\textup{II}} ]  +1 \right) \nonumber \\
   & \left. + \textup{Im} \left[  \frac{\lambda^2  - 1 }{ \lambda^2 - e^{2 i \phi_k}   }  e^{2 i \phi_k } \right] \left(  \cot k  -   \frac{1}{2} \frac{d \ln (s \bar{s})}{d k}  \right) \right\} .
  \label{qf}
\end{align}

From the transformation between the two gauges we obtain the relation
\begin{align}
    \textup{Im} [\chi_k^{\textup{I} \, \dagger} \frac{d}{d k} \chi_k^{\textup{I}}]  -\textup{Im} [\chi_k^{\textup{II} \, \dagger}  \frac{d}{d k}  \chi_k^{\textup{II}} ] = -1 - \frac{d \phi_k}{d k}.
    \label{id1_imp}
\end{align}
A less obvious identity
\begin{align}
s \bar{s} \sin^2 \phi_k = g_1^{\textup{I}}  \bar{f}^{\textup{II}}_{Z-1}  \sin^2 k \equiv \left( \prod_{j=1}^{Z-1} t_j^2\right) \sin^2 k
\label{id_non}
\end{align}
follows from the identifications
\begin{align}
      \frac{\bar{s}}{\sqrt{\bar{N}_k^{\textup{II}}}} e^{-i \phi_k } 
      &= \frac{f_1^{\textup{I}}  + g_1^{\textup{I}} e^{i k}}{\sqrt{N_k^{\textup{I}}}},  \\
 \frac{s}{\sqrt{N_k^{\textup{I}}}} e^{i \phi_k}  &= \frac{\bar{f}_{Z-1}^{\textup{II}} e^{-i k} + \bar{g}_{Z-1}^{\textup{II}} }{\sqrt{\bar{N}_k^{\textup{II}}}} ,
\end{align}
and the observation $g_1^{\textup{I}}=\bar{f}^{\textup{II}}_{Z-1} = \prod_{j=1}^{Z-1} t_j$ which can be made on the basis of expressions quoted in Ref.~\cite{pletyukhov_etal_long}.
Differentiating \eqref{id_non} with respect to $k$ yields
\begin{align}
     \cot k  -   \frac{1}{2} \frac{d \ln (s \bar{s})}{d k}  =
 \cot \phi_k \frac{d \phi_k}{d k} .
 \label{id2_imp}
\end{align}

With help of \eqref{id1_imp} and \eqref{id2_imp} we cast \eqref{qf} to 
\begin{align}
 Q_F +1 &=
   \int_{-\pi}^{\pi} \frac{d k}{2 \pi}  \left\{- \textup{Re} \left[  \frac{\lambda^2  - 1 }{ \lambda^2 - e^{2 i \phi_k}   }  e^{2 i \phi_k } \right] \frac{d \phi_k}{d k} \right. \nonumber \\
   & \left. + \textup{Im} \left[  \frac{\lambda^2  - 1 }{ \lambda^2 - e^{2 i \phi_k}   }  e^{2 i \phi_k } \right]  \cot \phi_k \frac{d \phi_k}{d k}  \right\}. \label{qfa}
\end{align}
Making the change of the integration variable $k \to \phi_k$ and accounting possible multiple windings of the phase $\phi_k$, which are quantified by the integer winding number $\textup{wn} [\phi_k] = \int_{-\pi}^{\pi} \frac{d k}{2 \pi i} e^{-i \phi_k} \frac{d}{d k} e^{i \phi_k}$, we express 
\begin{align}   
  Q_F  +1 &= \textup{wn} [\phi_k] \int_{-\pi}^{\pi} \frac{d \phi}{2 \pi}  \left\{- \textup{Re} \left[  \frac{\lambda^2  - 1 }{ \lambda^2 - e^{2 i \phi}   }  e^{2 i \phi } \right]  \right. \nonumber \\
   & \left. + \textup{Im} \left[  \frac{\lambda^2  - 1 }{ \lambda^2 - e^{2 i \phi}   }  e^{2 i \phi } \right]  \cot \phi \right\} \nonumber \\
   &= \textup{wn} [\phi_k] \, \textup{sign} ( \lambda^2 -1 ).
  \label{qfb}
\end{align}

For the two decoupled chains ($\lambda =0$), we obtain
\begin{align}
    Q_F^{(\alpha)} &= -1 - \textup{wn} [\phi_k^{(\alpha)}] \nonumber \\
    &= \textup{wn} [\varphi_k^{(\alpha)} (1) - \varphi_k^{(\alpha)} (Z)],
\end{align}
i.e. an integer number. This result persists in the whole range $0 \leq \lambda <1$.

We conclude that the total interface charge $Q_I(\lambda)$, which might also include integer edge state contributions, is a sum of integers for any $\lambda$, and therefore $Q_I(\lambda) = 0 \,\, \text{mod}(1)$. On the other hand, since $Q_B^R+Q_B^L = Q_I(\lambda=0)$ by definition, we find $Q_B^R+Q_B^L = Q_I(\lambda) \,\, \text{mod} (1)$, in agreement with \eqref{eq:QI_ICT} for the model discussed in this Appendix.

\section{Symmetries for single-channel and nearest-neighbor hopping models}
\label{app:single_channel_nearest_neighbor}

In this Appendix we prove the symmetry conditions (\ref{eq:Pi_condition_special}) and
(\ref{eq:S_condition_special}) for the special case of a tight-binding model with one channel $N_c=1$ and nearest-neighbor hopping $\delta=0,\pm 1$. In this case, the model is parametrized by $Z$ on-site potentials $v_m=v_{m+Z}$ and $Z$ nearest-neighbor hoppings $t_m=t_{m+Z}$ defined by 
\begin{align}
\label{eq:potential}
v_m = h_m(0) = v_m^* \quad,\quad t_m = - h_m(1) = - h_{m+1}(-1)^* \,.
\end{align} 
Without loss of generality one can choose all $t_m>0$ real and positive since possible phases can be gauged away by a unitary transformation (see, e.g., Appendix A in 
Ref.~\cite{pletyukhov_etal_long} for a proof). The unitary transformation $U_m$ must be a phase factor
\begin{align}
\label{eq:U_phase_factor}
U_m = e^{i\varphi_m}\quad,\quad \varphi_m = \varphi_{m+Z}\,.
\end{align}
Inserting these equations in the symmetry condition (\ref{eq:Pi_n_condition}) for $\Pi_n$ we find
\begin{align}
\label{eq:Pi_potential}
v_m &= v_{Z-m-n+1}\,,\\
\label{eq:Pi_hopping}
t_m &= e^{-i(\varphi_{Z-m-n}-\varphi_{Z-m-n+1})}t_{Z-m-n}\,.
\end{align}
Since $t_m$ and $t_{Z-m-n}$ are both positive this can only be fulfilled for $U_m=U_{m+1}$ which is just a homogeneous and trivial phase factor. Therefore, we can set $U_m=1$ and find the condition
(\ref{eq:Pi_condition_special}).

Considering the other symmetry condition (\ref{eq:S_n_condition}) for $S_n$ we find
\begin{align}
\label{eq:S_potential}
v_m &= - v_{m-n}\,,\\
\label{eq:S_hopping}
t_m &= - e^{-i(\varphi_{m-n+1}-\varphi_{m-n})}t_{Z-m-n}\,.
\end{align}
Since $t_m$ and $t_{Z-m-n}$ are both positive this requires $U_m=-U_{m+1}$ which, up to an unimportant common phase factor, is only realized for $U_m=(-1)^m$. This proves the condition (\ref{eq:S_condition_special}).

\section{Boundary charge for Dirac model}
\label{app:QB_dirac}

In this Appendix we determine all eigenstates of the semi-infinite Dirac model (\ref{eq:H_eff_semi_infinite}) and prove Eqs.~(\ref{eq:QB_edge}) and (\ref{eq:QB_scattering}). We start with solving the eigenvalue equation
\begin{align}
\label{eq:eigenvalue_semiinfinite}
[-iv_{F}\sigma_{3}\partial_{x} + |\Delta | (\sigma_{+}e^{i\alpha}+\sigma_{-}e^{-i\alpha})]\uline{\psi}(x)=\epsilon\uline{\psi}(x),
\end{align}
with $\sigma_\pm={1\over 2}(\sigma_x\pm i\sigma_y)$, $\uline{\psi}(x)=\begin{pmatrix}R(x), & L(x)  \end{pmatrix}^{T}$, and the boundary condition $R(0)+L(0)=0$. There are two distinct spectral regions: I) $ |\epsilon|< |\Delta |$, and II) $|\epsilon|> |\Delta |$. In the region $\text{I}$ we find a single bound state solution for $\sin\alpha>0$ at energy $\epsilon=-|\Delta | \cos\alpha$, whose wavefunction is given by
\begin{align}
\label{eq:psi_bound}
\uline{\psi}^{\text{I}}(x)=\sqrt{\kappa}\,\begin{pmatrix}1 \\ 
-1 \end{pmatrix}e^{-\kappa x}, 
\end{align}
with $\kappa=\frac{|\Delta| \,|\sin\alpha|}{v_{F}}$. 
In the second (II) spectral region we find a continuum of scattering states labeled by the momentum $k\in[0, \infty)$ and corresponding to the two bands with energies 
$\epsilon_{k, \pm}=\pm\sqrt{v_{F}^{2}k^{2}+|\Delta |^{2}} \equiv \pm\epsilon_{k}$. The eigenstates of the lower (valence) band have the following form
\begin{align}
\nonumber
\uline{\psi}_k(x)&=
{1\over\sqrt{2\pi N_k}}\left[
\begin{pmatrix}-|\Delta | e^{i\alpha} \\ v_{F}k+\epsilon_{k} \end{pmatrix} e^{ikx} \right.\\
\label{eq:psi_k}
&\left.\hspace{1cm} 
- s_k\, 
\begin{pmatrix}-|\Delta | e^{i\alpha} \\ -v_{F}k+\epsilon_{k} \end{pmatrix} e^{-ikx}\right]\,,
\end{align}
with the normalization factor 
\begin{align}
\label{eq:N_k}
N_{k}=|\Delta|^{2}+(v_{F}k+\epsilon_{k})^{2}=2\epsilon_{k}(\epsilon_{k}+v_{F}k)\,,
\end{align}
and  
\begin{align}
\label{eq:s_k}
s_k=\frac{|\Delta | e^{i\alpha}-v_{F}k-\epsilon_{k}}
{|\Delta | e^{i\alpha}+v_{F}k-\epsilon_{k}}\, .
\end{align}
We note the helpful properties
\begin{align}
\label{eq:s_k_properties}
|s_k|^2 = {\epsilon_k + v_F k \over \epsilon_k - v_F k}\quad,\quad
s_k s_{-k} = 1 \,.
\end{align}

Assuming that the chemical potential is located at the bottom of the conduction band, the bound state is occupied for $0<\alpha<\pi$, and all valence band states $\uline{\psi}_k$ are filled. Neglecting the strongly oscillating parts (providing unimportant corrections of $O({\Delta\over v_F k_F})\ll 1$), the contribution of each eigenstate to the density is given by  
\begin{align}
\label{eq:density}
\rho_{\psi}(x) = \uline{\psi}^\dagger(x)\uline{\psi}(x) = |R(x)|^2 + |L(x)|^2\,.
\end{align}
We denote the contributions of the eigenstates $\uline{\psi}^{\text{I}}$ and $\uline{\psi}_k$ to the physical density by $\rho_{\text{I}}(x)$ and $\rho_k(x)$, respectively. This gives for the total density relative to the average bulk density $\bar{\rho}$
\begin{align}
\label{eq:rho_total}
\rho(x) - \bar{\rho} &= \rho_{\text{I}}(x) + \delta\rho_{\text{II}}(x)\,,\\
\label{eq:delta_rho_II}
\delta\rho_{\text{II}}(x) &= \int_0^\infty dk \,\left[\rho_k(x)-{1\over\pi}\right]\,,
\end{align}
and, according to the definition (\ref{eq:QB_R}), the boundary charge follows from
\begin{align}
\label{eq:QB_total}
Q_B &= \int_0^\infty dx \,[\rho(x)-\bar{\rho}]f(x) 
= Q_B^{\text{I}} + Q_B^{\text{II}}\,,\\
\label{eq:QB_bound}    
Q_B^{\text{I}} &= \int_0^\infty dx \,\rho_{\text{I}}(x) f(x) \,,\\
\label{eq:QB_II}    
Q_B^{\text{II}} &= \int_0^\infty dx \,\delta\rho_{\text{II}}(x)f(x) \,.
\end{align}
For the envelope function $f(x)$ we choose the form  $f(x)=e^{-\eta{x}}$ with infinitesimally small $\eta \to 0^+$.  

The bound state is occupied for $0<\alpha<\pi$ and gives an integer contribution to the boundary charge
\begin{align}
\label{eq:QB_bound_result}
Q_{B}^{\text{I}} = \int_{0}^{\infty}dx|\uline{\psi}^{\text{I}}(x)|^{2} = \Theta_{0<\alpha<\pi}\,.
\end{align}
This proves Eq.~(\ref{eq:QB_edge}). 

To calculate the scattering part $Q_B^{\text{II}}$ to the boundary charge we use (\ref{eq:psi_k}), (\ref{eq:N_k}), (\ref{eq:s_k}), and (\ref{eq:s_k_properties}) and find after a straightforward calculation
\begin{align}
\nonumber
\delta\rho_{\text{II}}(x)&= -\frac{|\Delta |}{2\pi}\int_{-\infty}^{\infty}dk\,
\frac{e^{2 i k x}}{\epsilon_{k}} \nonumber \,\\
\label{eq:delta_rho_II_result}
& \hspace{1cm}
\times \frac{|\Delta |-\epsilon_{k}\cos\alpha  -i v_{F} k \sin\alpha}{\epsilon_{k} -|\Delta | \cos \alpha  }\,.
\end{align}
Inserting this result in (\ref{eq:QB_II}) and performing the integration over $x$ we obtain 
\begin{align}
Q^{\text{II}}_{B} &=-\frac{1}{4} -\frac{|\Delta |}{4\pi }\int_{-\infty}^{\infty}dk\,
 \frac{ v_{F}  \sin\alpha}{\epsilon_{k} (\epsilon_{k} -|\Delta | \cos \alpha )  }\, \nonumber \\
\label{eq:QB_II_result}
& =- \frac14+ \frac{\ln(-e^{i\alpha})}{2\pi{i}}.
\end{align}
This proves Eq.~(\ref{eq:QB_scattering}).

\section{Boundary charge at zero gap}
\label{app:zero_gap}

For the tight-binding model $H_0$, given by (\ref{eq:H_0}), restricted to the semi-infinite system $m>0$, the eigenfunctions are given by (we set $a=1$)
\begin{align}
\label{eq:H_0_eigenstates}
\psi_k(m) = {1\over \sqrt{2\pi}} (e^{ikm}-e^{-ikm})\,,
\end{align}
with $0<k<\pi$. For filling $\bar{\rho}=k_F/\pi$, this leads to the following charge $\rho(m)$ at site $m$
\begin{align}
\nonumber
\rho(m)&=\int_0^{k_F} dk\,|\psi_k(m)|^2\\
\label{eq:charge_m}
&=-{1\over 2\pi}\int_{-k_F}^{k_F}dk\,e^{2ikm}\,+\,\bar{\rho}\,.
\end{align}
Inserting this result in the formula (\ref{eq:QB_R}) for the boundary charge $Q_B\equiv Q_B^R$, we get \begin{align}
\label{eq:QB_1}
Q_B = -{1\over 4\pi}\int_{-k_F}^{k_F}dk \sum_{m=-\infty}^\infty e^{2ikm} f(m)\,+\,{k_F\over 2\pi}\,.
\end{align}
Choosing $f(m)=e^{-\eta |m|}$, we find $\sum_{m=-\infty}^\infty e^{2ikm} f(m) = \pi\delta(k)$ and obtain for the boundary charge of $H_0$ at zero gap
\begin{align}
\label{eq:QB_zero_gap}
Q_B = -{1\over 4} + {1\over 2}\bar{\rho}\,. 
\end{align}

This single-band model can be differently represented in terms of uniform unit cells with $Z$ sites. This is especially useful, if we have in mind to add a $Z$-periodic perturbation on top of $H_0$ \eqref{eq:H_0}. In the new representation, the single cosine band folds into $Z$ bands with the reduced Brillouin zone (RBZ) $[-\frac{\pi}{Z}, \frac{\pi}{Z})$, the adjacent bands touching each other either in the center or at the edges of the RBZ. Choosing $k_F/\pi$ of the original model to be rational, $\frac{k_F}{\pi} = \frac{\nu}{Z}$, we occupy $\nu$ bands in the folded representation, and \eqref{eq:QB_zero_gap} then reads
\begin{align}
\label{eq:QB_zero_gap_folded}
Q_B = -{1\over 4} +  \frac{\nu}{2 Z}\,. 
\end{align}

Adding a $Z$-periodic perturbation generically opens $Z-1$ gaps between all $Z$ bands. Having the chemical potential in the $\nu$th gap, we can evaluate the correction to \eqref{eq:QB_zero_gap_folded} due to the perturbation by means of the low-energy theory developed in Appendix \ref{app:QB_dirac}.

This consideration clarifies the physical meaning of Eq.~ \eqref{eq:QB_dirac_exact}.

\section{Interface charge for Dirac model}
\label{app:QI_dirac}

In this Appendix we consider an interface between two Dirac models according to the Hamiltonian (\ref{eq:H_eff_interface}), where the phase $\alpha(x)$ of the gap parameter depends on $x$. We will prove the Goldstone-Wilczek formula (\ref{eq:DQ5_II}) for the interface charge for the particular choice $\alpha(x)=\alpha_{R}\Theta(x)+\alpha_{L}\Theta(-x)$. We define the parameter $\delta\alpha=\alpha_R-\alpha_L$. The eigenstates follow from the equation
\begin{align}
[-iv_{F}\sigma_{3}\partial_{x} + |\Delta| (\sigma_{+}e^{i\alpha(x)}+\sigma_{-}e^{-i\alpha(x)})]\uline{\psi}(x)=\epsilon\uline{\psi}(x),
\end{align}
with $\sigma_\pm={1\over 2}(\sigma_x\pm i\sigma_y)$ and $\uline{\psi}(x)^T=\begin{pmatrix}R(x), & L(x)  \end{pmatrix}^{T}$.
Like in the case of the semi-infinite Dirac model discussed in Appendix~\ref{app:QB_dirac}, we separate the spectrum of the Hamiltonian into two regions: I) $|\epsilon|< |\Delta |$, and II) $|\epsilon|> |\Delta |$. 

The bound state solution appears for $\sin(\delta\alpha/2)>0$ with energy $\epsilon = - |\Delta | \cos (\delta \alpha/2)$, and is given by
\begin{align}
    \uline{\psi}^{\text{I}} (x) = \sqrt{\frac{\kappa}{2}} \left( \begin{array}{c} 1 \\ - e^{-i \frac{\alpha_R+\alpha_L}{2}} \end{array} \right) e^{- \kappa |x|} ,
    \label{eq:G17}
\end{align}
with $\kappa = \frac{|\Delta |}{v_F} \sin(\delta\alpha/2)$. If it is occupied it gives an integer contribution to the interface charge.

For each energy $|\epsilon_k | > |\Delta |$, the extended eigenstates can be chosen as scattering states within two scattering channels. The first one (denoted by the index $r$)
\begin{align}
\uline{\psi}_{k}^{(r)}(x)& =\frac{\Theta(-x)}{\sqrt{2\pi}}[\uline{\chi}_{L,k}e^{ikx}+r_{k}\,\uline{\chi}_{L,-k}e^{-ikx}] \nonumber  \\
\label{eq:G3}
&+\frac{\Theta(x)}{\sqrt{2\pi}}t_{k}\,\uline{\chi}_{R,k}e^{ikx}
\end{align}
represents the scattering of a wave incident on the interface from the left.
The second scattering eigenstate (denoted by the index $l$)
\begin{align}
\uline{\psi}_{k}^{(l)}(x) &=\frac{\Theta(x)}{\sqrt{2\pi}}[\uline{\chi}_{R,-k}e^{-ikx}+r_{k}'\,\uline{\chi}_{R,k}e^{ikx}] \nonumber \\
\label{eq:G5}
&+\frac{\Theta(-x)}{\sqrt{2\pi}} t_{k}'\,\uline{\chi}_{L,-k}e^{-ikx}
\end{align}
represents the scattering of a wave incident on the interface from the right.
In above expressions, $k\in[0,\infty)$ stands for the  momentum quantum number, and 
\begin{align}
\label{eq:G6}
&\uline{\chi}_{R/L, k}=\frac{1}{\sqrt{2 \epsilon(\epsilon-v_{F}k)}}\begin{pmatrix}-|\Delta | e^{i\alpha_{R/L}} \\ 
v_{F}k-\epsilon \end{pmatrix}
\end{align}
are the normalized Bloch eigenstates of the right-sided ($x>0$) and left-sided ($x<0$) bulk Hamiltonians with eigenenergies $\epsilon=\pm\epsilon_{k}$.

\begin{figure}
\centering
 \includegraphics[width=0.49\columnwidth]{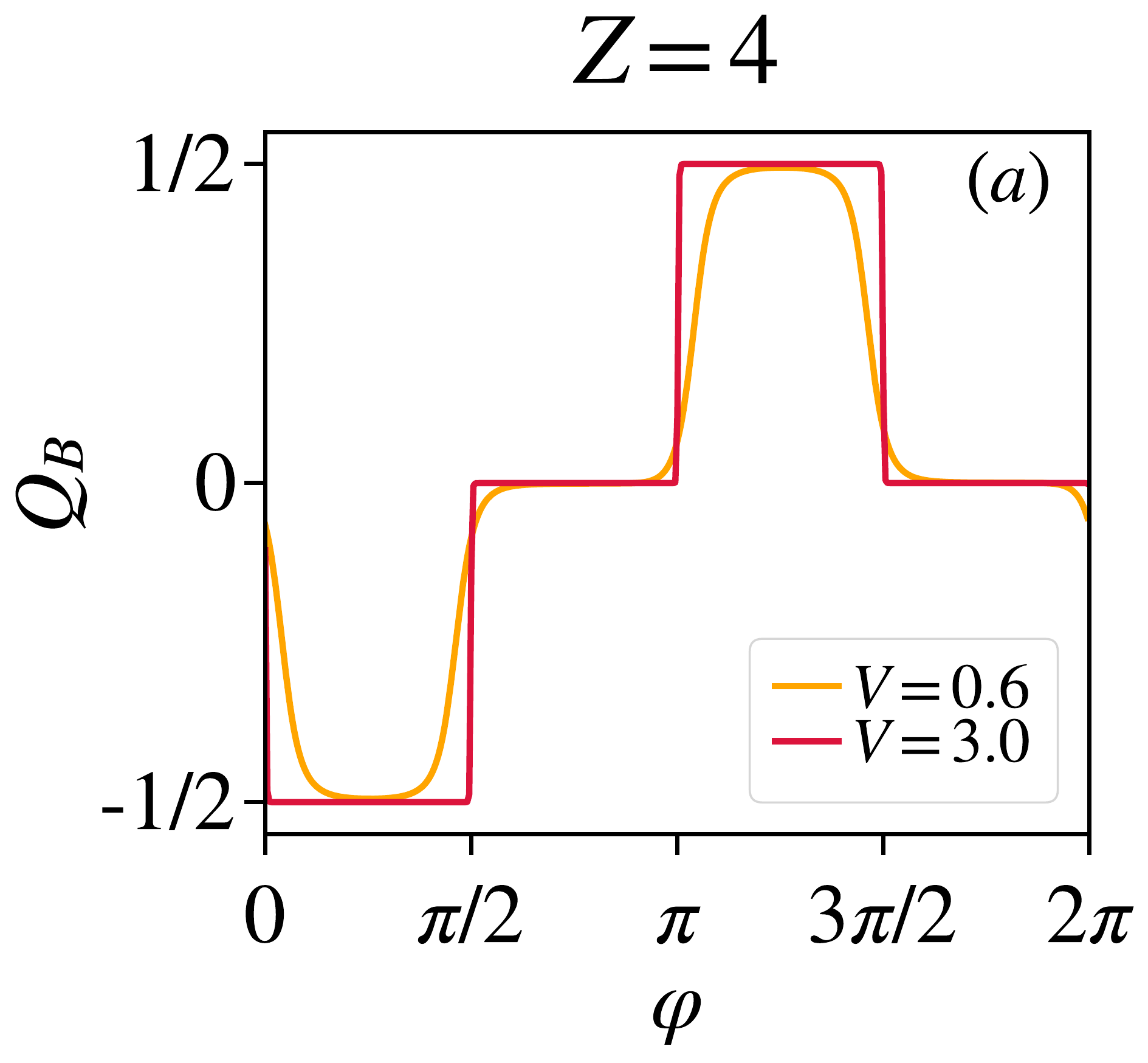}
 \includegraphics[width=0.49\columnwidth]{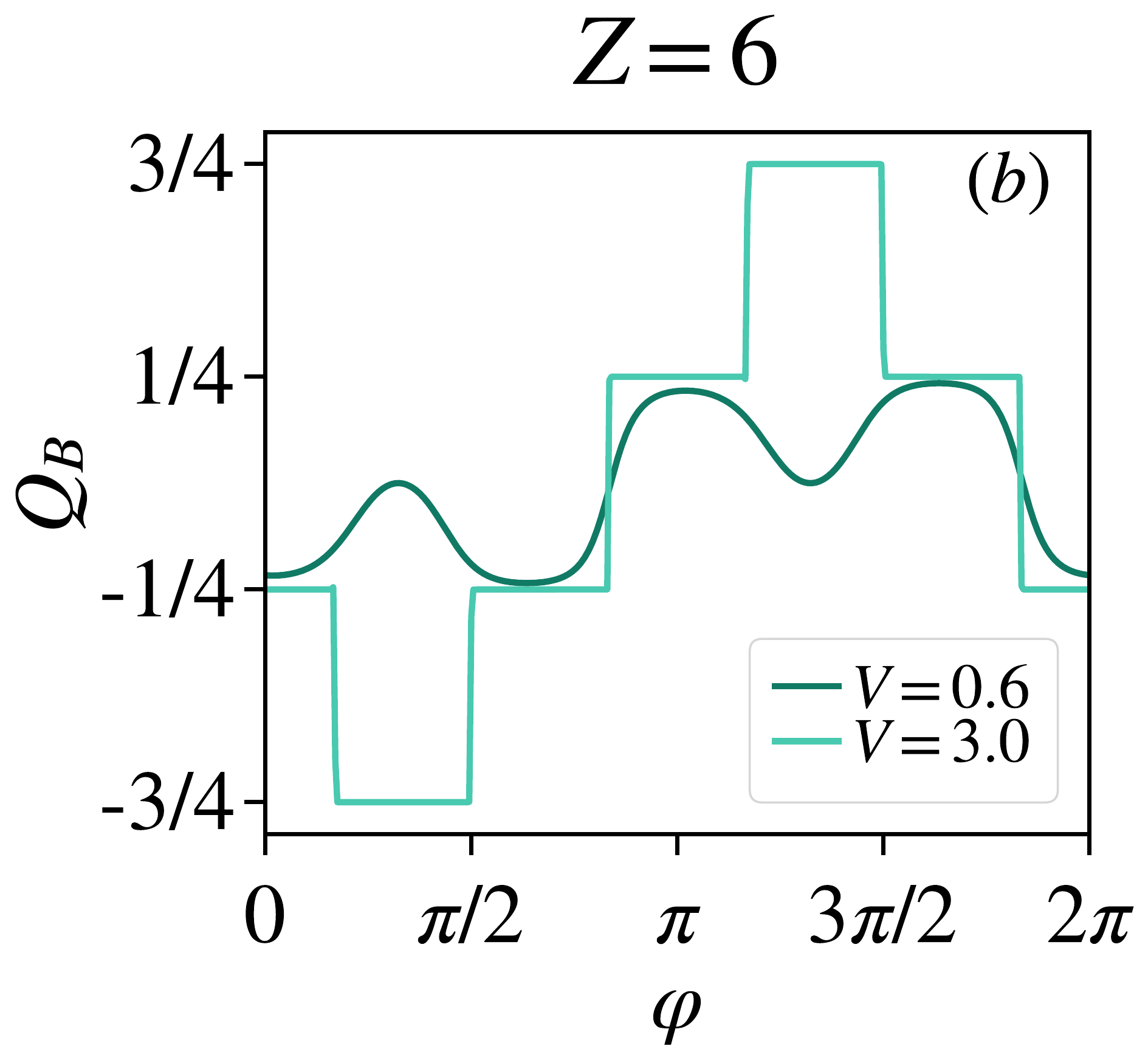}
 \caption{The same as the center row of Fig.~\ref{fig:weyl}, but for small $N=24$ and two values of the $V$.}
\label{fig:QB_small}
\end{figure}

The scattering amplitudes $r_{k}, t_{k}$ and $r_{k}', t_{k}'$ can be determined from the continuity condition at the interface
\begin{align}
\label{eq:G7}
&\uline{\psi}_{k}^{(r/l)}(0^{+})=\uline{\psi}_{k}^{(r/l)}(0^{-}).
\end{align}
This results in the expressions
\begin{align}
\label{eq:G8}
&r_{k}=r_{k}'=\frac{|\Delta | (e^{i\alpha_{R}}-e^{i\alpha_{L}})}{(\epsilon-v_{F}k)
e^{i\alpha_{L}}-(\epsilon+v_{F}k)e^{i\alpha_{R}}},\\
\label{eq:G9}
&t_{k}=\frac{2v_{F}ke^{i\alpha_{L}}}{(\epsilon+v_{F}k)e^{i\alpha_{R}}-(\epsilon-v_{F}k)e^{i\alpha_{L}}},\\
\label{eq:G10}
&t_{k}'=\frac{2v_{F}ke^{i\alpha_{R}}}{(\epsilon+v_{F}k)e^{i\alpha_{R}}-(\epsilon-v_{F}k)e^{i\alpha_{L}}}.
\end{align}
By an explicit calculation one can readily verify the fulfillment of the unitarity conditions
\begin{align}
\label{eq:G11}
&|t_{k}|^{2}+|r_{k}|^{2}=|t'_{k}|^{2}+|r_{k}'|^{2}=1,\\
\label{eq:G12}
&t^{*}r_{k}'+r_{k}^{*}t_{k}'=0.
\end{align}

For a filled valence band we choose $\epsilon=-\epsilon_{k}$ and identify the extended states' contribution to the interface charge
\begin{align}
Q_{I}^{\textup{II}}&=\int_{-\infty}^{\infty}dx f(x) \nonumber \\
& \times \int_{0}^{\infty}dk\Bigg(|\psi^{(r)}_{k}(x)|^{2} +|\psi^{(l)}_{k}(x)|^{2}-\frac{2}{\pi}\Bigg) \nonumber \\
&=\frac{|\Delta|}{\pi}\int_{0}^{\infty}dxf(x)\int_{-\infty}^{\infty}dk\frac{e^{2ikx}}{\epsilon_{k}}r'_{k}.
\label{eq:G15}
\end{align}
As one can conclude from \eqref{eq:G8}, this quantity periodically depends on $\delta \alpha = \alpha_R - \alpha_L$. 
Evaluating  \eqref{eq:G15} for $\delta \alpha \in (0, 2 \pi)$ with $f(x)=e^{-\eta |x|}, \ \eta\to 0^+,$ we obtain 
\begin{align}
    \label{QII_inter}
    Q_{I}^{\textup{II}} = \frac{\delta \alpha}{2 \pi}-1 .
\end{align}
Putting the chemical potential at the bottom of the conduction band, we receive an additional contribution $Q_I^{\textup{I}} =1$ from the edge state \eqref{eq:G17}, which is present for every value of $\delta \alpha$, and obtain the resulting expression \eqref{eq:DQ5_II} for the total interface charge.

\section{Finite smaller systems}
\label{app:finite_dot_array}

In this Appendix we show that the quantization of the boundary charge according to Fig.~\ref{fig:weyl} is already visible for a tight-binding chain of $\sim 20$ sites. As demonstrated in Fig.~\ref{fig:QB_small} for $N=24$ lattice sites the quantization can be demonstrated robustly as long as larger $V$ can be accessed such that the localization length becomes small compared to the lattice size. 

The results shown in Figs.~\ref{fig:weyl} and \ref{fig:QB_small} can be easily understood in the atomic limit $V \gg t$: The dominant contribution to $Q_B$ comes from the polarization charge $Q_P$ \eqref{eq:QP}, while an eventual integer-valued Friedel charge contribution \eqref{eq:QF} is exactly cancelled by edge state contributions. To compute $Q_P$, we use the elaborated expression (see Ref.~\cite{pletyukhov_etal_long} for details)
\begin{align}
    Q_P = - \frac{1}{Z} \sum_{\alpha=1}^{Z/2} \sum_{j=1}^Z j \left( |\chi^{(\alpha)} (j)|^2 - \frac{1}{Z} \right),
    \label{eq:QPP}
\end{align}
where the occupied bands $\epsilon^{(\alpha)}$ are approximately given by the potential components $v_{\tilde{j}} < 0$ (one can even associate $\tilde{j}$ with the band index $\alpha$ sorting $v_{\tilde{j}}$'s in the ascending order for each value of $\varphi$). The corresponding eigenstate $\chi^{(\alpha)} (j)$ possesses the only unity component $\chi^{(\alpha)} (\tilde{j}) =1$, while $\chi^{(\alpha)} (j \neq \tilde{j}) =0$. The plateau values in the discussed figures then immediately follow from \eqref{eq:QPP}.
[It can so happen that two eigenstates $v_{\tilde{j}_1} (\varphi)$ and $v_{\tilde{j}_2} (\varphi)$ become degenerate at some value of $\varphi$, and then it is necessary to consider $\frac{1}{\sqrt{2}} \{\chi^{(\alpha_1)} (j) \pm \chi^{(\alpha_2)} (j)\}$ for the eigenstates. This, however, does not alter the plateau value of $Q_B$.]

\end{appendix}


\begin{thebibliography}{99}
\bibitem{com_CDW_phase}
We note that $\varphi$ is the precise definition of the phase of the CDW, whereas $\alpha$ is the phase of the complex gap parameter entering the Dirac model. In general they are different. Away from half-filling, they are related by $\alpha=\pm\nu\varphi + \text{const}$, see Eq.~(\ref{eq:alpha_varphi}). However, at half-filling, $\alpha$ has a different meaning, it can be even pinned independent of $\varphi$, see Eq.~(\ref{eq:alpha_half_filling}).
\bibitem{com_LR}
In Ref.~\cite{pletyukhov_etal_long} it was shown that $Q_{B,0}^R+Q_{B,Z-1}^L=-1+{1\over Z}$. In addition, it was shown that $Q_{B,Z-1}^L=Q_{B,0}^L-{Z-1\over Z}-w$, where $w$ is the winding number of the phase difference of the Bloch wave function between the first and last site of the unit cell. This gives the result $Q_{B,0}^R+Q_{B,0}^L=w$.
\end{thebibliography}
\end{document}